\documentclass[pasp,numberedappendix]{emulateapj}

\usepackage{natbib}
\usepackage{graphicx}
\usepackage{hyperref}
\usepackage{amssymb}
\usepackage{amsmath}
\usepackage{xspace}
\usepackage{color}
\usepackage{longtable}
\usepackage{multirow}
\usepackage{listings}

\newcommand{\gws}{gravitational waves\xspace}
\newcommand{\gw}{gravitational wave\xspace}
\newcommand{\lalinf}{\texttt{LALInference}\xspace}
\newcommand{\lppenf}{{\scriptsize \texttt{lalapps\_pulsar\_parameter\_estimation\_nested}}\xspace}
\newcommand{\lppen}{\texttt{lppen}\xspace}
\newcommand{\lppef}{{\scriptsize \texttt{lalapps\_pulsar\_parameter\_estimation}}\xspace}
\newcommand{\lppe}{\texttt{lppe}\xspace}
\newcommand{\ee}[1]{\!\times\!10^{#1}}

\DeclareMathOperator{\erf}{erf}

\begin{document}

\title{A nested sampling code for targeted searches for continuous gravitational waves from pulsars
}
\begin{abstract}This document describes a code to perform parameter estimation and model selection in targeted
searches for continuous gravitational waves from known pulsars using data from ground-based \gw detectors.
We describe the general workings of the code and characterise it on simulated data containing both noise and
simulated signals. We also show how it performs compared to a previous MCMC and grid-based
approach to signal parameter estimation. Details how to run the code in a variety of
cases are provided in Appendix~\ref{app:usage}.\end{abstract}
\author{
M.~Pitkin,\altaffilmark{1}
M.~Isi,\altaffilmark{2}
J.~Veitch,\altaffilmark{1,3}
and G.~Woan\altaffilmark{1}
}

\altaffiltext{1}{SUPA, School of Physics \& Astronomy, University of Glasgow, Glasgow, G12 8QQ, United Kingdom}
\altaffiltext{2}{LIGO Laboratory, California Institute of Technology, Pasadena, CA 91125, USA}
\altaffiltext{3}{School of Physics \& Astronomy, University of Birmingham, Birmingham, B15 2TT, United Kingdom}

\email{matthew.pitkin@glasgow.ac.uk}

\date{\today}

\preprint{}

\maketitle

\section{Overview}

There are a variety of sources of \gws that are potentially observable using the current ground-based \gw detectors \citep{2015CQGra..32g4001L,2015CQGra..32b4001A}.
These include transient signals, such as those already observed from the merger of binary black hole systems \citep{2016PhRvL.116f1102A,2016PhRvL.116x1103A}, and also
continuous quasi-monochromatic signals, such as expected from assymetrically deformed rotating neutron stars \citep[see, e.g., Section~III of][]{2004PhRvD..69h2004A}.
Here we will describe a code used in searches for continuous \gw signals, in particular signals from sources that are seen as pulsars via electromagnetic observations.

One of the primary methods used in searches for \gws from known pulsars is based on two stages: a heterodyne
stage that pre-processes the calibrated \gw detector data by removing a signal's expected intrinsic phase
evolution (based on the observed solution from electromagnetic observations), whilst also low-pass filtering and
down-sampling (via averaging) the heterodyned data \citep{2005PhRvD..72j2002D}; and, a parameter estimation stage that takes the
processed data and uses it to estimate posterior probability distributions of the unknown signal parameters
\citep[e.g., using a Markov chain Monte Carlo (MCMC)][]{2010ApJ...713..671A}. This method has been variously
called the {\it Time Domain Method}, the {\it Heterodyne Method}, the {\it Bayesian Method}, the {\it Glasgow
Method}, or some combination of these names. Up to now this method has only been used to set upper limits
on signal amplitudes, but has included no explicit criteria for deciding on signal detection/significance. A further
limitation has been that the MCMC method used was inefficient\footnote{MCMC methods are not intrinsically inefficient for
searches such as our, but the particulat implementation we had used was not specically designed to be efficient at sampling the
given posterior volumes.} and required a lot of tuning. It also
did not straightforwardly have the ability to perform a search on the data over small ranges in the phase
evolution parameters, which is required if the \gw signal does not well match the known pulsar's phase evolution. A nested sampling
algorithm \citep{Skilling:2006}, in particular a method based on that described in \citet{Veitch:2010}, has been provided within
the \lalinf software library \citep{2015PhRvD..91d2003V} in \citet{lalsuite}. This has been used for parameter estimation and model
selection for both modelled transient sources, such as the observed black hole mergers \citep{2016PhRvL.116x1102A}, and unmodelled transient signal
searches \citep{2015arXiv151105955L}. Due to the above reasons, and to make use of the existing and well used library functions, the code has been
re-written to use the available nested sampling algorithm. This allows us to evaluate the {\it evidence} or {\it
marginal likelihood} for the signal model and compare it to other model evidences (i.e.\ the data is consistent with being just
Gaussian noise), whilst also giving posterior probability distributions for the signal parameters.

This code is called \lppenf (or \lppen for short through the rest of this document). For more detailed descriptions of how the algorithm works we refer
to \citet{Veitch:2010} and \citep{2015PhRvD..91d2003V}, whilst here we will provide some information on the specific proposal
distributions that can be used within the algorithm. This method has previously been briefly described in \citet{2012JPhCS.363a2041P}.
The previous code, called \lppef (or \lppe for short), used in, e.g., \citet{2010ApJ...713..671A}, could perform posterior evaluation
using two methods: an MCMC method, or, by gridding up the parameter space and explicitly evaluating the posterior at each
grid point. In \S\ref{sec:codeeval} we show comparisons of the various methods mainly to check \lppen for consistency.

The code can also be used to perform parameter estimation and model selection for signals for any generic
metric theory of gravity, however its use for this is discussed in more detail in a separate paper \citep{MaxCWpolariations}. When
searching over parameters that vary the phase evolution there is potential to speed up the analysis through
efficient likelihood evaluation via the {\it reduced order quadrature} method \citep[see, e.g.,][]{2014PhRvX...4c1006F,2015PhRvL.114g1104C},
but again that will be discussed in more detail in a future paper.

\subsection{Background knowledge}\label{sec:general}

The code calculates the Bayesian evidence, or {\it marginal likelihood}, for a particular signal model under
a set of assumptions. The evidence, $\mathcal{Z}$, for a given model, $M$, defined by a set of parameters, $\vec{\theta}$, is given by
\begin{equation}\label{eq:evidence}
\mathcal{Z}_M = p(d|M,I) = \int^{\vec{\theta}} p(d|\vec{\theta},M,I)p(\vec{\theta}|M,I) {\rm d}\vec{\theta},
\end{equation}
where $p(d|\vec{\theta},M,I)$ is the likelihood function for the data $d$ given the
model and its set of defining parameters, $p(\vec{\theta}|M,I)$ is the prior probability distribution on
$\vec{\theta}$, and $I$ represents any additional prior assumptions. During nested sampling this integral is evaluated by transforming it into the one dimensional
sum
\begin{equation}\label{eq:nestedsampev}
\mathcal{Z}_M = \sum_i^N p(d|\vec{\theta}_i,M,I) w_i,
\end{equation}
where $w_i = p(\vec{\theta}_i|M,I) \Delta\vec{\theta}_i$ is the ``prior weight''
(i.e.\ the fraction of the prior occupied by point $i$).

By default the signal model evidence is compared to the evidence that the data consists of only Gaussian noise to form the odds for the two models
\begin{align}\label{eq:oddsratio}
\mathcal{O} &= \frac{p(d|M,I)}{p(d|\text{noise},I)}\frac{p(M|I)}{p(\text{noise}|I)} \nonumber \\
&= \frac{\mathcal{Z}_M}{\mathcal{Z}_{\text{noise}}},
\end{align}
where on the right hand side we have explicitly set the prior odds for the two models to
$p(M|I)/p(\text{noise}|I) = 1$.

Other than this evidence value and odds, the code also returns the samples accumulated during
the nested sampling process. These samples are not drawn from the posterior probability distribution as in
an MCMC method, but they can be resampled to generate a subset of samples that are drawn from the posterior
distribution. This resampling \citep[performed using the {\tt lalapps\_nest2pos} {\tt python} script within
{\tt lalapps}][]{lalsuite} uses the
value of $L_i w_i$ for each point, normalised by $(L_iw_i)_{\rm max}$, to give values proportional to the
posterior probability, and then accepts a point with a probability given by its value, i.e.\ a point is
accepted with the probability
\begin{equation}\label{eq:postaccept}
P_{\text{accept}} = \frac{L_i w_i}{(L_iw_i)_{\rm max}}.
\end{equation}

In our case the data $d$ in equation~\ref{eq:oddsratio} is the heterodyned, low-pass filtered and down-sampled data
from a detector, or set of detectors, each giving a single data stream or two data streams depending on
whether the heterodyne was performed at one or both potential signal frequencies near the rotation frequency
and/or twice the rotation frequency. Here we will use $\mathbf{B}$ to represent a vector of these heterodyned
data values \citep{2005PhRvD..72j2002D}, for which a single time sample is often referred to using the jargon term $B_k \equiv B(t_k)$ (``{\it B of k}''), although we will
not consistently use $k$ as the time index for the data. Throughout this document if we refer to ``heterodyned'' data we mean
data that has been processed, or simulated as if processed, by this method.

\section{Core functions}

Here we will describe the core parts of the code defining the signal model and the probability functions used
for the inference. These assume that the data comprises a number of calibrated narrow-band complex-heteroyned time
series. These time series data streams can be from different detectors, and/or at different heterodyne
frequencies. For example you could have a pair of times series from both the LIGO Hanford (H1) and LIGO
Livingston (L1) detectors, with each pair containing one produced with a heterodyne at twice the rotation frequency of a known
pulsar and the other produced with a heterodyne at the rotation frequency.

\subsection{The signal model}\label{sec:model}

Our code assumes that the electromagnetically observed rotational phase evolution of a pulsar is represented by its Taylor expansion
\begin{equation}
\phi(t) = 2\pi\left(fT(t) + \frac{1}{2}\dot{f}T(t)^2 + \frac{1}{6}\ddot{f}T(t)^3 \ldots \right)
\end{equation}
where $T$ is the time, corrected to an inertial reference frame (the solar system barycentre
for isolated pulsars, or the binary system barycentre for pulsars in binaries or multiple component systems), and the $f$'s give
the observed rotation frequency and its time derivatives. The value of $T = (t+\tau(t)-t_0)$ where $t$ is the
time at a detector, $\tau(t)$ is the time dependent correction to the inertial frame and $t_0$ is the epoch.
The code can accept frequency derivatives of any order. We assume the
calibrated detector data, $d(t)$, has been heterodyned such that
\begin{equation}
B'(t) = d(t)e^{-i\Phi_{i,\text{het}}(t)},
\end{equation}
where $\Phi_{i,\text{het}}(t)$ is the assumed phase evolution for a given data stream $i$, and we produce $B(t)$ by
low-pass filtering and resampling (via averaging) values of $B'(t)$.

Under the standard assumption that the general theory of relativity (GR) is correct, the code uses the most general form of
the signal model defined in Equations~51--54 of \citet{2015arXiv150105832J}, which, when heterodyned and assuming low-pass
filtering, gives a signal at a pulsar's rotation frequency (where $\Phi_{1,{\text{het}}}(t) = \phi(t)$) of
\begin{widetext}
\begin{equation}\label{eq:hf}
h_f(t) =  e^{i\Delta\phi_1(t)}\left(-\frac{C_{21}}{4}F^D_{+}(\psi,t)\sin{\iota}\cos{\iota}\,e^{i\Phi_{21}^C} +
i\frac{C_{21}}{4}F^D_{\times}(\psi,t)\sin{\iota}\,e^{i\Phi_{21}^C} \right)
\end{equation}
and at twice the pulsar's rotation frequency (where $\Phi_{2,{\text{het}}}(t) = 2\phi(t)$) of
\begin{equation}\label{eq:h2f}
h_{2f}(t) =  e^{i\Delta\phi_2(t)}\left(-\frac{C_{22}}{2}F^D_{+}(\psi,t)[1+\cos{}^2\iota]e^{i\Phi_{22}^C} +
iC_{22}F^D_{\times}(\psi,t)\cos{\iota}\,e^{i\Phi_{22}^C} \right).
\end{equation}
\end{widetext}
The $F^D_{+}$ and $F^D_{\times}$ values are the detector ($D$) dependent antenna patterns (see, e.g., \citet{1978PhRvD..17..379F,1987MNRAS.224..131S,2009PhRvD..79b2002F}, or
\citet{1998PhRvD..58f3001J} for the convention used within this document in, e.g., equation~\ref{eq:antenna}), which are a function of the
detector position, source sky position and source polarisation angle $\psi$ (in equations~\ref{eq:hf} and \ref{eq:h2f} we only
explicitly note the $\psi$ dependence as for our sources the sky location is known). The $C_{21}$, $C_{22}$,
$\Phi_{21}^C$ and $\Phi_{22}^C$ values are convenient ways of representing the waveform in terms of an
amplitude and phase of the signal for the $l=2$, $m=1$ harmonic and $l=m=2$ harmonic respectively. The
$\Delta\phi(t)$ values represent any time dependent phase deviation between the phase used in the heterodyne
and the true signal phase \citep[which does not necessarily have to precisely follow the electromagnetically observed rotational phase, see discussions in, e.g.,][]{2008ApJ...683L..45A}, so
$\Delta\phi_1(t) = (\phi_{1,{\text{true}}}(t)-\Phi_{1,{\text{het}}}(t))$ and $\Delta\phi_2(t) = (\phi_{2,{\text{true}}}(t)-\Phi_{2,{\text{het}}}(t))$. More generally,
for emission at an arbitrary scaling of the rotation frequency, $\mathcal{K}$ (where $\Phi_{\mathcal{K},{\text{het}}}(t) = \mathcal{K}\phi(t)$), an even more generic signal model (still assuming GR is correct) would be
\begin{equation}\label{eq:hkf}
h_{\mathcal{K}f}(t) =  \frac{e^{i\Delta\phi_{\mathcal{K}}(t)}}{2}\left(H_+F^D_{+}(\psi,t)e^{i\Phi_0} +
iH_{\times}F^D_{\times}(\psi,t)e^{i\Phi_{0}} \right),
\end{equation}
where $H_+$ and $H_{\times}$ are the amplitude components for the `$+$' and `$\times$' polarisations, and $\Phi_0$ is the signal's initial phase.

To calculate the $\Delta\phi_{\mathcal{K}}$ values using up to the $(n-1)^{\text{th}}$ frequency derivative, and to avoid numerical overflow issues when dealing
with large phases, we use
\begin{widetext}
\begin{equation}\label{eq:deltaphi}
\Delta\phi_{\mathcal{K}}(t) = 2\pi \mathcal{K} \sum_{k=1}^n \left( \frac{\left(f^{(k-1)}_{{\text{true}}} - f^{(k-1)}_{{\text{het}}}\right)}{k!}(t+\delta t_{\text{het}})^k + \frac{f^{(k-1)}_{{\text{true}}}}{k!} 
\sum_{i=0}^{i<k} \left(\begin{array}{c}k \\ i\end{array} \right) (\delta t_{\text{true}}-\delta t_{\text{het}})^{k-i} (t+\delta t_{\text{het}})^i \right),
\end{equation}
\end{widetext}
where $f^{(n)}$ is the $n^{\text{th}}$ frequency derivative (for both the `true' and `het' the frequency/frequency dervitative values used here
are those prior to scaling by $\mathcal{K}$ for a given signal model), and $\delta t$ is the combination of any solar system barycentring and binary system barycentring
time delays.

By default the code assumes emission just from the $l=m=2$ mode, i.e.\ there is only a signal at twice the
rotation frequency. In this case $C_{22}$ and $\Phi_{22}^C$ can be related to the more familiar physical
$h_0$ and $\phi_0$ values via $h_0 = -2C_{22}$ \citep[where the minus sign maintains consistency of equation~\ref{eq:h2f} with the form given in][]{1998PhRvD..58f3001J} and pulsar rotational 
phase $\phi_0 = \Phi_{22}^C/2$. For the more general case the
relations between the waveform amplitude and phase parameters and physical source parameters are given in
\citet{2015arXiv150105832J} and \citet{2015MNRAS.453.4399P}. In general, for previous searches we have often assumed that we track the true
signal phase perfectly with the heterodyne, and as such $\Delta\phi_{\mathcal{K}}(t) = 0$. In such cases the only time
varying components of the signal are the antenna pattern functions, which allows great speed increases in the
signal generation and likelihood calculations (see \S\ref{sec:fastlike}).

\subsubsection{General tensor-vector-scalar signal model}

A more generic {\it heterodyned} model, assuming gravity is described by an arbitrary metric theory, but not necessarily GR, is given by
\citep[see, e.g.][]{2015PhRvD..91h2002I,MaxCWpolariations}
\begin{widetext}
 \begin{equation}\label{eq:hnongr}
    h_{\mathcal{K}f}^{\text{tvs}}(t) =  \frac{e^{i\Delta\phi_{\mathcal{K}}(t)}}{2}\left[ e^{i\Phi_{\text{t}}}\left(F_+^DH_+ + F_{\times}^DH_{\times}e^{i\psi_{\text{t}}}\right)
    + e^{i\Phi_{\text{v}}}\left(F_{\text{x}}^DH_{\text{x}} + F_{\text{y}}^DH_{\text{y}}e^{\psi_{\text{v}}}\right) + e^{i\Phi_{\text{s}}}\left(F_{\text{b}}^DH_{\text{b}} + F_{\text{l}}^DH_{\text{l}}e^{i\psi_{\text{s}}} \right)\right],
 \end{equation}
\end{widetext}
where $H_{p}$ (for $p \in \{+,\times,\text{x},\text{y},\text{b},\text{l}\}$) are the amplitudes for the tensor (`$+$' and `$\times$'), vector (`x' and `y'),
and scalar (`b' and `l') \gw amplitude components, $F_{p}$ are their associated time dependent antenna responses
\citep[see, e.g.][]{2009PhRvD..79h2002N,2012PhRvD..85d3005B,2015PhRvD..91h2002I}, and $\Phi_{\text{t,v,s}}$ and $\psi_{\text{t,v,s}}$ are angles related to the initial
phase of, and rotation within, the tensor, vector and scalar components. For networks of quadrupolar antennas, like a standard \gw interferometer, the two scalar components are entirely
degenerate \citep{2014LRR....17....4W}, and as such only one of them needs to be defined \citep[e.g., only the `b' mode is used in Equation~19 of][]{MaxCWpolariations}. 
When searching for a generic signal such as this the orientation of the wave-frame (given by $\psi$ and $\iota$) used to define the polarisation modes is arbitrary:
one is free to pick any right-handed frame one wants and define the modes there, but we usually make a choice that we know simplifies things for a given signal.
Hence, for a generic search, we are free to arbitrarily set our frame such that $\psi=0$ and the values $F_{p}$ given in equation~\ref{eq:hnongr} can be defined
by, e.g., equations~24--29 of \citet{2015PhRvD..91h2002I}. However, when we have a specific prediction of what a given waveform should look like, it is always in a
{\it particular} frame. By convention, this is usually a frame in which the x-axis is aligned with the intersection of the equatorial plane of the source with the plane
of the sky (though the convention is not universal). In other words, if a theory predicts certain amounts of polarisation with some specific phase evolution (say,
that given by GR) then that is a frame dependent statement, and $\psi$ in required to orient the model frame accordingly. In such a case \citep[e.g., GR or some other
theory such as G4v in][]{2015arXiv150304866M} the antenna patterns could be obtain from those with $\psi=0$ using equations~30--35 (and $\psi_{\text{t}}$ and $\psi_{\text{v}}$
from equations 46--47, and 49--50, for GR and G4v respectively) of \citet{2015PhRvD..91h2002I} would be required.

\subsection{The likelihood functions}\label{sec:likelihood}

Our code can make use of two different likelihood functions. The default is a
Student's {\it t}-likelihood function in which it is assumed that the standard deviation of the noise in the
data is unknown, but can be marginalised over. A Gaussian likelihood function can also be used, for
which the code can either take in estimates of the noise standard deviation at each data point, or calculate
these internally based on stationary stretches of data. For the Student's {\it t}-likelihood function, and if
calculating noise standard deviations for the Gaussian likelihood function internally, the code needs to break
up the data into segments for which the noise can be assumed to have been drawn from independent Gaussian distributions.
The method for doing this is given in \S\ref{sec:splitting}.

\subsubsection{Student's {\it t}-likelihood}\label{sec:stlikelihood}

A full derivation of the Student's {\it t}-likelihood function \citep[see, e.g.,][]{2005PhRvD..72j2002D} is given
in Appendix~\ref{app:likelihood}, but the final form of the joint likelihood (and its natural logarithm,
which is actually used within the code to maintain precision) for multiple detectors and data streams is
given by
\begin{widetext}
\begin{align}\label{eq:stlikelihood}
p(\mathbf{B}|\vec{\theta}) &= \prod_{i=1}^{N_{\text{dets}}} \prod_{j=1}^{N_{\text{s}}} \prod_{k=1}^{M_{i,j}}
\frac{(m_{i,j,k}-1)!}{2\pi^{m_{i,j,k}}}
\left(\sum_{n=n_{i,j,0}}^{n_{i,j,0}+(m_{i,j,k}-1)} |B_{i,j,n}-y(\vec{\theta})_{i,j,n}|^2\right)^{-m_{i,j,k}},
\nonumber \\
\ln{p(\mathbf{B}|\vec{\theta})} &= \sum_{i=1}^{N_{\text{dets}}} \sum_{j=1}^{N_{\text{s}}}
\sum_{k=1}^{M_{i,j}} \left( \mathcal{A}_{i,j,k} - m_{i,j,k}\ln{
\left\{\sum_{n=n_{i,j,0}}^{n_{i,j,0}+(m_{i,j,k}-1)} |B_{i,j,n}-y(\vec{\theta})_{i,j,n}|^2\right\}}
\right),
\end{align}
\end{widetext}
where $N_{\text{dets}}$ is the number of detectors used, $N_{\text{s}}$ is the number of data streams (e.g.,
heterodyned data from both the rotation frequency and twice the rotation frequency) per detector, $M_{i,j}$ is
the total number of independent data chunks for detector $i$ and data stream $j$ with lengths $m_{i,j,k}$, and
$n_{i,j,0} = \sum_{l=1}^{k} 1+m_{i,j,l-1}$ (with $m_{i,j,0} = 0$) being the index of the first data point in
each chunk. The signal model $y(\vec{\theta})$ is that given by Eqns.~\ref{eq:hf} and/or \ref{eq:h2f} (or Eqn.~\ref{eq:hkf})
depending on which data streams are being analysed. For notational convenience we have made the substitution
$\mathcal{A}_{i,j,k} = \ln{\left([m_{i,j,k}-1]!\right)} - \ln{2} - m_{i,j,k}\ln{\pi}$.

\subsubsection{Gaussian likelihood}\label{sec:glikelihood}

The Gaussian likelihood, and its natural logarithm, are similarly given by
\begin{widetext}
\begin{align}\label{eq:gausslikelihood}
p(\mathbf{B}|\vec{\theta}) &= \prod_{i=1}^{N_{\text{dets}}} \prod_{j=1}^{N_{\text{s}}} \prod_{k=1}^{L_{i,j}}
\frac{1}{2\pi\sigma_{i,j,k}^2}\exp{\left(-\frac{|B_{i,j,k}-y(\vec{\theta})_{i,j,k}|^2}{2\sigma_{i,j,k}^2}
\right)}, \nonumber \\
\ln{p(\mathbf{B}|\vec{\theta})} &= \sum_{i=1}^{N_{\text{dets}}} \sum_{j=1}^{N_{\text{s}}}
\sum_{k=1}^{L_{i,j}} \left(\mathcal{B}_{i,j,k} -
\left[\frac{|B_{i,j,k}-y(\vec{\theta})_{i,j,k}|^2}{2\sigma_{i,j,k}^2 } \right] \right)
\end{align}
\end{widetext}
where $L_{i,j}$ is the length of each dataset and $\mathcal{B}_{i,j,k} = -\ln{(2\pi\sigma_{i,j,k}^2)}$. Note
that normalisation factor reflects the fact that exponential is already the product of the real and imaginary data components.

\subsubsection{The null likelihood}\label{sec:nulllike}

As we will most often want to perform model comparison for the signal model against one for which the data just contains
noise (the null hypothesis in this case) we can define the null likelihoods for both of the above likelihoods
by setting $y(\vec{\theta}) = 0$, giving
\begin{align}\label{eq:nulllike}
\ln{p(\mathbf{B}|y=0)} &= \sum_{i=1}^{N_{\text{dets}}} \sum_{j=1}^{N_{\text{s}}}
\sum_{k=1}^{M_{i,j}} \Bigg( \mathcal{A}_{i,j,k} - \nonumber \\
&m_{i,j,k}\ln{
\left\{\sum_{n=n_{i,j,0}}^{n_{i,j,0}+(m_{i,j,k}-1)} |B_{i,j,n}|^2\right\}}
\Bigg),
\end{align}
for the Student's {\it t}-likelihood, and
\begin{equation}
\ln{p(\mathbf{B}|y=0)} = \sum_{i=1}^{N_{\text{dets}}} \sum_{j=1}^{N_{\text{s}}}
\sum_{k=1}^{L_{i,j}} \left(\mathcal{B}_{i,j,k} -
\left[\frac{|B_{i,j,k}|^2}{2\sigma_{i,j,k}^2 } \right] \right)
\end{equation}
for and Gaussian likelihood.

If we were only interested in comparing models calculated using equivalent likelihood functions we could in
general ignore the factors that do not depend on the data or the model, as they would cancel in any odds
ratio. But, in this code we keep them for cases when such a comparison is not performed, e.g., if we
want to compare the joint multi-detector likelihood for a signal with the incoherent product of likelihoods
from each detector then we would need these factors to still be present.

\subsubsection{Fast likelihood evaluations}\label{sec:fastlike}

In cases when the only time varying components of the model are the antenna pattern functions (i.e.\ when 
$\Delta \phi_j(t)$ in equation~\ref{eq:hf} or \ref{eq:h2f} is zero) the likelihood
evaluation can be greatly sped-up by pre-calculating the components in the internal summations. For a given sky
position and detector, $D$, and using the standard reference frame, the antenna patterns can be written as \citep[following the form of]{1998PhRvD..58f3001J}
\begin{align}\label{eq:antenna}
F_+^D(t,\psi) &= \sin{\zeta}\left[a(t)\cos{2\psi} + b(t)\sin{2\psi}\right], \nonumber \\
F^D_{\times}(t,\psi) &= \sin{\zeta}\left[b(t)\cos{2\psi} - a(t)\sin{2\psi}\right],
\end{align}
where $\psi$ is the \gw polarisation angle, $\zeta$ is the known angle between the detector arms (generally
$90^{\circ}$), and $a(t)$ and $b(t)$ are the time dependent functions for a given detector position and
source sky location that vary over a sidereal day.\footnote{The functions $a(t)$ and $b(t)$ are just the antenna patterns
in a frame defined with $\psi=0$. The actual definition of the antenna pattern functions is given in, e.g., Equation~23 of
\citet{2015PhRvD..91h2002I}.} These functions can be precomputed at a set of times over
a sidereal day and used, via look-up table interpolation, to give the value at any other time (our code
defaults to calculate $a(t)$ and $b(t)$ at 2880 points over a sidereal day). However, the vast majority of the
speed-up comes from pre-computing summations over the combinations of the data and the $a(t)$ and $b(t)$ functions.
This is described in more detail in Appendix~\ref{app:fle}.

In the case where we want to search over phase evolution parameters, meaning that $\Delta\phi_{\mathcal{K}}(t) \ne 0$ in the model, then
we cannot perform this pre-summing. However, {\it reduced order quadrature} methods
\citep[e.g.][]{2014PhRvX...4c1006F, 2015PhRvL.114g1104C} may help in these cases. Such a method is implemented in our
code, but will be discussed in a separate paper.

\subsection{The prior functions}\label{sec:priorfuncs}

An important part of any Bayesian inference method is the choice of parameter prior probability functions. Here we describe the prior
functions allowed in the code for any of the required parameters.

Certain parameters are physically not allowed to take negative values, so it is hardcoded that the following parameters
have zero probability below zero: gravitational wave amplitude, $h_0$; $l=m=2$ mass quadrupole moment, $Q_{22}$; the pulsar
distance; the pulsar parallax; the speed of gravitational waves; the projected semi-major axis of a binary orbit; the total
binary system mass; and, the companion mass in a binary system.

There are also other hardcoded priors for waveform amplitudes in two particular cases when searching for emission at both once
and twice the rotation frequency: if searching for a signal from a biaxial star then the two waveform amplitudes, $C_{21}$ and $C_{22}$,
must either both be positive or both be negative; if using the ``source'' model then the two amplitudes $I_{21}$ and $I_{31}$, must
have $I_{31} \geqslant I_{21}$ \citep[see, e.g.,][]{2015MNRAS.453.4399P}.

\subsubsection{Uniform prior}\label{sec:uniformprior}

A parameter can be given a uniform (or flat, or top-hat) prior, in which the probability is constant within a given
range and zero outside that range, i.e.
\begin{equation}
p(x|I) = \begin{cases}
             C & \text{if } x_{\text{min}} < x < x_{\text{max}}, \\
             0 & \text{otherwise},
            \end{cases}
\end{equation}
where for parameter $x$ the upper and lower bounds on the parameter values are $x_{\text{min}}$ and $x_{\text{max}}$. This
prior can be normalised by setting $C = 1/(x_{\text{max}}-x_{\text{min}})$, although within nested sampling the normalisation
is not specifically required. The prior requires a lower and upper bound to allow it to be normalisable and to enable
an initial set of samples to be drawn from it. However, it should be noted that the choice of range will have an
effect on evidences that are calculated as it affects the volume of parameter space allowed by the model.

A histogram of a set of samples produced by the code when drawing from a uniform prior distribution are shown in Figure~\ref{fig:prioruniform},
along with the true prior function.

Some parameters for which the uniform prior is often considered appropriate are angle parameters, such as the $\Phi_{22}^C$ and $\psi$
parameters in Equation~\ref{eq:h2f}, which can be restricted to a specific non-degenerate range \citep[see, e.g., Table~1 in][]{2015MNRAS.453.4399P}. 
In previous gravitational wave pulsar searches, such as \citet{2010ApJ...713..671A,2014ApJ...785..119A} a uniform prior was often used for
the gravitational wave amplitude parameter $h_0$. For those searches the prior upper bound was based on inspection of the data to find a
value that was very large compared to the sensitivity suggested by the data. A uniform prior was used, as opposed to the more
uninformative choice for such a scale parameter of the log-uniform prior (see \S\ref{sec:loguniform}), due to the fact that the main
result was in producing upper limits, and wanting these to be trivially related to the likelihood, rather than being heavily influenced
by the prior.

\begin{figure}[!phtb]
\begin{center}
\includegraphics[width=1\columnwidth]{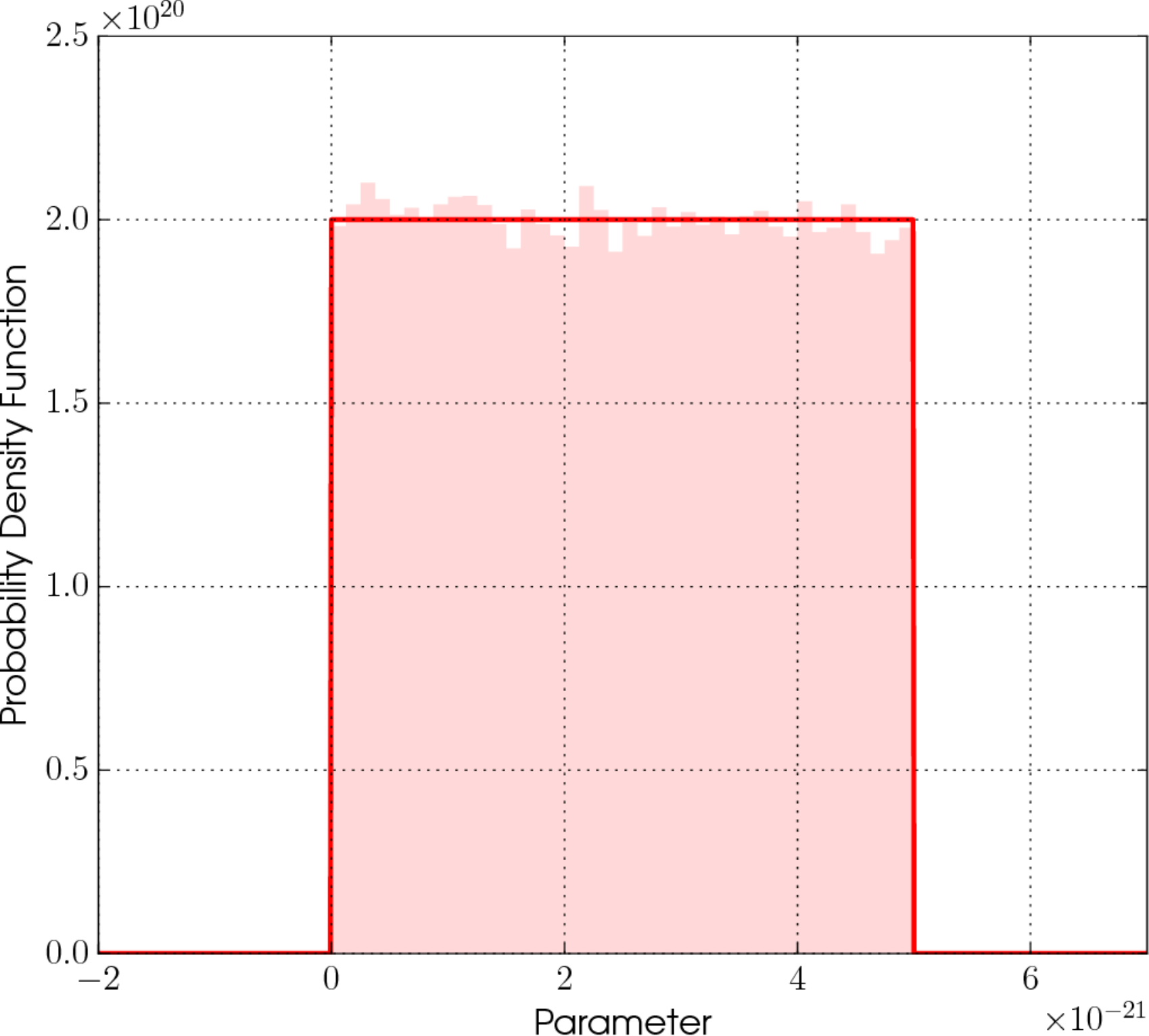}
\caption{ \protect\input{./figures/priors/uniform/caption}}
\end{center}
\end{figure}

\subsubsection{Gaussian prior}\label{sec:gaussianprior}

A parameter can be given a prior with a Gaussian distribution, defined by its mean, $\mu_x$, and standard deviation, $\sigma_x$, i.e.\
\begin{equation}
 p(x|I) = \frac{1}{\sqrt{2\pi\sigma_x^2}}\exp{\left[-\frac{(x-\mu_x)^2}{2\sigma_x^2}\right]}.
\end{equation}

A histogram of a set of samples produced by the code when drawing from a Gaussian prior distribution are shown in Figure~\ref{fig:priorgaussian},
along with the true prior function.

As well as single parameters being given Gaussian priors, we can define a set of $k$ parameters, $\vec{x}$, to have a multi-variate Gaussian prior
\begin{equation}
 p(\vec{x}|I) = \frac{\exp{\left(-\frac{1}{2}[\vec{x}-\vec{\mu_x}]'C^{-1}[\vec{x}-\vec{\mu_x}] \right)}}{(2\pi)^{k/2}|C|^{1/2}},
\end{equation}
where $\vec{\mu_x}$ are the means and $C$ is the covariance matrix. In reality we input the correlation coefficient matrix and individual
parameter standard deviations rather than the covariance matrix, but it is easy to convert between the two.

\begin{figure}[!phtb]
\begin{center}
\includegraphics[width=1\columnwidth]{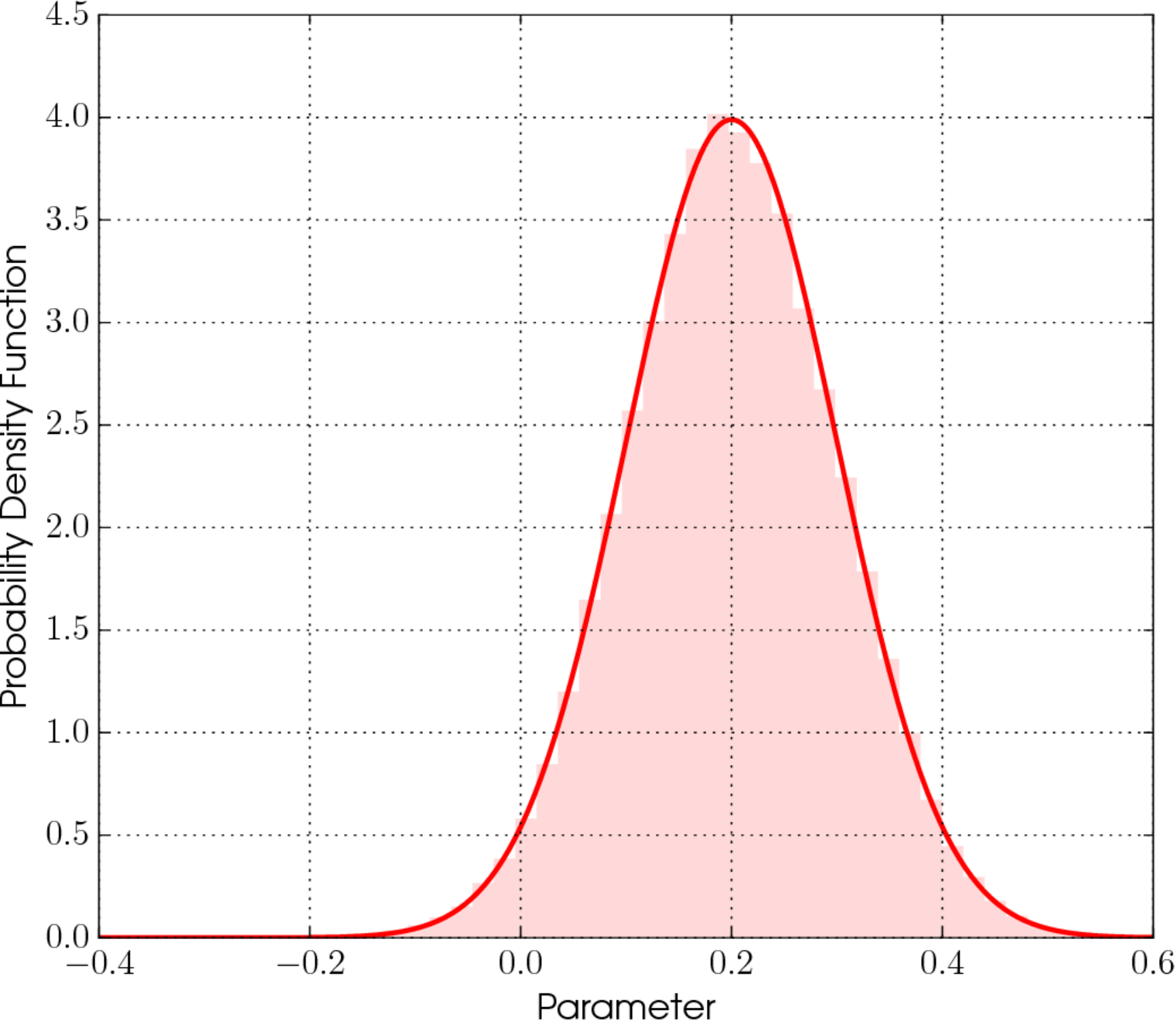}
\caption{ \protect\input{./figures/priors/gaussian/caption}}
\end{center}
\end{figure}

\subsubsection{Gaussian Mixture Model prior}\label{sec:gmmprior}

A parameter, or any set of parameters, can be given a prior distribution composed of a superposition of Gaussian distributions (a
Gaussian Mixture Model), with each component defined by a set of means for each parameter, a covariance matrix, and a weight
describing the relative probability concentrated in it. For an $n$ component mixture on a set of $k$ parameters, $\vec{x}$, the prior
function will be
\begin{equation}
 p(\vec{x}|I) = \sum_{i=1}^n w_i\frac{\exp{\left(-\frac{1}{2}{\Delta\vec{x}_i}'C_i^{-1}\Delta\vec{x}_i\right)}}{(2\pi)^{k/2}|C_i|^{1/2}},
\end{equation}
where, for component $i$, $w_i$ is the weight, $\Delta\vec{x_i}= \vec{x}-\vec{\mu}_{x_i}$, $\vec{\mu}_{x_i}$ are the means, and
$C_i$ is the covariance matrix.

Such a model can be used, for example, if there is a bimodal Gaussian prior on a particular parameter, as is the case for the
inclination angle $\iota$ if it is calculated from fits to pulsar wind nebulae \citep[see, e.g., Appendix B in][]{2017arXiv170107709T}.
In such a case the Gaussian Mixture Model can define
two Gaussian distributions of equal standard deviations and weights. A more complex use of this prior would be for reconstructing a smooth
probability distribution using samples drawn from it, i.e.\ taking posterior samples from a search for a particular pulsar, and
reconstructing a smooth model of those samples (for, say, the two parameters $h_0$ and $\cos{\iota}$) to then be input as a prior when looking
at future data for the same source.

Two examples of histograms of samples drawn from one-dimensional Gaussian Mixture Models with two and three modes respectively
are shown in Figure~\ref{fig:priorgmm}.

\begin{figure}[!phtb]
\begin{center}
\includegraphics[width=1\columnwidth]{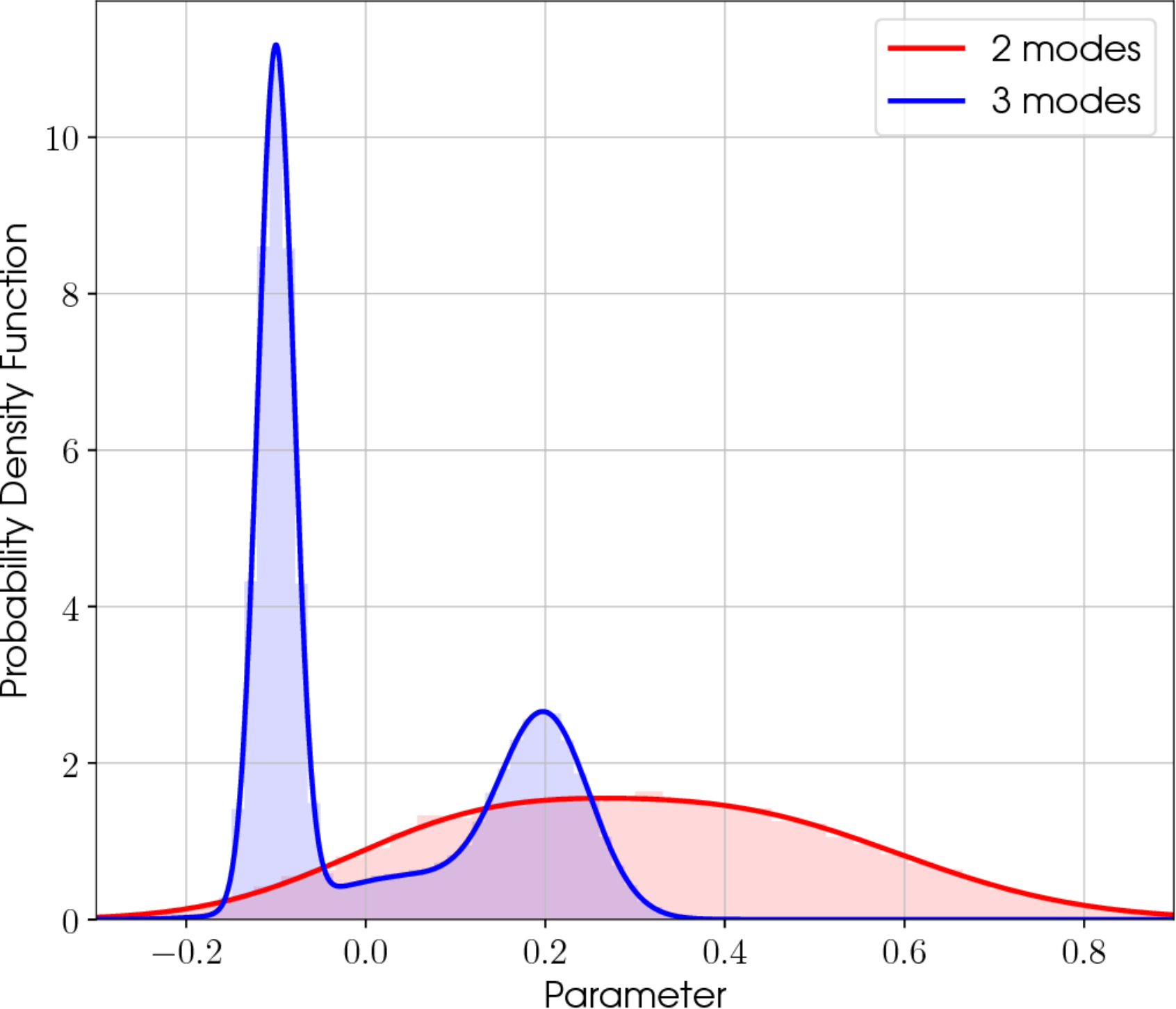}
\caption{ \protect\input{./figures/priors/gmm/caption}}
\end{center}
\end{figure}

\subsubsection{Log-Uniform prior}\label{sec:loguniform}

A parameter can be given a prior distribution that is uniform in the logarithm of the parameter, i.e.
\begin{equation}
 p(x|I) = \begin{cases}
             \left(\ln{x_{\text{max}}}-\ln{x_{\text{min}}}\right)^{-1}\frac{1}{x} & \text{if } x_{\text{min}} < x < x_{\text{max}}, \\
             0 & \text{otherwise}.
            \end{cases}
\end{equation}
This prior requires minimum and maximum values to be specified to make it normalisable and allow samples to be drawn from the
distribution. It should be noted that the choice of range, and in particular the lower bound, will have some effect on
the calculated evidence \citep[see, e.g., Appendix~B of][]{MaxCWpolariations}.

A histogram of a set of samples produced by the code when drawing from a log-uniform prior distribution, with a range between
$10^{-3}$ and 1 are shown in Figure~\ref{fig:priorloguniform}, along with the true prior function.

\begin{figure}[!phtb]
\begin{center}
\includegraphics[width=1\columnwidth]{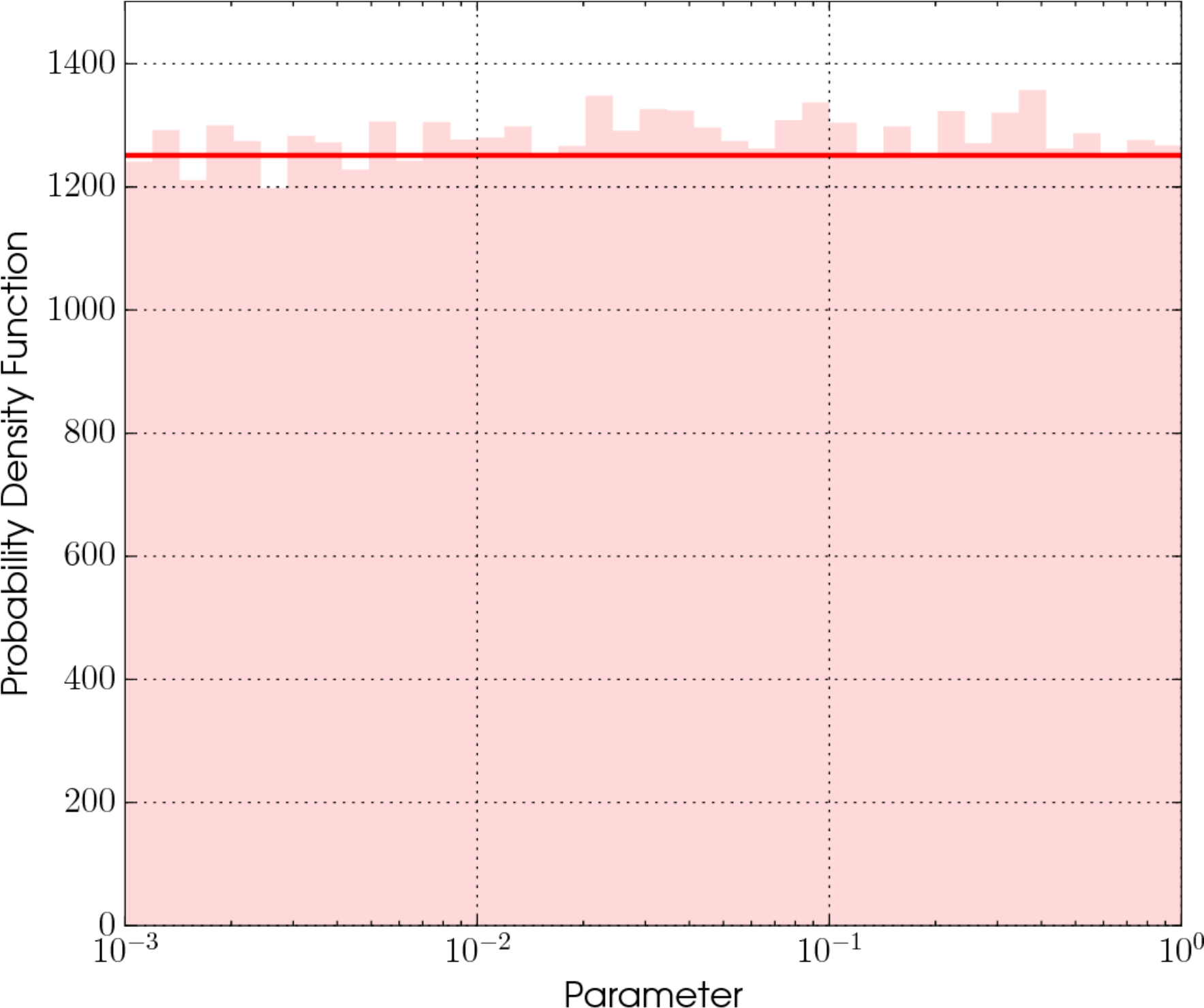}
\caption{ \protect\input{./figures/priors/loguniform/caption}}
\end{center}
\end{figure}

\subsubsection{Fermi-Dirac prior}\label{sec:fdprior}

A Fermi-Dirac distribution prior function was inspired by that used in \citet{Middleton_2015}. The prior has a sigmoid, or logistic,
shape, although starts at large values and slopes downwards and is restricted to positive values. The form of the function, for
parameter $x$, is
\begin{equation}\label{eq:fermidirac}
 p(x|\sigma, \mu, I) = \begin{cases}\frac{1}{\sigma\ln{\left(1+e^{\mu/\sigma} \right)}}\left(e^{(x-\mu)/\sigma} + 1\right)^{-1} & \text{if } x \geqslant 0, \\
                        0 & \text{otherwise},
                       \end{cases}
\end{equation}
and is defined by two parameters $\sigma$ and $\mu$. $\mu$ defines the point at which the distribution falls to 50\% of
its maximum value, whilst $\sigma$ defines the range over which the attenuation happens. We also make use of a parameter
$r = \mu/\sigma$. The band over which the probability falls from 97.5\% to 2.5\% of its maximum value is given by $\mu \pm Z\mu/2r$, where
$Z\approx7.33$. Therefore, if we want, for example, the distribution to have this particular fall-off over a range that is 10\% of $\mu$,
we would have $Z/2r = 0.1$ and $\sigma = 0.027\mu$.

Three different examples of Fermi-Dirac priors, and samples drawn from them by the code, are shown in Figure~\ref{fig:priorfermidirac},
for a selection of $\mu$ and $\sigma$ values.

\begin{figure}[!phtb]
\begin{center}
\includegraphics[width=1\columnwidth]{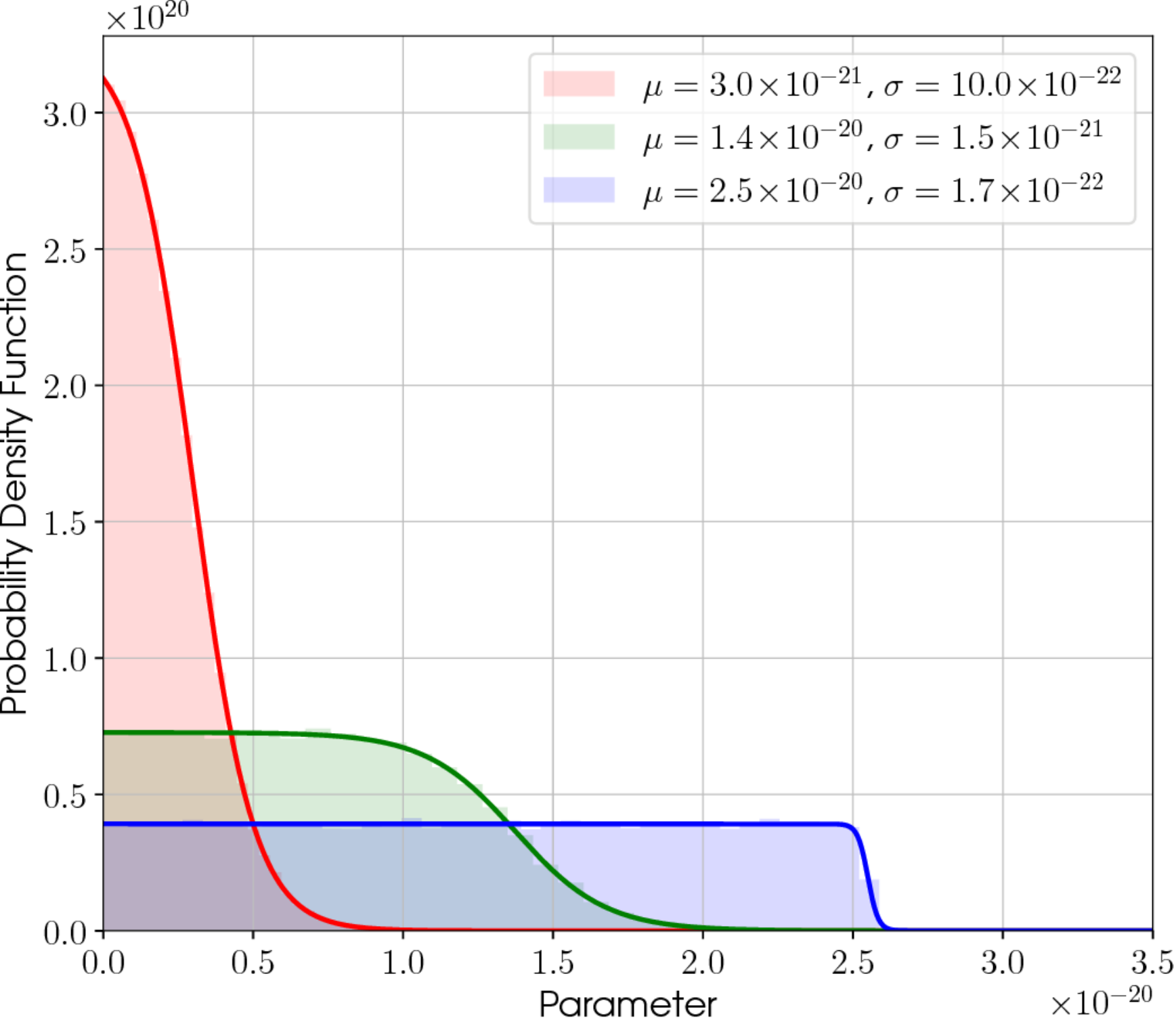}
\caption{ \protect\input{./figures/priors/fermidirac/caption}}
\end{center}
\end{figure}

The sampling from the prior uses inverse transform sampling. The cumulative distribution function (CDF) of Equation~\ref{eq:fermidirac}
is
\begin{widetext}
\begin{align}
 C(X_0) &= \int_0^{X_0} \frac{1}{\sigma\ln{\left(1+e^{r} \right)}}\left(e^{(x-\mu)/\sigma} + 1\right)^{-1} \text{d}x, \nonumber \\
 &= \frac{1}{\ln{\left(1+e^{r}\right)}}\left[\frac{X_0}{\sigma} + \ln{\left(1+e^{-r} \right)} - \ln{\left( 1+e^{(X_0-\mu)/\sigma}\right)} \right],
\end{align}
which can be inverted to give
\begin{equation}
X_0 = -\sigma \ln{}\left(-e^{-r} + \left(1+e^{r} \right)^{-C(X_0)} + e^{-r}\left(1+e^{r} \right)^{-C(X_0)} \right).
\end{equation}
\end{widetext}
Thus, drawing CDF samples uniformly between 0 and 1, and inverting via the above equation gives values drawn from Equation~\ref{eq:fermidirac}.

This prior can use useful for amplitude parameters, where a uniform prior may have previously been used. It has the property of being roughly
flat at low values of the parameter and with an exponential fall-off at large values. This gives the advantage over a uniform prior with a
sharp upper cut-off in that the distribution is continuous rather than (perhaps) arbitrarily truncated.

This form of prior was used for the \gw amplitude parameter, $h_0$, in \citet{2017arXiv170107709T} as opposed to the earlier use of
a uniform prior with arbitrary upper cut-off used in, e.g., \citet{2014ApJ...785..119A}.

\subsection{Splitting the data}\label{sec:splitting}

The Student's $t$-likelihood function (\S\ref{sec:stlikelihood}) requires us to have an idea about the timescales on which the data is
stationary, i.e.\ periods when the noise in the data is best described as being drawn from a single
Gaussian distribution. We therefore use a scheme similar to the {\it BayesianBlocks} change point algorithm
\citep{1998ApJ...504..405S}, or {\it BlockNormal} \gw burst finding algorithm \citep{2004CQGra..21S1705M}, to find points at which the statistics of the
noise changes. This is done with a top-down iterative {\it divide and conquer} approach \citep{2000physics...9033S}.

We start by taking a full complex time series data set for a detector and subtracting a running median from both
the real and imaginary components. This
subtraction is performed to try and remove the effect of any very strong signals in the data, as we want to
assess the properties of the noise rather than any signal.\footnote{In reality this is only likely to be an issue for strong
simulated signals injected into the data, as real signals will most likely have a low signal-to-noise ratio in short stretches of data.} The running median is calculated over a running window
of 30 data points (or a minimum of 15 points at the start or end of the data), but not accounting for gaps in the data.
For the standard data sample rate of
1 per 60 seconds, this means that half an hour of data is used, over which time the signal (through the varying antenna patterns) will not change
very much (unless there is a gap in the data). However, it should be noted that, if using a much more slowly sampled data set,
then this hardcoded 30 sample window may not work well for very strong signals.

The next thing we do is calculate the evidence that the whole (running-median-removed) dataset is drawn from a single
Gaussian distribution with unknown variance. To do this we just use Equation~\ref{eq:nulllike}, with $N_{\text{dets}} =1$,
$N_{\text{s}}=1$, $M_{1,1}=1$, $n_{1,1,0}=1$ and, given a dataset of length $N$, set $m_{1,1,1}=N$. We will call the natural logarithm
of this value $\ln{\mathcal{Z}_{\text{single}}}$. We want to calculate the odds between this evidence and that for the data containing
{\it any} single change point (i.e. point at which the data seems to be drawn from a different Gaussian distribution). So, for
one single change point, at index $i$, we calculate $\ln{\mathcal{Z}_{\text{cp,i}}}$, by using Equation~\ref{eq:nulllike} and setting
$N_{\text{dets}} =1$, $N_{\text{s}}=1$, $M_{1,1}=2$, $n_{1,1}=\{1,i\}$ and $m_{1,1} = \{i,N-i\}$. We also impose a minimum allowed data
chunk size of $n_{\text{min}}$ (which defaults to a value of 5 data points in the code), so that $i \geqslant n_{\text{min}}$. As we want
the evidence for {\it any} single change point, we must take the sum of all the single change point evidences
\begin{equation}
 \ln{\mathcal{Z}_{\text{any-cp}}} = \ln{\left(\sum_{i=n_{\text{min}}}^{N-n_{\text{min}}} \mathcal{Z}_{\text{cp,i}} \right)}.
\end{equation}
Finally, we are left with the odds (assuming equal prior probabilities for each hypothesis)
\begin{equation}\label{eq:cpodds}
 \ln{\mathcal{O}^{\text{any-cp}}_{\text{single}}} = \ln{\mathcal{Z}_{\text{any-cp}}} - \ln{\mathcal{Z}_{\text{single}}},
\end{equation}
for which values above zero favour there being a change point in the data. We could simply use this as the criterion on which to
split the data into two chunks at the change point value $i$ with the largest evidence. However, in practice this leads to
data drawn from a single Gaussian distribution actually being split far too often, and also there being a dependence on the
data length. So, instead, we empirically calculate an odds threshold above which the value of Equation~\ref{eq:cpodds} must be to
impose a split in the data.\footnote{A different, but equivalent, approach would be to apply different priors to each hypothesis, although we have just taken
the empirical approach.} The empirical threshold we use is worked out by finding the 99\% upper limit on the value of
Equation~\ref{eq:cpodds} when data is purely drawn from a single Gaussian distribution (i.e. it gives a 1\% false alarm probability)
as a function of the data length $N$. The results of doing this for 30\,000 instances of data for a range of $n$ values, and three
different false alarm probabilities (1\%, 0.5\% and 0.1\%), are shown in Figure~\ref{fig:changepoint}. Fitting a line to the 1\% false alarm
probability points gives an odds ratio threshold dependence of
\begin{equation}
 T = 4.07 + 1.33\log{}_{10}{N}.
\end{equation}

\begin{figure}[!phtb]
\begin{center}
\includegraphics[width=1\columnwidth]{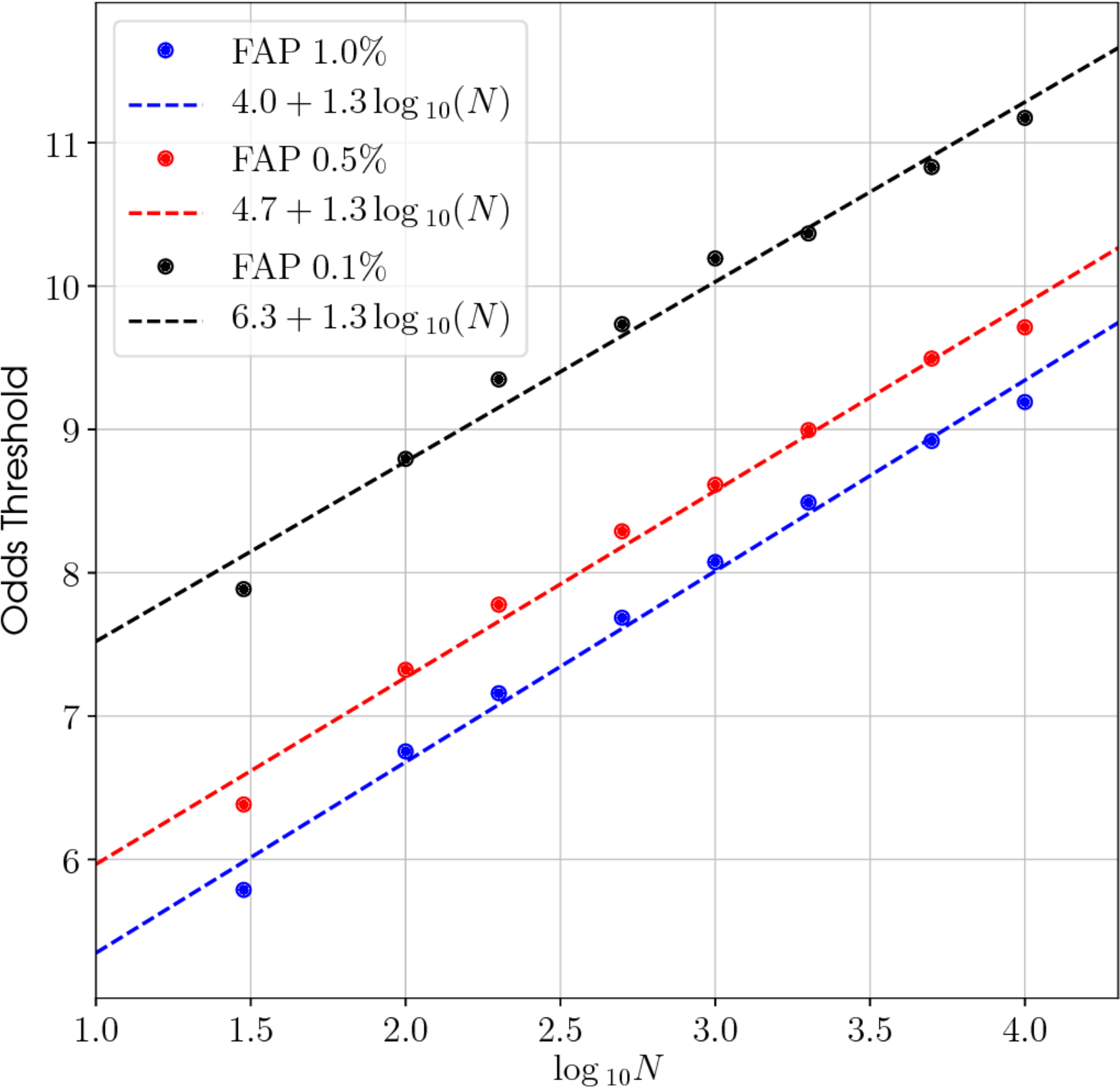}
\caption{ \protect\input{./figures/changepoint/caption}}
\end{center}
\end{figure}

The above process just splits the data once, so it must be applied iteratively to split the data further. When a change point is
found then the process is repeated on the two split data chunks, and stops when either no change point is found within a chunk,
or the split would give a segment smaller than $n_{\text{min}}$.

\subsection{Signal-to-noise ratio calculation}\label{sec:snr}

Whether a signal is found or not, the code will generate a signal-to-noise ratio (SNR). The calculated SNR uses the set of signal
parameters, $\vec{\theta}_{\text{ML}}$, that give the maximum value for the likelihood, which are then used to create the best-fit
complex signal template $y(\vec{\theta}_{\text{ML}})$. We then use calculations of the
noise standard deviation for each data chunk after removal of a running median from the data (the splitting and running median are
described in \S\ref{sec:splitting}). The {\it coherent} SNR is then calculated as
\begin{equation}\label{eq:snr}
 \rho_{\text{coh}} = \sqrt{\sum_{i=1}^{N_{\text{dets}}} \sum_{j=1}^{N_{\text{s}}}
\sum_{k=1}^{M_{i,j}} \sum_{n=n_{i,j,0}}^{n_{i,j,0}+(m_{i,j,k}-1)}\left(\frac{|y(\vec{\theta}_{\text{ML}})_{i,j,n}|^2}{\sigma_{i,j,k}^2}\right)},
\end{equation}
using the notation given in \S\ref{sec:likelihood}.\footnote{If using data input from the spectral interpolation code \citep{2017CQGra..34a5010D}
the noise standard deviations are actually passed to the code (see \S\ref{app:examples}), so in Equation~\ref{eq:snr} $M_{i,j}$ is replaced
by $L_{i,j}$ and $y(\vec{\theta}_{\text{ML}})_{i,j,n}$ is replace by $y(\vec{\theta}_{\text{ML}})_{i,j,k}$ as in \S\ref{sec:glikelihood}.}

A plot of the residual between the recovered single-detector and joint detector SNRs and their true values\footnote{The true values are calculated
by using the actual simulated signal model in equation~\ref{eq:snr}, rather than the recovered maximum likelihood model.} versus the true signal SNRs,
as calculated from Equation~\ref{eq:snr}, for a set of simulated signals (described in \S\ref{sec:simsignal}) are shown in Figure~\ref{fig:snrplot}.
The standard deviation of the residuals is found to be one, with a mean around zero, and the distribution does not appear to change as a function of SNR (except at the
lowest SNRs, where the noise in the data will generally give rise to a maximum likelihood model with a larger amplitude than the true signal, and thus
an overestimate of the recovered SNR).

\begin{figure}[!phtb]
\begin{center}
\includegraphics[width=1\columnwidth]{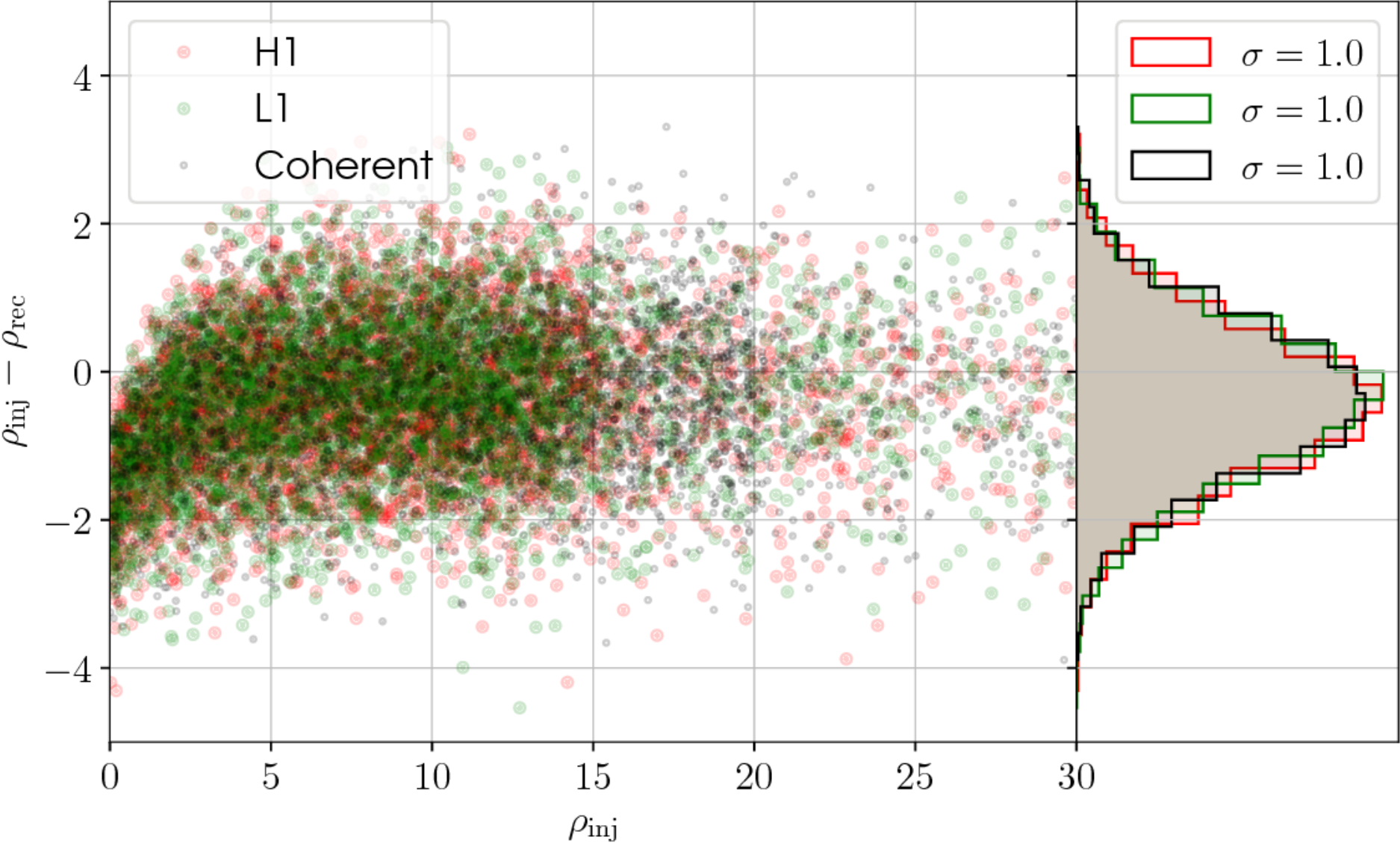}
\caption{ \protect\input{./figures/codeeval/stats/snrs/caption}}
\end{center}
\end{figure}

\subsection{Odds values}\label{sec:odds}

This is not explicitly a core function of the code, but here we will define some model selections (odds between competing models)
\begin{equation}
 \mathcal{O}_{M_1/M_2} = \frac{\mathcal{Z}_{M_1}}{\mathcal{Z}_{M_2}}\frac{p(M_1|I)}{p(M_2|I)},
\end{equation}
that can be produced from the output of the code. The code outputs two (natural logarithm) evidence values: the evidence for a coherent signal
($\mathcal{Z}_{\text{S}}$)
in all the datastreams (multi-detector and multi-frequency-band), and the evidence ($\mathcal{Z}_{\text{noise}}$) for the data consisting of
Gaussian noise (in sets of stationary segments as described in \S\ref{sec:splitting}). These can be combined in various ways for different model
selection situations, and we will define three such situations here \citep[see, e.g.,][for two of these]{2017arXiv170107709T}:
\begin{enumerate}
 \item the odds for a (multi-detector and multi-frequency-band) coherent signal versus (non-stationary) Gaussian noise, $\mathcal{O}_{\text{S}/\text{N}}$;
 \item the odds for a (multi-detector and multi-frequency-band) coherent signal versus incoherent (between detector) signals, $\mathcal{O}_{\text{S}/\text{I}_{\text{simple}}}$;
 \item the odds for a (multi-detector and multi-frequency-band) coherent signal versus combinations of incoherent (between detector) signals or noise,
 $\mathcal{O}_{\text{S}/\text{I}}$ \citep[cf.\ Equation~A6 in][]{2017arXiv170107709T}.
\end{enumerate}

The first of these (actually also given in Equation~\ref{eq:oddsratio}) just requires the two evidence values, $\mathcal{Z}_{\text{S}}$ and $\mathcal{Z}_{\text{noise}}$
\begin{align}\label{eq:sigvsnoise}
 \mathcal{O}_{\text{S}/\text{N}} &= \frac{\mathcal{Z}_{\text{S}}}{\mathcal{Z}_{\text{noise}}}, \nonumber \\
 \ln{\left(\mathcal{O}_{\text{S}/\text{N}}\right)} &= \ln{\left(\mathcal{Z}_{\text{S}}\right)} - \ln{\left({\mathcal{Z}_{\text{noise}}}\right)}.
\end{align}
The second requires that the code has been run for a multi-detector analysis, and also individually for each detector, $i$, to give $\mathcal{Z}_{\text{S}_i}$
and $\mathcal{Z}_{\text{N}_i}$, and is given by
\begin{align}\label{eq:cohvincoh1}
 \mathcal{O}_{\text{S}/\text{I}_{\text{simple}}} &= \frac{\mathcal{Z}_{\text{S}}}{\prod_i^{N_{\text{dets}}} \mathcal{Z}_{\text{S}_i}}, \nonumber \\
 \ln{\left(\mathcal{O}_{\text{S}/\text{I}_{\text{simple}}}\right)} &= \ln{\left(\mathcal{Z}_{\text{S}}\right)} - \sum_i^{N_{\text{dets}}}\ln{\left(\mathcal{Z}_{\text{S}_i}\right)}.
\end{align}
Such an odds is useful in vetoing strong incoherent signals that are therefore not astrophysical, e.g.\ instrumental lines \citep[see, e.g., the similar
line-robust statistic defined in][]{2014PhRvD..89f4023K}. In both these cases it is assumed
that the prior odds ($p(M_1|I)/p(M_2|I)$) between the two models are set to be equal and therefore are not explicitly given.

The third odds makes use of a compound model in the denominator with all evidences for all permutations of Gaussian noise or incoherent signal being used.
If we have the set of single detector signal and Gaussian noise models $M_j = \{\text{S}_j, \text{N}_j\}$, then \citep[see, e.g., Equation~50 of][]{MaxCWpolariations}
\begin{equation}\label{eq:cohvincoh2}
 \mathcal{O}_{\text{S}/\text{I}} = \frac{\mathcal{Z}_{\text{S}}p(\text{S}|I)}{\prod_{j=1}^{N_{\text{dets}}}\left[ Z_{\text{S}_j}p(\text{S}_j|I) + Z_{\text{N}_j}p(\text{N}_j|I) \right]},
\end{equation}
where $p(\text{S}|I)$ is the prior on the coherent signal model, and $p(\text{S/N}_j|I)$ are the priors for the signal/noise hypotheses for each individual detector.
As there are two possible models for data in each detector, signal or noise, the number of logical {\it or} combined sub-hypotheses \citep[see, e.g., the four sub-hypotheses
used in Equation~A6 of][where the priors in the denominator are the combinations that would be given from expanding out equation~\ref{eq:cohvincoh2}]{2017arXiv170107709T}
in the compound model is just given by $2^{N_{\text{dets}}}$. In \citet{2017arXiv170107709T}, and later in this 
document, we have assigned the equal probabilities to the coherent signal prior
and to each of the sub-hypothesis priors in the denominator, so that they cancel out. However, other priors could be used.

It is also worth noting that in all the above cases the models (and sub-hypotheses) are mutually exclusive. This may not at first seem obvious, as,
for example, the Gaussian noise model is contained within the signal model provided the prior on the signal amplitude allowed a value of zero. But, as noted in
\citet{2012PhRvD..85h2003L, MaxCWpolariations}, within the signal noise hypothesis having an amplitude of exactly zero (as in the Gaussian
noise hypothesis) occupies an infinitesimally thin slice of the parameter space, and therefore has no weight. The same can be said for the situation of a coherent
signal model within the incoherent signal model.

\section{Nested sampling setup}

Here we will briefly describe some of the core functions used for the Nested Sampling algorithm \citep{Skilling:2006} within our code. We will
not describe nested sampling itself, over-and-above the very brief summary given in \S\ref{sec:general}, but instead refer
the reader to \citet{Veitch:2010}. We also refer the reader to \citet{2015PhRvD..91d2003V} for a more thorough description
of some of the concepts we will discuss below.

One main point about the nested sampling algorithm is that it starts by randomly drawing a certain number of samples (the number of
which we call the number of live points, $N_{\text{live}}$, which will be the constant number of ``active'', or ``live'', samples),  
where a sample is a vector, $\vec{\theta}$, containing a value for each parameter drawn from their respective prior distributions (see \S\ref{sec:priorfuncs}).
The samples must be independent draws. For each of these samples the likelihood will be calculated and the sample $\vec{\theta}_{\text{min}}$
corresponding to the minimum likelihood, $L_{\text{min}} = p(\mathbf{B}|\vec{\theta}_{\text{min}})$, will be noted. As the algorithm
progresses the number of live points stays the
same, but at each iteration a new sample will be added in place of the one with the minimum likelihood, and therefore the
minimum likelihood will continuously be updated to be the minimum value for the current set of points.

\subsection{Proposal functions}\label{sec:proposals}

The nested sampling algorithm requires a way to draw new samples from the parameter space that is being explored. The
particular requirement for nested sampling is that any new sample is an independent draw from a constrained version
of prior probabilty function for each parameter. The constraint is that the new sample's likelihood must be greater
than the current value of $L_{\text{min}}$, e.g. for a single parameter $x$, the probability of drawing a point at $x_i$ 
would be
\begin{equation}
 p(x=x_i|I)_{\text{constrained}} = \begin{cases}
             p(x=x_i|I) & \text{if~} p(\mathbf{B}|x_i) > L_{\text{min}}, \\
             0 & \text{otherwise},
            \end{cases}
\end{equation}
where $p(x|I)$ is the prior on $x$.

There are various methods to sample from the constrained prior, the simplest of which is just to sample from the full
prior and only accept a sample if it meets the likelihood constraint. However, such a simple method soon becomes very
inefficient if the likelihood gets more and more tightly concentrated within the prior volume (which can happen rapidly
for high-dimensional problems). One popular nested sampling method is MultiNest \citep{2009MNRAS.398.1601F}, which performs
the sampling by calculating a set of ellipsoids bounding the live points and drawing points uniformly from within them.
Various other approaches exist such as Diffusive Nested Sampling \citep{Brewer2011,2016arXiv160603757B} and POLYCHORD \citep{2015MNRAS.450L..61H}.

The version of nested sampling used in the \lalinf library, as used by our code, makes use of a Markov chain Monte Carlo (MCMC) approach to
drawing new samples \citep{Veitch:2010}. The MCMC attempts to sample from the constrained prior distributions in an efficient manner, with
the final sample of the chain being a statistically independent sample from the rest of the live points. To sample most efficiently
the MCMC requires a proposal distribution that closely resembles the underlying constrained prior that it is trying to sample from.
\lalinf has a variety of potential proposal distributions \citep{2015PhRvD..91d2003V}, but our code currently allows five different
proposals, that can be used in different proportions:
\begin{description}
 \item[Ensemble walk] This is one of the affine invariant ensemble proposals described by \citet{GoodmanWeare}, which adopt the
 scale and shape of the current set of samples. The hard-coded setup in \lalinf is that three samples randomly drawn from a cache
 of samples are used to define the ``walker'' samples. This is the main default proposal of our code, and is used as the proposal
 75\% of the time.
 \item[Ensemble stretch] This is another affine invariant ensemble proposal described by \citet{GoodmanWeare}, and makes use of
 two samples randomly drawn from the sample cache. By default this proposal is not used at all.
 \item[Differential evolution] This is similar to the stretch proposal in that it uses two samples randomly drawn from the sample
 cache. By default this proposal is not used at all.
 \item[Uniform proposal] This proposal just draws points from their full prior for parameters that have a uniform prior specified.
 Other parameters are not evolved. This proposal is used 25\% of the time by default.
 \item[Frequency jump] This proposal tries jumping between Fourier frequency bins if frequency is one of the required parameters.
 This proposal has not been tested and, by default, is not used at all.
\end{description}
The cache of samples mentioned above, and used by the nested sampling algorithm, is the set of live points, but this cache is
only updated on iterations of the algorithm that are a factor of 10\% of the number of live points (e.g., for 1000 live points
the cache is updated after every 100 new iterations).

The length of each MCMC run can be chosen by the user, but by default it is automatically set within the code. To do this,
at the same rate as the cache is updated, the autocorrelation length of the parameters is calculated. A random sample from the current
live points in chosen, and evolved via the MCMC, and the autocorrelation length of each parameter, $\vec{{\text{acl}}}$, is calculated. The number
of MCMC iterations, $n_{\text{MCMC}}$, is then chosen as $n_{\text{MCMC}} = \text{min}[\text{max}(\vec{\text{acl}}), 5000]$, where 5000 is
a hardcoded maximum length.

\subsubsection{Testing the proposals}\label{sec:proposaltesting}

We would like to test whether some of the above proposals lead to accurate calculations of the evidence integral as given in Equation~\ref{eq:evidence}.
It is especially interesting to do this for a variety of prior volumes, starting of with volumes similar to the posterior volume, but then diverging to be
much larger than the posterior volume (or in other words, as the information gain, $H$, or Kullback-Leibler divergence, increases). To do this our
code has the option of setting a one-dimensional Gaussian distribution as the likelihood function, using a flat prior bounded between $A$ and $B$,
and working out the evidence. Analytically the evidence is given by
\begin{align}\label{eq:testgaussev}
 \mathcal{Z} &= \int_A^B \frac{1}{\sqrt{2\pi}\sigma}\exp{\left(-\frac{(x-\mu)}{2\sigma^2} \right)}\frac{1}{(B-A)} \text{d}x, \nonumber \\
   &= \frac{1}{2(B-A)}\left(\erf{\left[\frac{(\mu-A)}{\sqrt{2}\sigma}\right]}-\erf{\left[\frac{(\mu-B)}{\sqrt{2}\sigma}\right]}\right).
\end{align}
As we are often interested in setting upper limits of, for example, the \gw amplitude $h_0$, we can also use the test to see if the
upper limits we obtain from the posterior samples match the expected upper limit obtained (via root finding) from the CDF of the distribution
\begin{equation}\label{eq:testcdf}
 \text{UL}(x) = \frac{1}{2a}\left(\erf{\left[\frac{(\mu-A)}{\sqrt{2}\sigma}\right]} - \erf{\left[\frac{(\mu-x)}{\sqrt{2}\sigma}\right]}\right),
\end{equation}
where $a = (\erf{\left[(\mu - A)/\sqrt{2}\sigma\right]} - \erf{\left[(\mu - B)/\sqrt{2}\sigma\right]})/2$ is the area under the distribution.
We can calculate the true Kullback-Leibler divergence via
\begin{widetext}
\begin{align}\label{eq:kldiv}
H =& \int_A^B p(x|\mu,\sigma,I) \ln{\left(\frac{p(x|\mu,\sigma,I)}{p(x|I)}\right)} \text{d}x, \nonumber \\ 
 =& \int_A^B \frac{p(\mu,\sigma|x,I)p(x|I)}{\mathcal{Z}} \ln{\left(\frac{p(\mu,\sigma|x,I)p(x|I)}{\mathcal{Z}p(x|I)}\right)} \text{d}x, \nonumber \\
 =& \frac{p(x|I)}{\mathcal{Z}} \int_A^B p(\mu,\sigma|x,I) \left[\ln{\left( p(\mu,\sigma|x,I) \right)} - \ln{\mathcal{Z}}\right] \text{d}x, \nonumber \\
 =& -\frac{p(x|I)}{4\mathcal{Z}}\Bigg[ \left(1+2\left(\ln{\mathcal{Z}} 
+\ln{\left[\sqrt{2\pi}\sigma\right]}\right)\right)\left(\erf{\left(\frac{\mu-A}{\sqrt{2}\sigma}\right)}-\erf{\left(\frac{\mu-B}{\sqrt{2}\sigma}\right)} \right) + \nonumber \\
 & \frac{2}{\sqrt{2\pi}\sigma}\left((A-\mu)\exp{\left(-\frac{(A-\mu)^2}{2\sigma^2}\right)} - (B-\mu)\exp{\left(-\frac{(B-\mu)^2}{2\sigma^2}\right)}  \right) \Bigg]
\end{align}
\end{widetext}
where $p(x|\mu,\sigma,I)$ is the posterior, $p(\mu,\sigma|x,I)$ the likelihood, and $p(x|I)$ is the prior (with $p(x|I) = (B-A)^{-1}$), and $\mathcal{Z}$ is the
evidence calculated via Equation~\ref{eq:testgaussev}. For the case of $\mu=0$ and $A=0$, and provided $B \gg \sigma$, this generally simplifies
to
\begin{equation}
H \approx -\frac{1}{4\mathcal{Z}B}\left[1+2\left(\ln{\mathcal{Z}} +\ln{\left(\sqrt{2\pi\sigma^2}\right)}\right)\right].
\end{equation}
A final thing that we can check, on top of the upper limit value, is whether the posterior samples appear to be drawn from the expected
posterior distribution. To do this we use a Kolmogorov-Smirnov test comparing the CDF of the samples with the analytical CDF and calculating the
two-sided $p$-value for the null hypothesis that the two distributions are the same.

In the tests below we have run the code on this Gaussian likelihood with a Gaussian of zero mean and $\sigma = 10^{-24}$, truncated at zero
by fixing the lower prior bound to $A=0$. This is similar to the form of the likelihood seen in pulsar searches when no signal is present.
We have then set a range of upper bounds, $B$, from $10^{-23}$ to $10^{-13}$ increasing by a factor of ten each time, and a range of numbers of
live points from $2^9$ to $2^{13}$ increasing by powers of two each time. For each combination of upper bound and number of live points we have
run the code on the test Gaussian likelihood 100 times to see the statistical spread of the output.

In the first test we check how using only the ensemble walk proposal (which is {\it not} the default configuration described above) fairs at
estimating the evidence, upper limit, and posterior distribution. Figure~\ref{fig:walkpropevs} shows a variety of information, but it
primarily shows the distribution of the ratio of evidences output from our code, $\mathcal{Z}$, to the true evidence calculated via
Equation~\ref{eq:testgaussev} as a function of the value of the upper prior bound $B$ (or equivalently on the top axis the
information gain, in \href{https://en.wikipedia.org/wiki/Nat_(unit)}{natural units of information} (nats), calculated via Equation~\ref{eq:kldiv}).
This is shown for the range of different numbers of live points used.
It is clear that as the information gain increases (i.e.\ the posterior is constrained within a successively smaller part of the
prior range) we are systematically more biased towards overestimating the evidence. It can also be seen that the bias increases with
increasing number of live points. Linear fits (in $\ln{(\mathcal{Z}/\mathcal{Z}_{\text{true}})}$ and $H$, the information gain) for different numbers
of live points are shown in Table~\ref{tab:klvsz}. It is interesting to note that the statistical spread of values (if ignoring the
systematic offset) are consistent with the expected range given by the error bars, and calculated using $\sqrt{\langle H \rangle/N_{\text{live}}}$ \citep{Skilling:2006}, where here
$\langle H \rangle$ is the mean of the {\it information gain} (in nats) returned by the nested sampling algorithm for each run.

\begin{table}[h]
\caption{Linear coefficients in fits of $\ln{(\mathcal{Z}/\mathcal{Z}_{\text{true}})} = mH + c$ from the results
shown in Figure~\ref{fig:walkpropevs} for differing numbers of live points.\label{tab:klvsz}}
\begin{center}
\begin{tabular}{l c c}
\hline
\hline
$N_{\text{live}}$ & $c$ & $m$ \\
\hline
512 & $-0.27$ & 0.045 \\
1024 & $-0.24$ & 0.058 \\
2048 & $-0.18$ & 0.065 \\
4096 & $-0.12$ & 0.068 \\
8192 & $-0.09$ & 0.070 \\
\hline
\end{tabular}
\end{center}
\end{table}

\begin{figure}[phtb]
\begin{center}
\includegraphics[width=1\columnwidth]{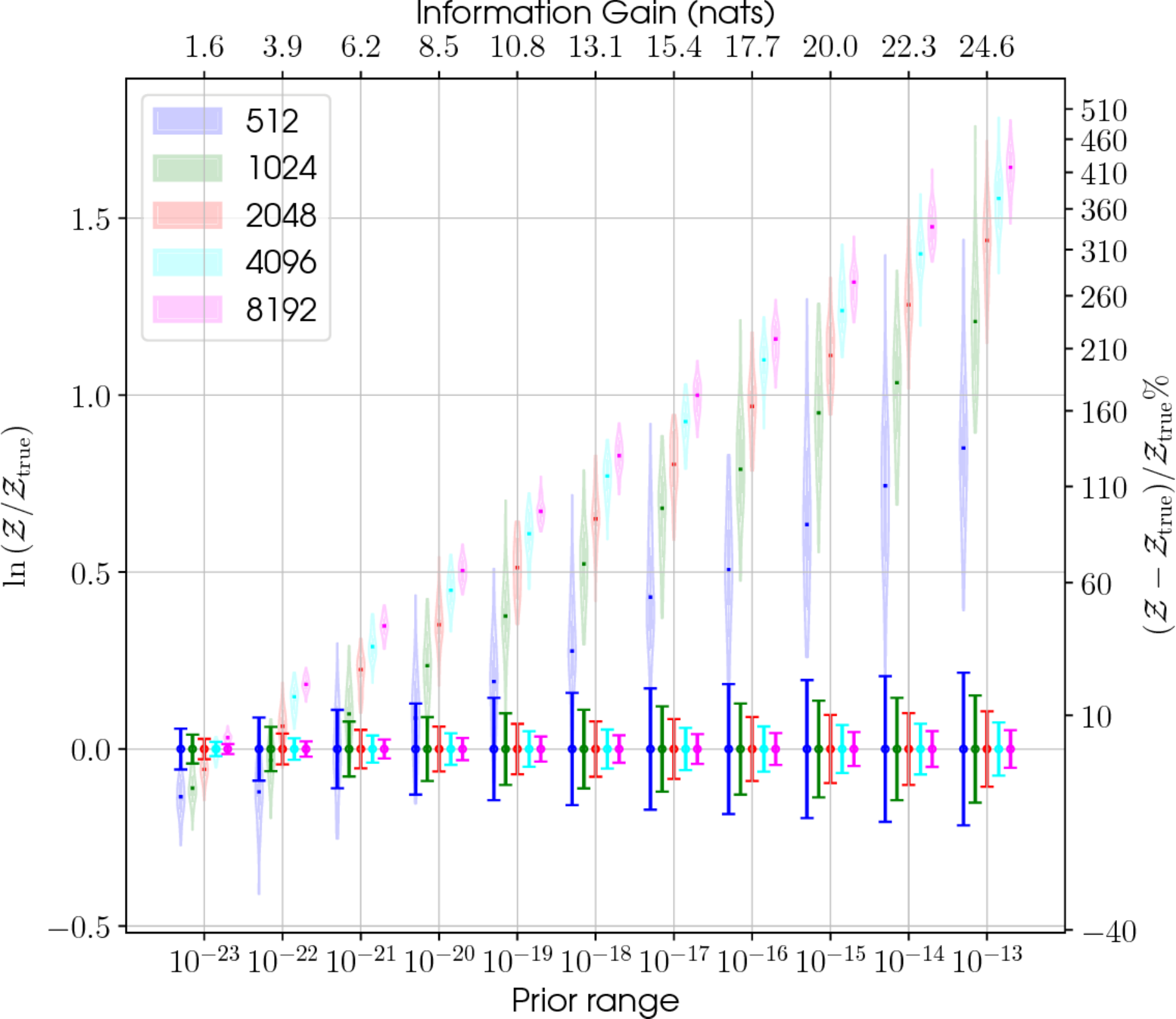}
\caption{ \protect\input{./figures/proptesting/walk_prop/evidences/caption}}
\end{center}
\end{figure}

\begin{figure}[phtb]
\begin{center}
\includegraphics[width=1\columnwidth]{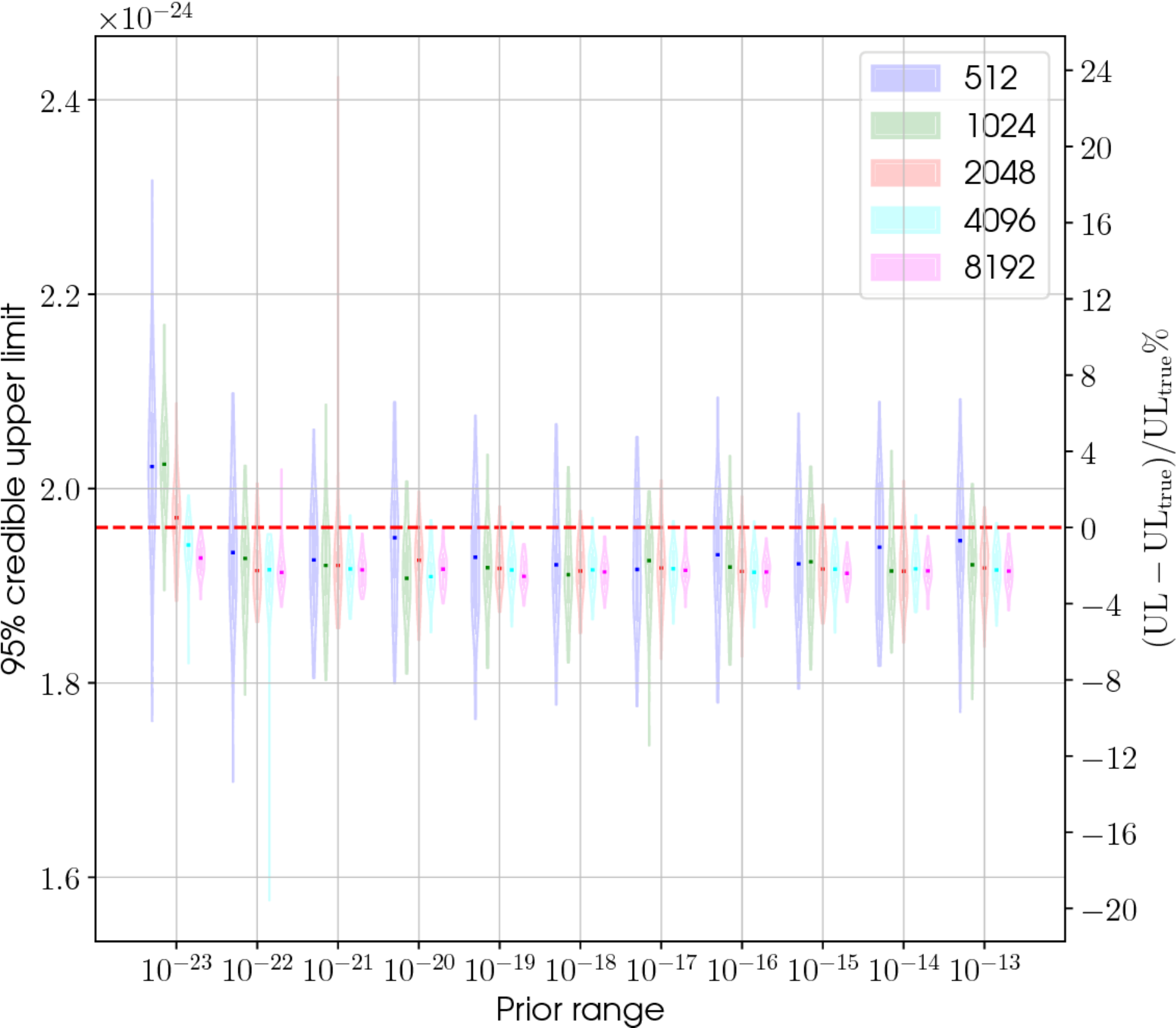}
\caption{ \protect\input{./figures/proptesting/walk_prop/upperlimits/caption}}
\end{center}
\end{figure}

\begin{figure}[phtb]
\begin{center}
\includegraphics[width=1\columnwidth]{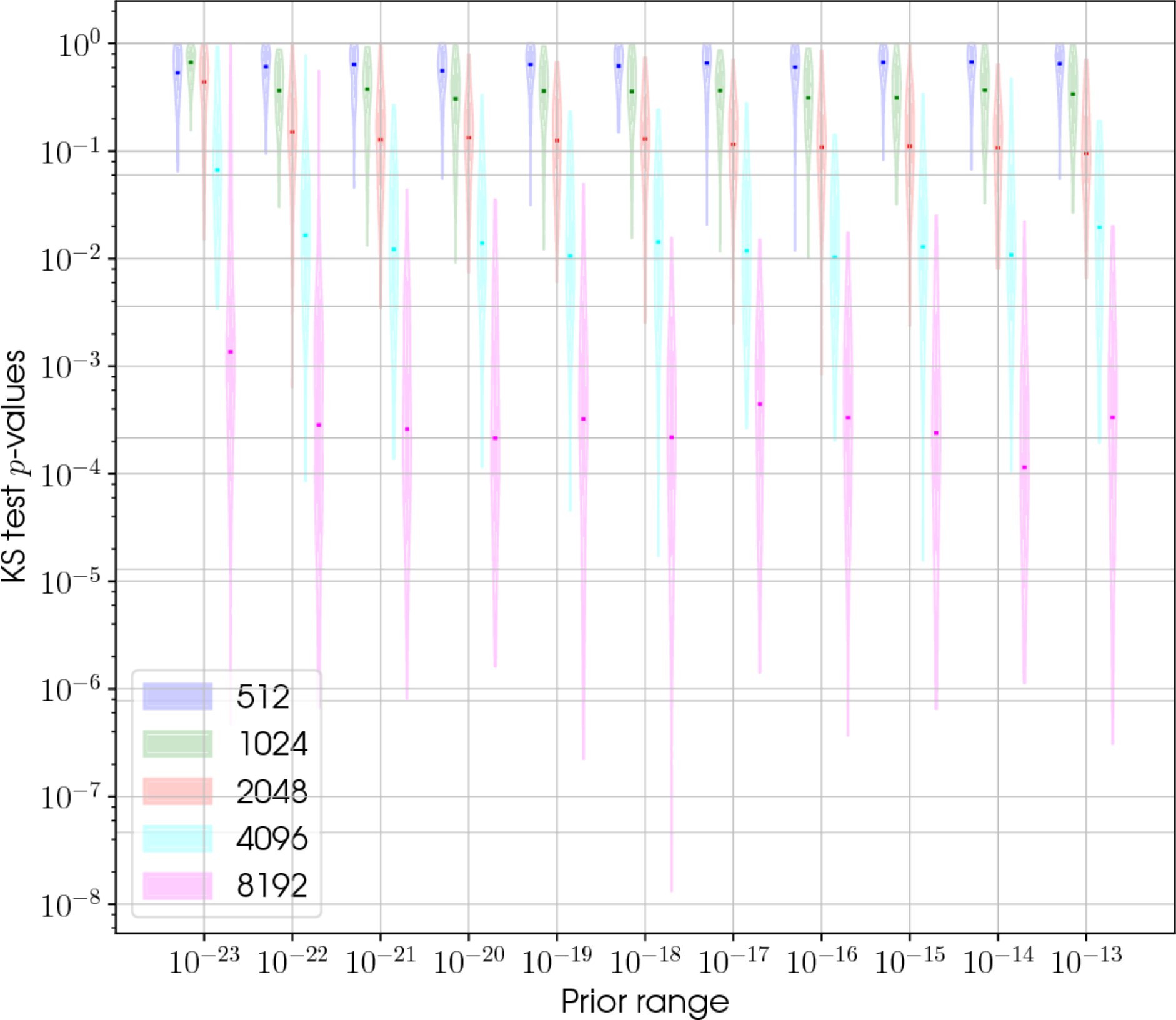}
\caption{ \protect\input{./figures/proptesting/walk_prop/kstest/caption}}
\end{center}
\end{figure}

Figure~\ref{fig:walkpropuls} shows the distribution of 95\% upper limits on the $x$ parameter in Equation~\ref{eq:testcdf} (calculated by solving
the equation for $\text{UL}(x) = 0.95$). It can be seen that, other than for the narrowest prior range, there is a systematic bias towards the
upper limit being underestimated by $\sim 2\%$. Figure~\ref{fig:walkpropks} show the distribution of two-sided $p$-values from a Kolmogorov-Smirnov
test comparing the CDF of posterior samples against the analytical CDF in Equation~\ref{eq:testcdf}. The $p$-value null hypothesis is that both
distributions are the same, and it should be noted that we performed 100 trials for each prior range value and number of live points. For the
lower numbers of live points the distributions appear to be consistent, however, for larger numbers of live points the distributions are rather
inconsistent a large amount of the time. This is most likely due to differences in distributions showing up more obviously when there are more
posterior samples to use.

The systematic biases in evidence values and upper limits may not be a major problem\footnote{The evidence biases at the level seen are unlikely to sway model comparisons as
they will most likely show up when one model is very highly favoured over another. The observed biases in the upper limits are at a level that is currently less
than expected uncertainties due to instrumental calibration.}, but does point towards there being some issue with the
implementation of the ensemble walk proposal.

In the second test we have checked what happens with the code's current default proposal settings: using the ensemble walk proposal 75\%
of the time and the uniform proposal 25\% of the time. It should be noted that this default is not arbitrary, but came about from tests
similar to these when just using the ensemble walk proposal.

Figure~\ref{fig:walkunipropevs} shows the evidence distributions equivalent to those in Figure~\ref{fig:walkpropevs}. However, what can
be seen in this case is that there is no obvious systematic bias as a function of $H$, and the statistical uncertainty is
consistent with expectations (as shown by the error bars). Similarly, the upper limits and Kolmogorov-Smirnov $p$-values shown in
Figures~\ref{fig:walkunipropuls} and \ref{fig:walkunipropks} respectively are consistent. The reason why this combination of proposals
seems to work whereas the ensemble walk proposal on its own does not is currently not understood, but we plan more work on figuring this
out.

\begin{figure}[!phtb]
\begin{center}
\includegraphics[width=1\columnwidth]{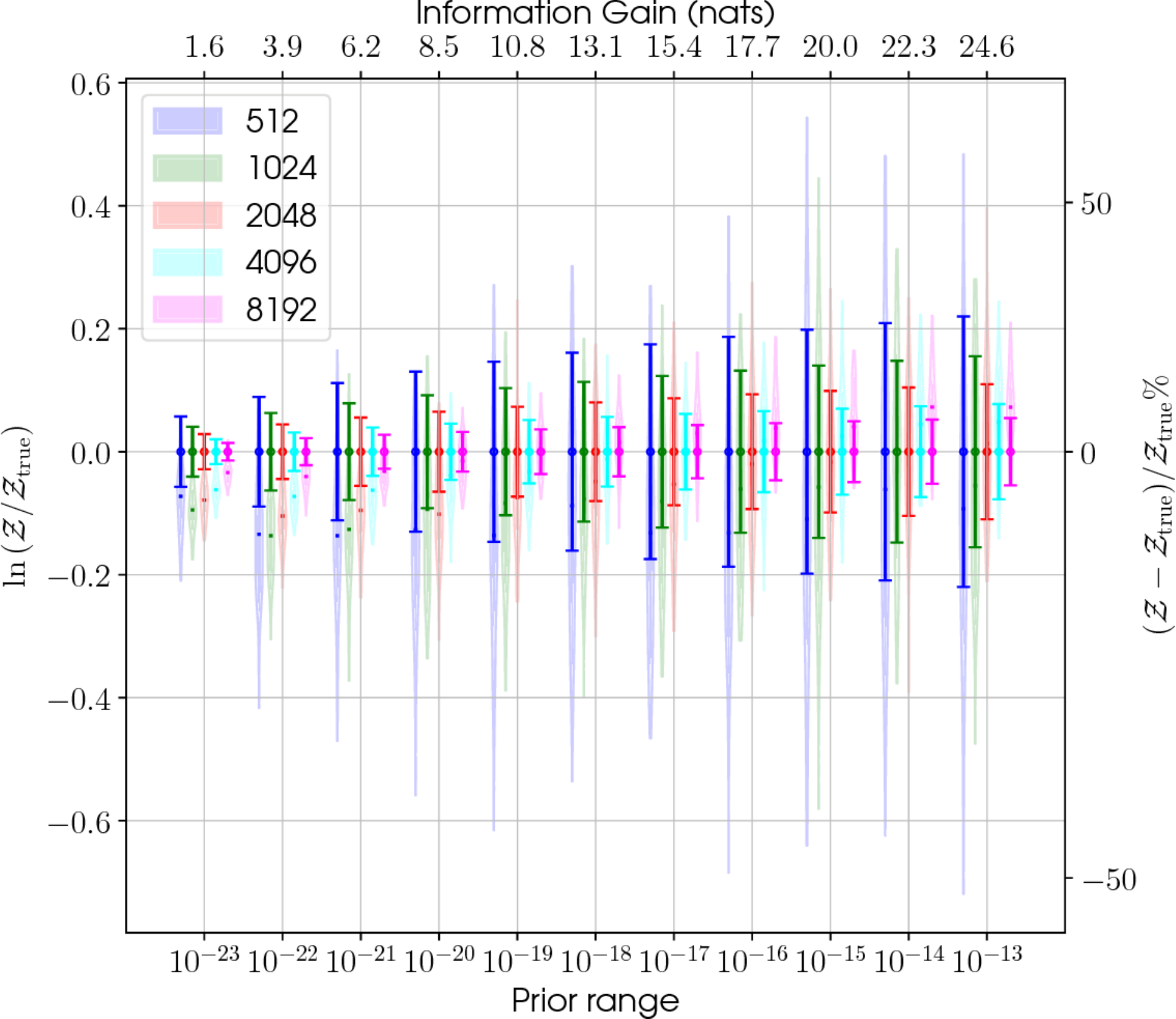}
\caption{ \protect\input{./figures/proptesting/walk_uniform_prop/evidences/caption}}
\end{center}
\end{figure}

\begin{figure}[!phtb]
\begin{center}
\includegraphics[width=1\columnwidth]{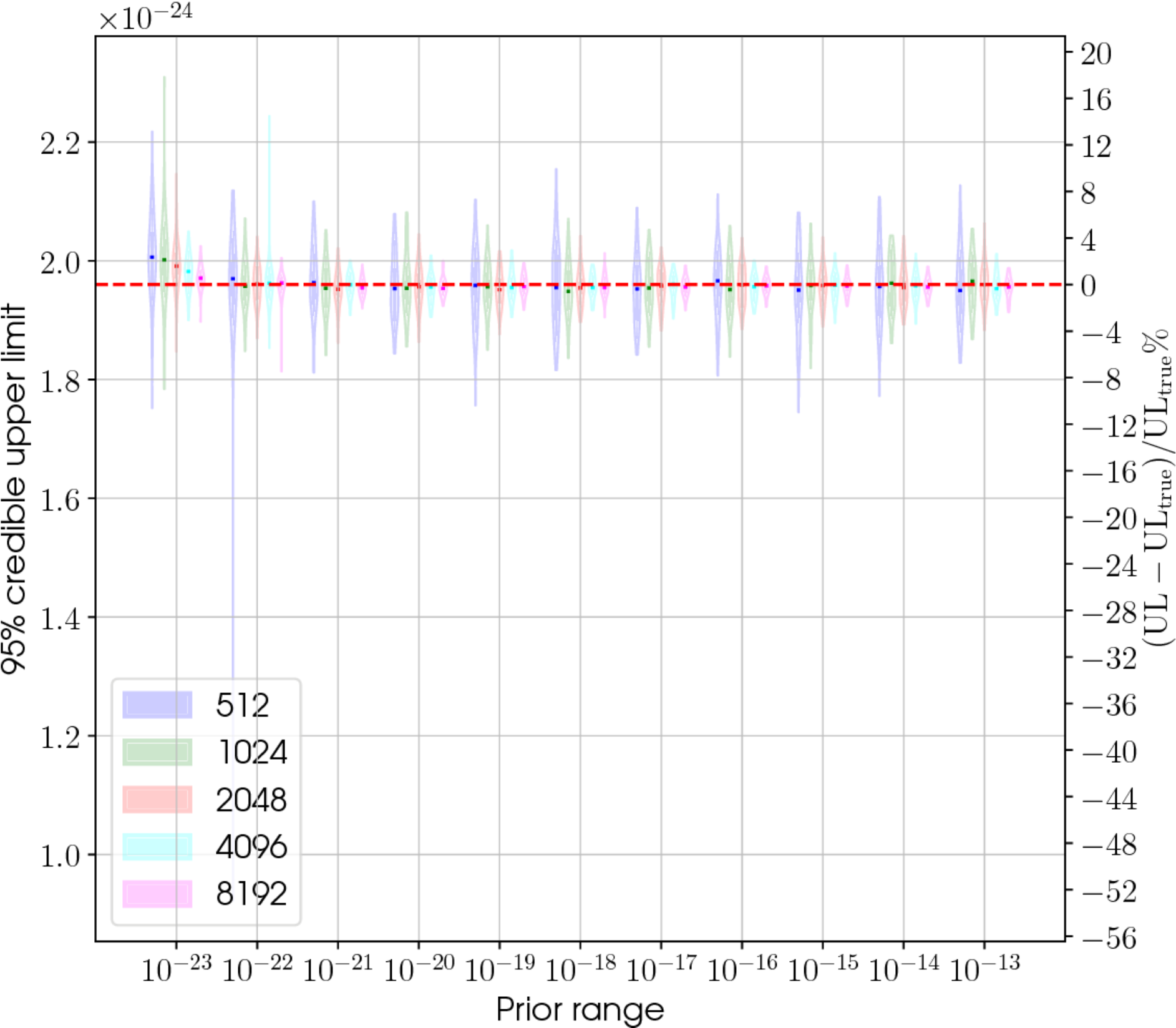}
\caption{ \protect\input{./figures/proptesting/walk_uniform_prop/upperlimits/caption}}
\end{center}
\end{figure}

\begin{figure}[!phtb]
\begin{center}
\includegraphics[width=1\columnwidth]{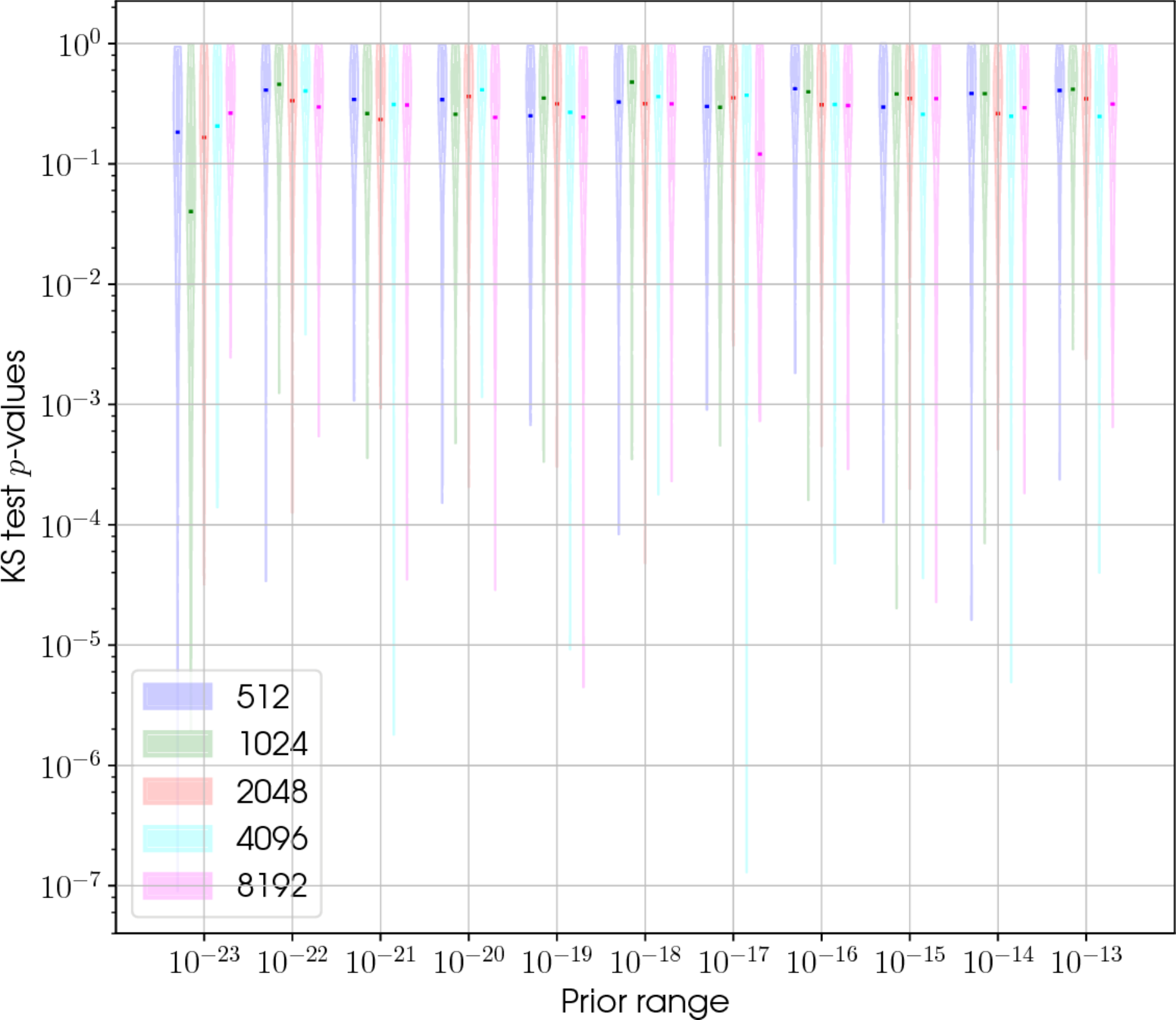}
\caption{ \protect\input{./figures/proptesting/walk_uniform_prop/kstest/caption}}
\end{center}
\end{figure}

\subsection{Posterior sample generation}\label{sec:postsamps}

As discussed in \S\ref{sec:general} nested sampling codes, such as \lppen, do not intrinsically output samples that can be histogrammed to give
representations of marginalised posterior distributions for the model parameters. They instead output an ascending likelihood ordered list of samples,
each of which occupies a known approximate amount prior volume. Samples must be drawn from these, with appropriate weighting (see Equation~\ref{eq:postaccept}),
to give a new set of samples that does represent the posterior distribution. During this process it is also possible to appropriately combine samples from parallel
independent nested sampling runs (weighting each based on their calculated evidences) to recalculate the evidence and generate posterior samples.

What is described above is not a core function of the \lppen code, but is an integral part to generating results with it, and has been used to
produce the posterior distributions discussed earlier in the section, and later in this document. The number of posterior samples generated by the code is
dependent on the number of live points used when running nested sampling. The actual dependence of the number of posterior samples on $N_{\text{live}}$ can be seen in
Figure~\ref{fig:numposts} (this uses the simulations discussed in \S\ref{sec:reseval}). It can be seen empirically, that the mean number of posterior
samples goes roughly as $\langle N_{\text{post}} \rangle \approx 2.4N_{\text{live}}$, although there is a lower tail extending to $N_{\text{post}} \approx
0.01N_{\text{live}}^{1.14}$.

\begin{figure}[!phtb]
\begin{center}
\includegraphics[width=1\columnwidth]{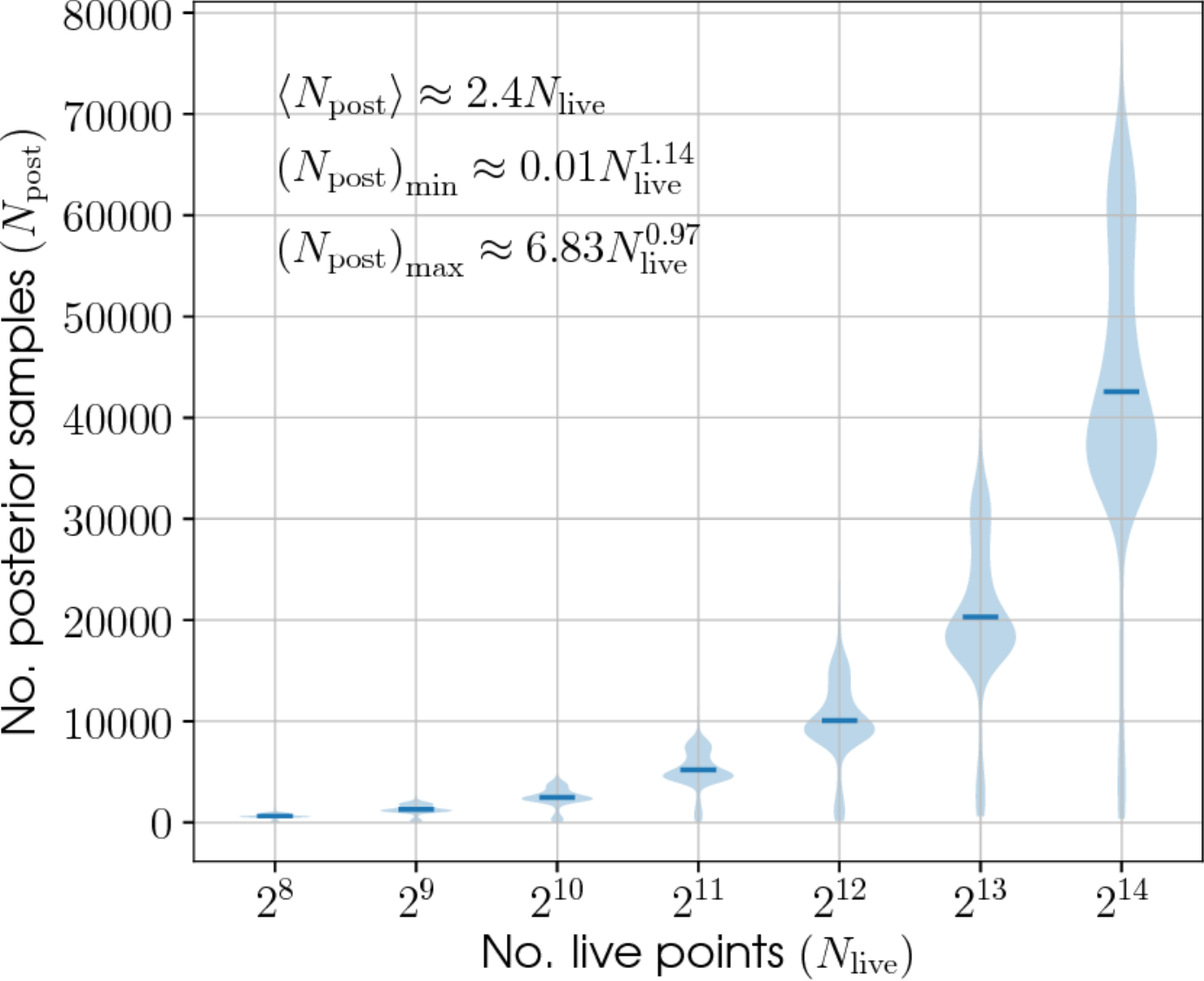}
\caption{ \protect\input{./figures/codeeval/stats/numposts/caption}}
\end{center}
\end{figure}

\subsection{Code timing}\label{sec:timing}

How long the code takes to runs depends on two main factors: the number of live points used and the information gain going from prior to posterior. Increases
in both of these will generally mean that the code takes longer to run. Using the simulations from \S\ref{sec:proposaltesting} we have assessed the code
run time for the simple testing Gaussian likelihood function. This allows us to quantify the code run time as a function of both number of live points and information
gain.

We find that the median calculation time for the simple Gaussian likelihood function used in \S\ref{sec:proposaltesting} (on an Intel ``Haswell'' processor,
specifically Intel CPU E5-2680 v3 @ 2.50GHz), including some standard overheads\footnote{Without the overheads it is roughly $\sim 3.7\ee{-7}$ seconds.} is 
$\sim 5.9\ee{-6}$ seconds. The median time for calculation of the likelihood (including model calculation time) for a standard search (Eqn.~\ref{eq:stlikelihood})
over only non-phase evolution parameters, and for one chunk of data, is $8.1\ee{-5}$ seconds.\footnote{Without the additional overheads this is a similar time of
$4.9\ee{-5}$ seconds.}

The median run time per likelihood evaluation for the nested sampling part of the code for the standard likelihood function (using 1024 live points) 
is $8.5\ee{-4}$ seconds, which is $\sim 10$ times greater than the single likelihood evaluation time (this ratio will change as a function of the number
of live points). The median run time per likelihood evaluation for the nested sampling part of the code for the test Gaussian likelihood function
(using 1024 live points) is $1.8\ee{-4}$ seconds, which is $\sim 30$ times greater than the single likelihood evaluation time.

In Figure~\ref{fig:timings} we show the time taken for the runs in \S\ref{sec:proposaltesting},
using the default proposals, as a function of number of live points and information gain. The code run time is divided by $\mathcal{T}_{L}= 5.9\ee{-6}$ to
provide a way to scale it for other values of $\mathcal{T}_{L}$, e.g.\ that found from the code's standard likelihood function. If we perform a 2D linear
fit to the natural logarithm of the median run time, $R/\mathcal{T}_L$, we get the relation
\begin{align}\label{eq:runtime}
 \ln{\left(\frac{R}{\mathcal{T}_L}\right)} &\approx 1.57 \ln{N_{\text{live}}} + 1.05 \ln{H} + 3.60, \\
 \frac{R}{\mathcal{T}_L} &\approx 36.4 N_{\text{live}}^{1.57} H^{1.05}.
\end{align}
This relation should be considered as a rough lower limit to the run time for any particular likelihood function. Above, we saw that for our standard likelihood
function the nested sampling algorithm was roughly three times quicker per likelihood evaluation than the simple test function. In both cases the internal MCMC, which
draws new samples within the algorithm, required very similar small lengths to give uncorrelated samples, so there must be other internal bottlenecks that make the
simpler likelihood relatively slower. However, in general, longer MCMC chains will be required to give uncorrelated samples, so the number of likelihood evaluations
will increase.

\begin{figure}[!phtb]
\begin{center}
\includegraphics[width=1\columnwidth]{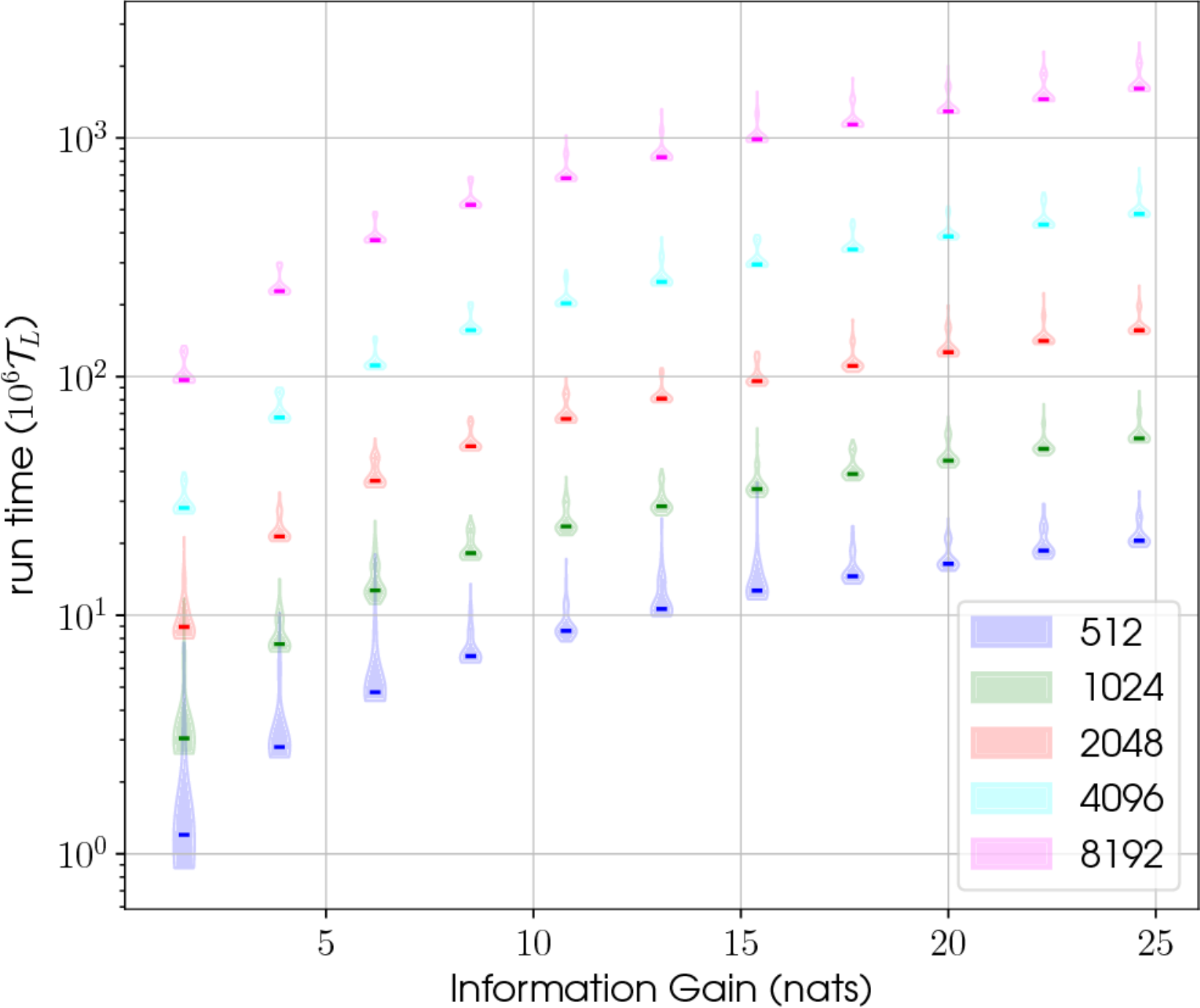}
\caption{ \protect\input{./figures/proptesting/walk_uniform_prop/timing/caption}}
\end{center}
\end{figure}

\section{Evaluating the code}\label{sec:codeeval}

Here we will evaluate the code in a variety of ways. We compare a selection of posterior probability distributions generated using \lppen,
over a range of signal parameters, with those produced using a previous code. We also use a selection of many simulated signals to show how the
odds comparing a signal model and a noise only model change with signal-to-noise ratio, along with comparing coherent verses incoherent signals
(for coherent simulated signal and incoherent simulations). The simulations also allow us to evaluate the posterior probability distributions
produced by the code using what is colloquially known as 
\href{https://en.wikipedia.org/wiki/P\%E2\%80\%93P\_plot}{``P-P plots''} (see \S\ref{sec:ppplots})

\subsection{Code comparison}

In this section we will show how the posterior parameter distributions that we can generate from the output
of our nested sampling code compare to evaluating them with a previous implementation of the analysis code
\citep[\lppef, or for the rest of this document shortened to \lppe, used in, e.g.,][]{2014ApJ...785..119A} using a (generally inefficient) MCMC algorithm
and over a grid in parameters.\footnote{For the MCMC examples in this section, we have used ``burn-in'' periods of 100\,000 iterations, followed by
500\,000 posterior sample iterations, from which there have been roughly 1000 independent samples used to produce the posterior plots. For the grid-based examples
the grid sampling used has been $N_{h_0}\times N_{\phi_0} \times N_{\cos{\iota}} \times N_{\psi} = 80\times40\times40\times40$.} We will show comparisons of the codes in two different regimes: no signal is present in the data, and an obvious signal is
present in the data. We do this using simulated data both containing purely Gaussian noise and containing simulated signals
added to Gaussian noise.

In each of the tests below we will be running our code (\lppen) using the default proposal distributions discussed in \S\ref{sec:proposals}
and with the number of live points fixed to be 2048. We will also assume a source at a fixed sky position and that the signal model is purely
that for the $l=m=2$ harmonic (Equation~\ref{eq:h2f}), and that we work in terms of the signal amplitude $h_0$ and signal phase
$\Phi_{22}^C$ (this is for consistency with the old code which uses $\Phi_{22}^C$ rather than $\phi_0$, which we have earlier defined as the
rotational phase). The simulated data in all cases 
will be generated as the complex heterodyned
time series sampled once per minute. When comparisons between the current code and previous codes are 
made we run the current code in a way that splits datasets into 30 point chunks, rather than using the algorithm described in \S\ref{sec:splitting},
as this makes it consistent with the old code (\lppe). We also purely use the uniform prior ranges for the amplitude parameter to be
consistent between codes.

\subsubsection{Simulated noise}\label{sec:simnoise}

An initial test of the code is whether the output posterior probability distributions match those produced when evaluating the
posterior over a fixed grid in the four-dimensional parameter space of $\vec{\theta} = \{h_0, \cos{\iota}, \psi, \phi_{22}^C\}$, when
using purely Gaussian noise.\footnote{It is worth noting that in many of these tests the polarisation angle prior covers $-\pi/4$ to $\pi/4$
rather than the 0 to $\pi/2$ range given in, e.g., \citet{2015MNRAS.453.4399P}, due to this being the range required by the older
parameter estimation MCMC code. Although this range is degenerate to rotations over $\pi/2$, so spans the same range of waveform models.}
A comparison of the marginalised posteriors for individual parameters, and pairs of parameters, are shown when using simulated data lasting
ten days from a single detector (in this case assumed to be the LIGO Hanford detector, H1) in Figure~\ref{fig:simnoise_single}.

\begin{figure}[!phtb]
\begin{center}
\includegraphics[width=1\columnwidth]{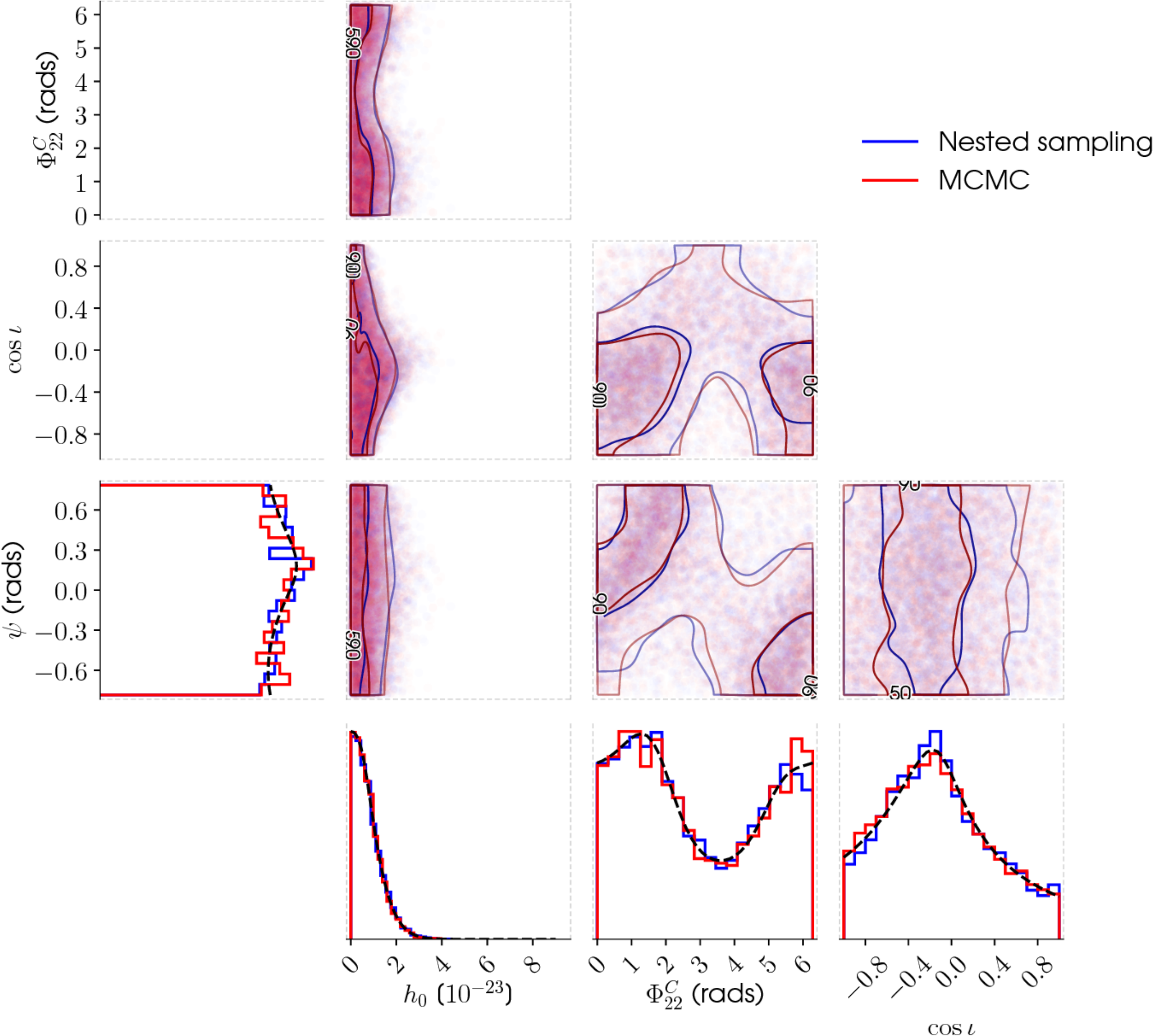}
\caption{ \protect\input{./figures/codeeval/simulations/noise/caption}}
\end{center}
\end{figure}

It can be seen that the output posteriors look qualitatively consistent. However, we can also quantify some aspects of consistency by
comparing the evidence output from the nested sampling code with that estimated from the grid-based method from \lppe, the upper limits
on $h_0$ produced by the codes, and performing Kolmogorov-Smirnov consistency tests between the nested sampling posterior samples
and MCMC posterior samples (giving a $p$-value for the null hypothesis that the samples are from the same distributions).
These are shown in Table~\ref{tab:codeeval} and show very good consistency between the codes, although it should be noted that
these are for one particular run and some statistical fluctuations in the exact values for different runs are to be expected (see, e.g.,
the variations in evidence values in \S\ref{sec:proposaltesting}). A {\tt jupyter} notebook with this test can be found \href{https://github.com/mattpitkin/CW_nested_sampling_doc/blob/master/figures/codeeval/simulations/noise/SimulatedNoiseTestsPaper.ipynb}{here}.

\begin{table*}[hptb]
\caption{Consistency tests between outputs of the new code, \lppen, and the old code, \lppe, when running on simulated data
and searching over the four parameters $\{h_0, \cos{\iota}, \psi, \Phi_{22}^C\}$.\label{tab:codeeval}}
\begin{center}
\begin{tabular}{l c c | c c c c}
\hline
\multirow{2}{*}{Simulation} & \multirow{2}{*}{$\ln{\left(\frac{Z_{\text{nested}}}{Z_{\text{grid}}}\right)}$} & \multirow{2}{*}{$\frac{(h_0^{95\%})_{\text{nested}}}{(h_0^{95\%})_{\text{grid}}}$} & 
\multicolumn{4}{c}{K-S $p$-value} \\ \cline{4-7}
 &  &  & $p(h_0)$ & $p(\Phi_C^{22})$ & $p(\cos{\iota})$ & $p(\psi)$ \\                      
\hline
\hline
Noise (single detector)  & $-0.07$ & $1.009$ & 0.404 & 0.026 & 0.010 & 0.703 \\
Noise (two detectors)    & $-0.05$ & $0.992$ & 0.141 & 0.564 & 0.538 & 0.493 \\
Signal (single detector) & 0.245   & $1.004$ & 0.237 & 0.291 & 0.125 & 0.175 \\
Signal (two detectors)   & 0.240   & 0.991   & 0.722 & 0.052 & 0.067 & 0.032 \\
\hline
\end{tabular}
\end{center}
\end{table*}

We see a very similar situation, in terms of agreement between the codes, when running on simulated data assumed to be from two detectors (the LIGO
H1 and L1 sites) as shown in Figure~\ref{fig:simnoise_multi} and the second line of Table~\ref{tab:codeeval}. A {\tt jupyter} notebook with this test
can be found \href{https://github.com/mattpitkin/CW_nested_sampling_doc/blob/master/figures/codeeval/simulations/noise_multidet/SimulatedNoiseTestsMultidetPaper.ipynb}{here}.

\begin{figure}[!phtb]
\begin{center}
\includegraphics[width=1\columnwidth]{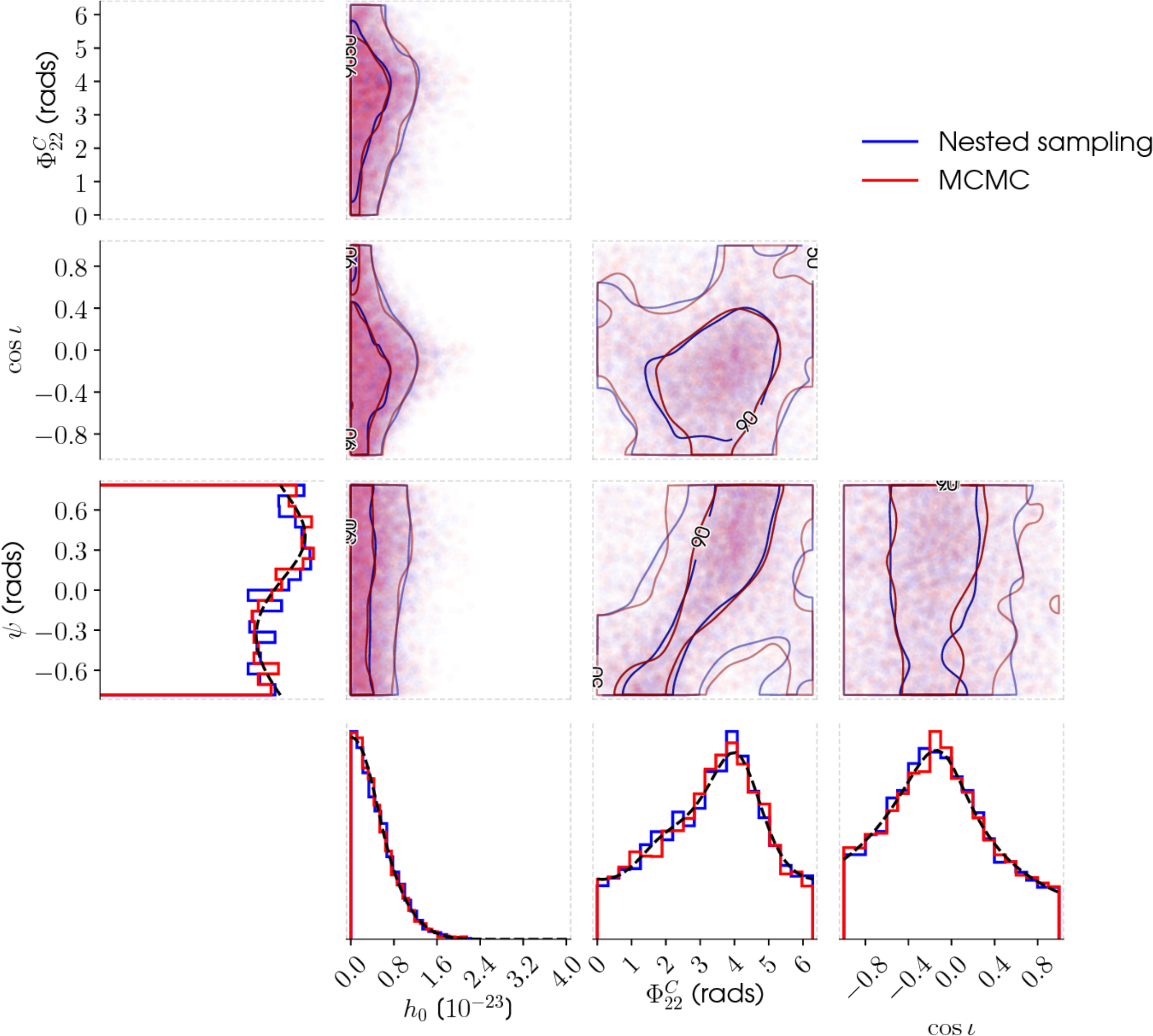}
\caption{ \protect\input{./figures/codeeval/simulations/noise_multidet/caption}}
\end{center}
\end{figure}

The above tests using simulated noise, and with searches covering the four standard unknown \gw parameters, show that the code appears to
be working as expected, and is in agreement with our previously used code. However, more generally we can perform parameter estimation
over a larger number of signal parameters, and can again check for consistency with the previous code. We produce one day of Gaussian
noise for a single detector and search over the four \gw parameter with the same priors as before, but then also use a multi-variate Gaussian prior (\S\ref{sec:gaussianprior})
over the {\it phase} parameters: rotational frequency and first frequency derivative, and binary parameters of binary period, time of periastron, angle of
periastron and its first derivative, projected semi-major axis, and eccentricity ($\{f,\dot{f},P_{\text{b}}, T_0, \omega_0, \dot{\omega}_0, a\sin{i}, e\}$).
This gives a total of 12 parameters in the search. The multi-variate prior is defined such that all parameters are uncorrelated except, in this case,
the parameter pairs $[T_0, \omega_0]$ and $[P_{\text{b}}, \dot{\omega}_0]$ for which we use a very high correlation of $0.9999$ (we do not set them to
be fully correlated due to numerical issues inverting such matrices).

The one-and-two dimensional posteriors (expressed as offsets from the assumed heterodyne parameter values) output for this case when run with both
\lppe and \lppen can be seen in Figure~\ref{fig:noise_multiparam},
which also overlays the marginalised priors on top. It can be seen qualitatively that the codes are consistent\footnote{It is worth mentioning that in performing
these tests a bug was discovered in \lppe in which Equation~\ref{eq:deltaphi} was being applied with the wrong sign, leading to parameter estimates
having the wrong sign. However, this bug was fixed for the example shown here.} and for the {\it phase} parameters the priors are recovered, except for
$\dot{f}$, which has some structure due to the specifics of the noise realisation. The Kolmogorov-Smirnov test $p$-values testing the null hypothesis
that the posterior samples from each code are drawn from the same distribution are shown in Table~\ref{tab:noisemultiks}, and suggest that the
distributions are consistent.

\begin{table*}[hptb]
\caption{Kolmogorov-Smirnov test $p$-values testing the null hypothesis that the samples output by \lppen and \lppe, when running on simulated
noise data and searching over the twelve parameters $\{h_0, \cos{\iota}, \psi, \Phi_{22}^C, f,\dot{f},P_{\text{b}}, T_0, \omega_0, \dot{\omega}_0, a\sin{i}, e\}$,
are drawn from the same distributions.\label{tab:noisemultiks}}
\begin{center}
\begin{tabular}{l | c c c c c c c c c c c c}
\hline
Parameter & $h_0$ & $\Phi_C^{22}$ & $\cos{\iota}$ & $\psi$ & $f$ & $\dot{f}$ & $P_{\text{b}}$ & $T_0$ & $\omega_0$ & $\dot{\omega}_0$ & $a\sin{i}$ & $e$ \\                      
\hline
\hline
$p$-value  & 0.13 & 0.04 & 0.78 & 0.23 & 0.44 & 0.26 & 0.85 & 0.28 & 0.29 & 0.86 & 0.52 & 0.28 \\
\hline
\end{tabular}
\end{center}
\end{table*}

\begin{figure*}[!phtb]
\begin{center}
\includegraphics[width=0.98\textwidth]{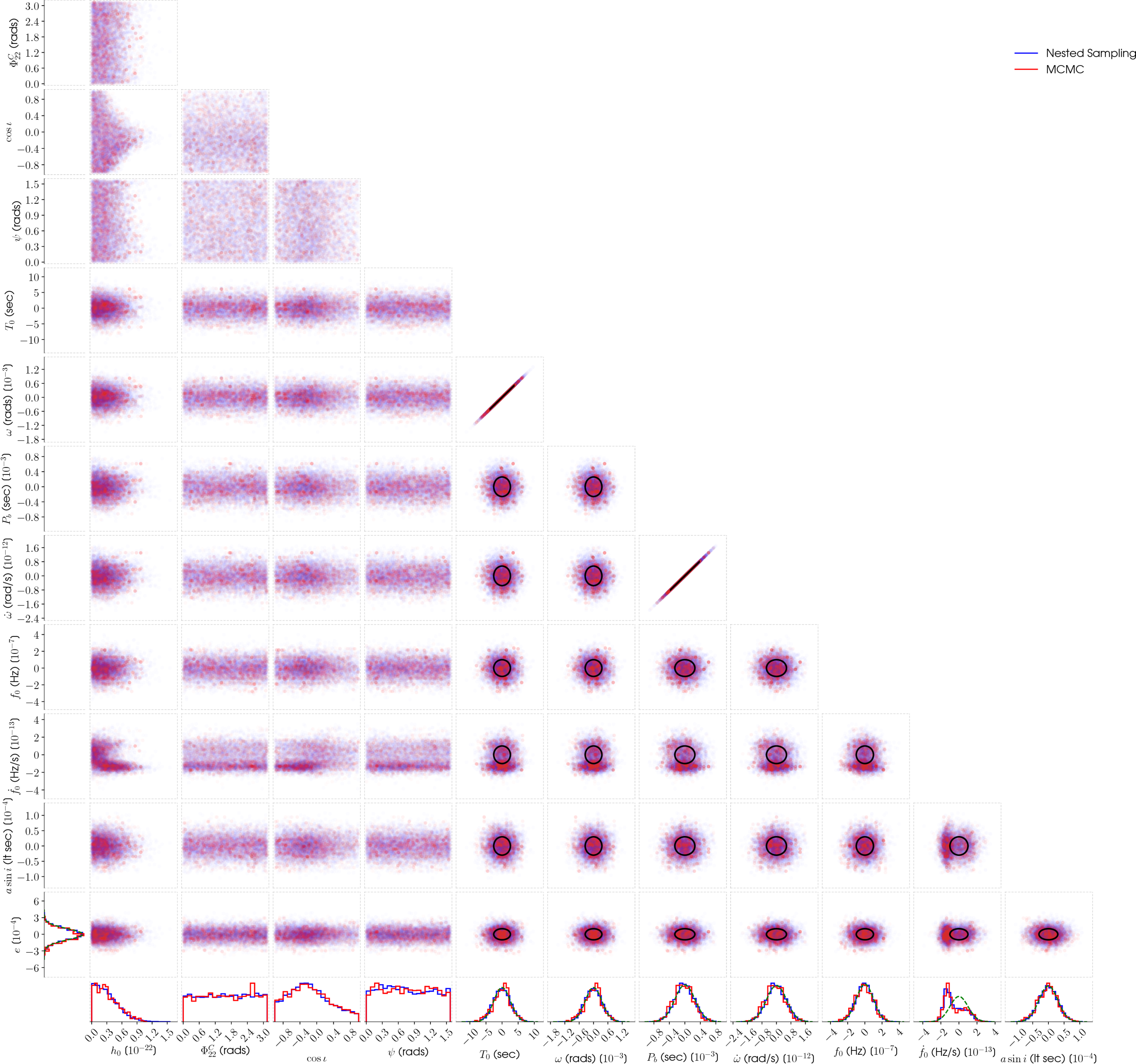}
\caption{ \protect\input{./figures/codeeval/simulations/noise_multiparam/caption}}
\end{center}
\end{figure*}

\subsubsection{Simulated signals}\label{sec:simsignal}

The next test is checking how the codes compare when there is a signal present in the data. When generating the simulated signals\footnote{Simulated signals
used in this work have either been generated directly using \lppen (see, e.g., Appendix~\ref{sec:example2}), or using functions within the {\tt pulsarpputils.py} 
{\tt python} module in {\tt lalapps} within LALSuite \citep{lalsuite}.} we initially
assume that the signal's phase evolution is perfectly known and has been removed via the heterodyne described in \S\ref{sec:model}, and therefore
$\Delta\phi_2 = 0$ from Equation~\ref{eq:deltaphi}. As in \S\ref{sec:simnoise} we search over the four-dimensional parameter space of
$\vec{\theta} = \{h_0, \cos{\iota}, \psi, \Phi_{22}^C\}$.

We create a simulated signal in the H1 detector spanning ten days of data with parameters ($h_0 = 6.2\ee{-23}$, $\Phi_{22}^C = 2.4$\,rad, $\cos{\iota} = 0.3$
and $\psi = 0.1$\,rad) that produce a signal-to-noise ratio of 8 when injected into
Gaussian noise. The posterior probability distributions estimated for the parameters (expressed as offsets from the assumed heterodyne parameter values) are
shown in Figure~\ref{fig:simsignal_single}, which show consistency
between the old and new codes and with the injected signal parameters. The quantitative consistency between the codes can be seen in Table~\ref{tab:codeeval},
where is should be noted that the evidence ratio between the codes is larger than statistical uncertainty would expect, although the fractional difference between
the values ($\lesssim 1\%$) is still small enough to not greatly effect conclusions drawn from either values if used in model selection.

\begin{figure}[!phtb]
\begin{center}
\includegraphics[width=1\columnwidth]{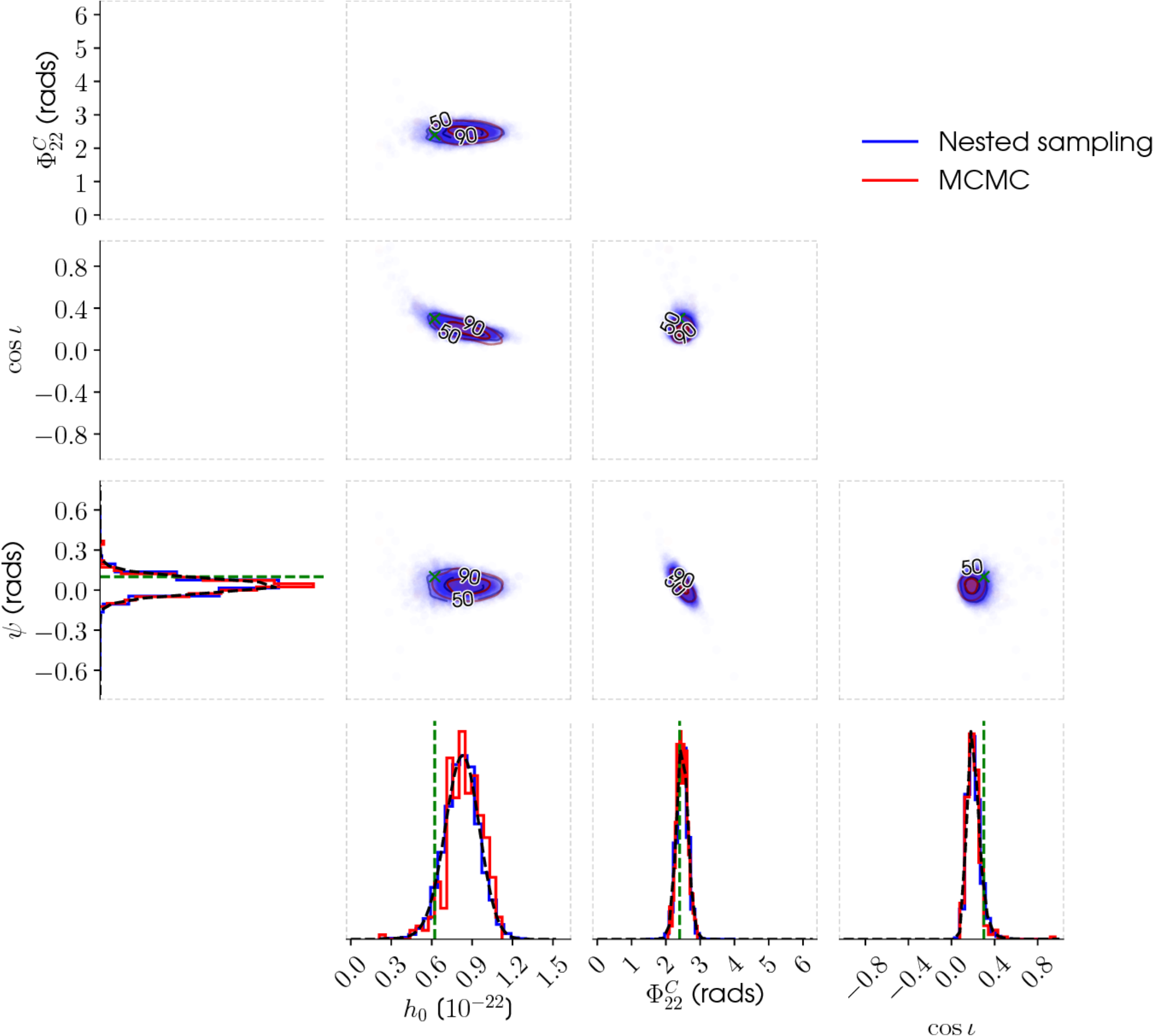}
\caption{ \protect\input{./figures/codeeval/simulations/signal/caption}}
\end{center}
\end{figure}

Similarly, using a signal with a coherent multi-detector signal-to-noise ratio of 8 (for the same parameters as above except with $h_0 = 3.5\ee{-23}$) when
simulated and injected into Gaussian noise for two detectors (H1
and L1) we see the posteriors shown in Figure~\ref{fig:simsignal_multi}, and consistency checks in Table~\ref{tab:codeeval}. The notebooks for the above two tests
can be found \href{https://github.com/mattpitkin/CW_nested_sampling_doc/blob/master/figures/codeeval/simulations/signal/SimulatedSignalTestsPaper.ipynb}{here} and 
\href{https://github.com/mattpitkin/CW_nested_sampling_doc/blob/master/figures/codeeval/simulations/signal_multidet/SimulatedSignalMultidetTestsPaper.ipynb}{here}.

\begin{figure}[!phtb]
\begin{center}
\includegraphics[width=1\columnwidth]{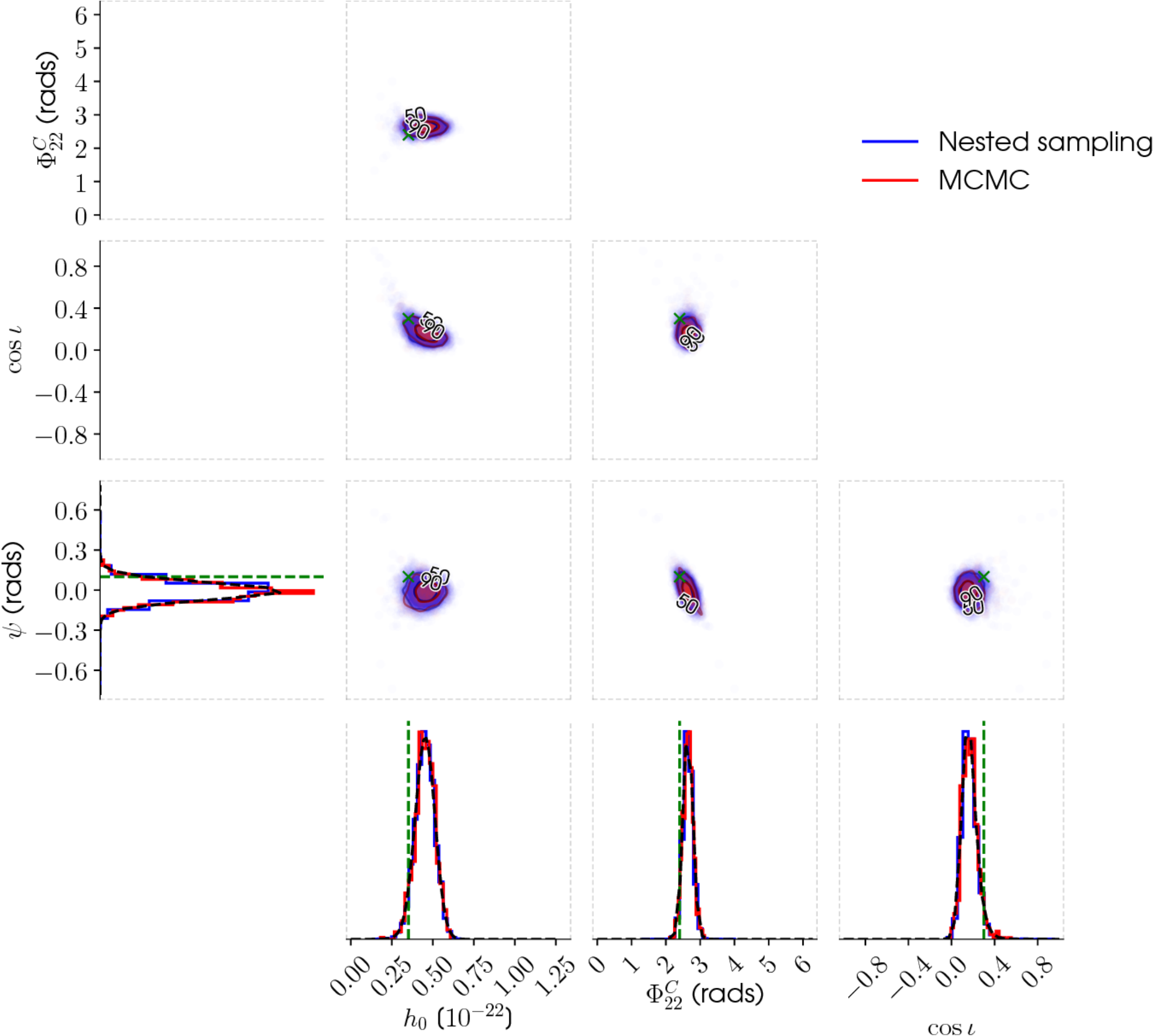}
\caption{ \protect\input{./figures/codeeval/simulations/signal_multidet/caption}}
\end{center}
\end{figure}

A set of simulated signals with sky location and orbital parameters set to those of the Low-mass X-ray binary Sco X-1 were injected into
Gaussian noise for the study in \citet{2015PhRvD..92b3006M}. As a test of our code when recovering a signal from a source in a binary system
we have recovered one of these signals using approximately 1.4 days of simulated data. We have purposely performed the heterodyne data processing stage
with values of several of the phase parameters ($f_0$, $\alpha$, $T_0$, $a\sin{i}$ and $P_{\text{b}}$) set to {\it not} match the known simulated signal
values. Therefore, to recover the signal, in addition to the four \gw parameters we have had to allow the code to search over these extra parameters.
For these additional parameters Gaussian priors were used, with the prior means set to be the parameter values used for heterodyning (not those for
the actual signal), with standard deviations wide enough to encompass the true signal values. For $h_0$ a Fermi-Dirac prior was used, whilst flat prior were used for
the other angle parameters, covering their minimal allowed ranges. The recovered signal posterior distributions can be seen
in Figure~\ref{fig:scox1_inj}, which show that the true parameters are correctly recovered (the plot shows recovered phase parameters as offsets from
the heterodyne parameter values, i.e.\ the centres of the Gaussian priors). It can be seen that for $f_0$ and $P_{\text{b}}$ the posteriors contain the
correct value, but are peaked well away from the value used for heterodyning (which would be at zero in the plots), showing the code's ability to explore
the prior space in these parameters. The one-dimensional marginal distribution for $\phi_0$ spans the full prior range, however, this is due to it being
highly correlated with $f_0$, and, although not visible on the plot, this very strong correlation is present in the two-dimensional posterior for these
parameters. Finally, for $\alpha$ and $T_0$, it can be seen that the posteriors just fill the prior ranges as the data provides no additional information
about these parameters. For completeness, we find that the signal was recovered with SNRs of $\sim 20$ in both H1 and L1 individually, with a coherent
SNR of $\sim 28$, and odds values for the multi-detector analysis of $\log{}_{10}\left(\mathcal{O}_{\text{S}/\text{N}}\right) = 152$
and $\log{}_{10}\left(\mathcal{O}_{\text{S}/\text{I}}\right) = 7.8$.

\begin{figure*}[!phtb]
\begin{center}
\includegraphics[width=0.9\textwidth]{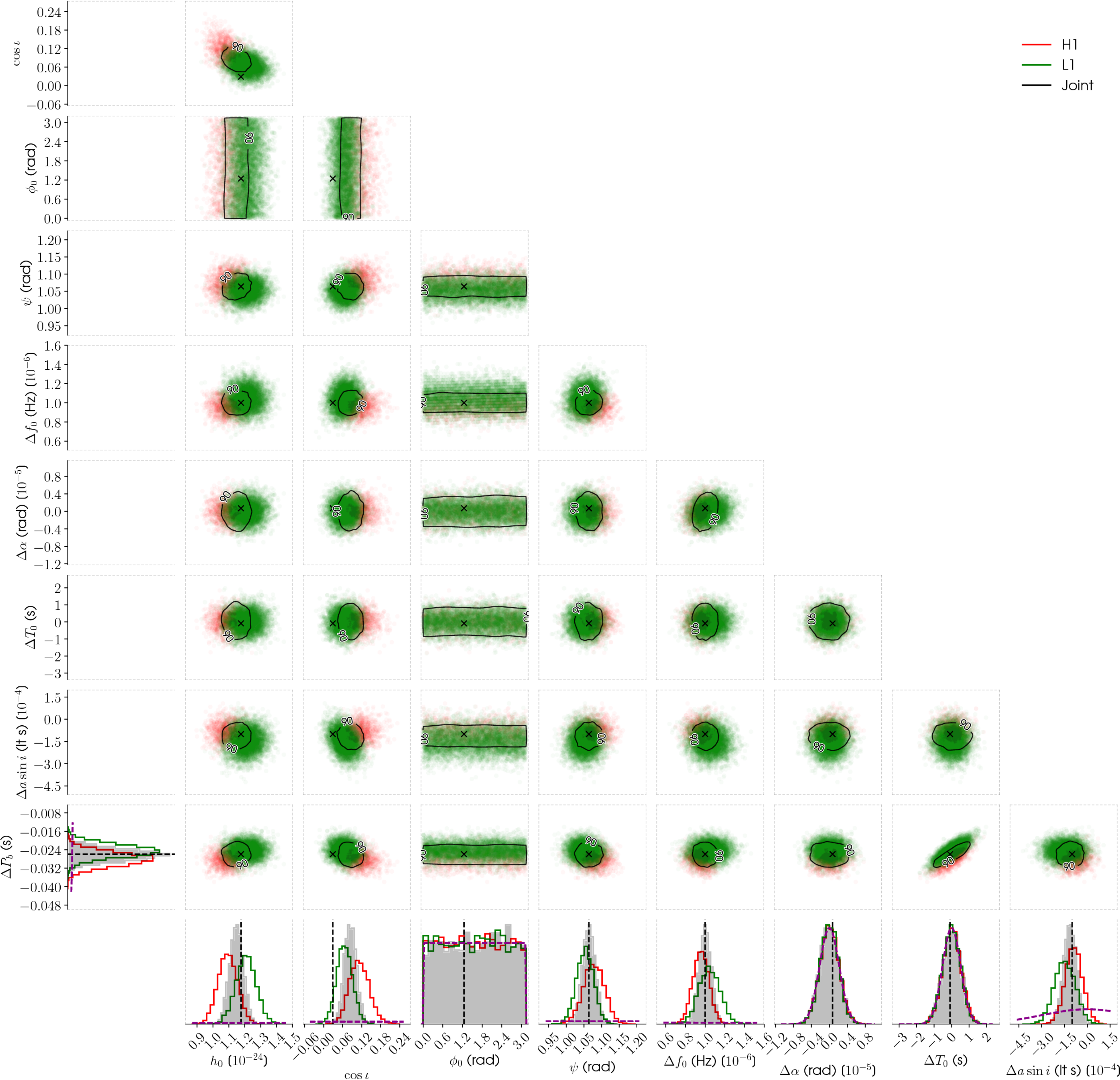}
\caption{ \protect\input{./figures/codeeval/simulations/scox1_inj/caption}}
\end{center}
\end{figure*}

\subsubsection{Results evaluation}\label{sec:reseval}

The above results have shown good qualitative, and quantitative, agreement between the old and new codes for individual cases, but it is also useful
to see how they compare for a large number of noise realisations when assessing two important quantities: the 95\% upper limit on \gw amplitude $h_0$,
and the signal model evidence. It is also useful to see how these compare as a function of the number of live points used by the code. To this end we
have chosen numbers of live points from 256 to 16384, increasing in a power of two for each step, and for each number generated 500 realisations of
complex Gaussian noise with 1440 points over one day. Choosing a random source sky location for each realisation, and using uniform priors for the
parameters $h_0$, $\cos{\iota}$, $\Phi_{22}^C$ and $\psi$, we have performed parameter estimation using both \lppen and \lppe in its grid-based mode.
For \lppe the integrals required for calculating the signal evidence and 95\% upper limit on $h_0$ are performed using the trapezium rule, whilst
for the output of \lppen the upper limit has been calculated using a greedy-binning approach bounded from zero. This assessment is very similar to what
we did in \S\ref{sec:proposaltesting}, but has not just used a fixed Gaussian likelihood function in each case. We will make the assumption that the
grid-based results are a representation of the true values for the evidence and upper limit, however, it should be noted that there will in fact be
errors (and potentially biases) from the trapezium rule approximation.

In Figure~\ref{fig:nest_evs} we see a comparison between the signal evidences evaluated using nested sampling and the trapezium rule. The left-hand
axis shows the logarithm of the ratio of the two values, whilst the right hand axis shows the actual percentage difference between the evidences. These
are plotted as a function of the number of live points used in the nested sampling evaluation by \lppen. We find that the distribution of evidences
very closely follows the theoretically expected distribution calculated as $\sigma_{\mathcal{Z}} = \sqrt{H/N_{\text{live}}}$ (see
\S\ref{sec:proposaltesting}), where given our prior range set up we found $\langle H \rangle \approx 2.45$. It can also be seen that there is a slight
bias towards the nested sampling evidence being underestimated by a few percent, as was also observed for the simple Gaussian likelihood function case
in Figure~\ref{fig:walkunipropevs}.

\begin{figure}[!phtb]
\begin{center}
\includegraphics[width=1\columnwidth]{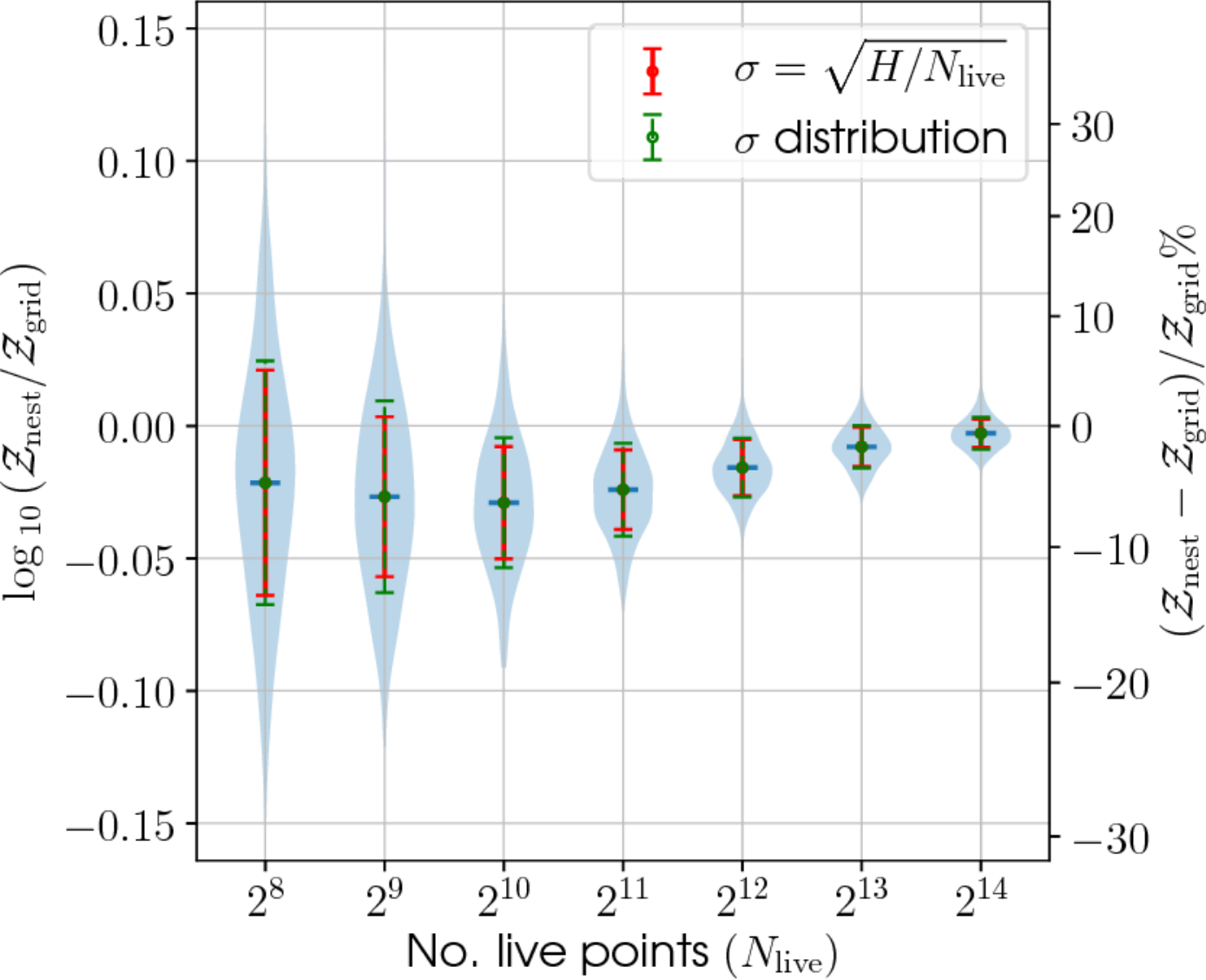}
\caption{ \protect\input{./figures/codeeval/stats/nest_evs/caption}}
\end{center}
\end{figure}

In Figure~\ref{fig:uls} we see a comparison between the 95\% upper limits on $h_0$ as a function of the number of live points. The core distribution of upper
limits produced via nested sampling are centred around the true (or grid-based) values, with the distribution decreasing in width with increasing numbers of live points.
The standard deviation on the distribution roughly follows $2^{6.8}N_{\text{live}}^{-1/2}$. The figure shows some extreme outliers in the distributions, but it
is found that these outliers occur in the cases when the number of posterior samples drawn from the nested samples is extremely low (the lower tails of the
distributions in Figure~\ref{fig:numposts}).

\begin{figure}[!phtb]
\begin{center}
\includegraphics[width=1\columnwidth]{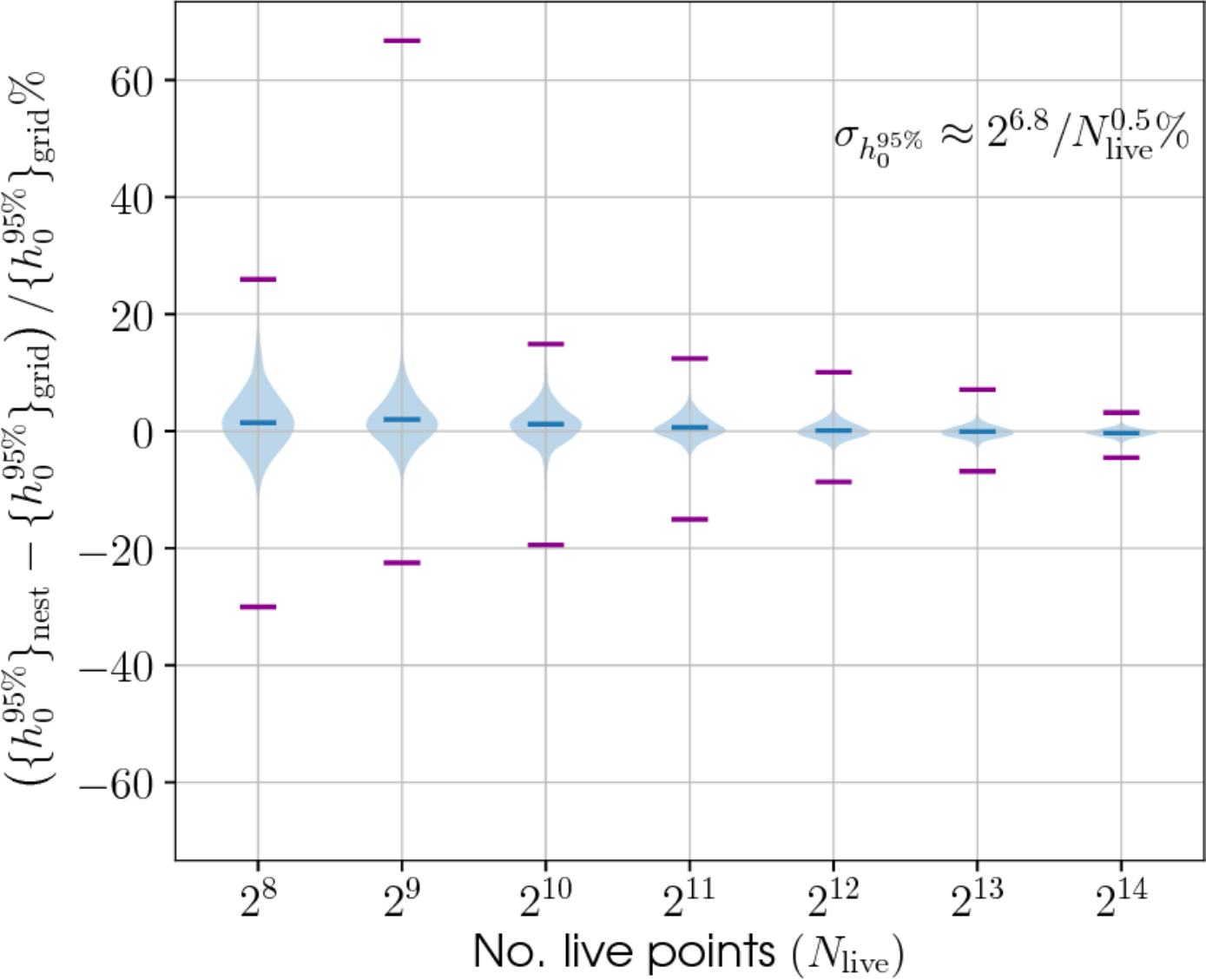}
\caption{ \protect\input{./figures/codeeval/stats/uls/caption}}
\end{center}
\end{figure}

\subsection{Injections into real data}

During all bar the first initial LIGO science run, simulated signals from a variety of sources have been physically ``injected'' into the detector
datastreams. These are known as ``hardware injections'' \citep[see][for a general discussion of hardware injections, in particular relating to their
use, and extraction, in advanced LIGO's first observing run]{2016arXiv161207864B}. These have included a range of continuous wave signals of the form given
by, e.g., Equation~2 of \citet{2017arXiv170107709T}, which in past have been searched for using a variety of analysis pipelines, including the one
described in this document using \lppe \citep[see, e.g., Appendix~B of][]{2007PhRvD..76d2001A}.

Here we have used \lppen to perform parameter estimation on the parameters $\vec{\theta} = \{h_0, \cos{\iota}, \psi, \phi_{0}\}$ for two hardware injections added
to intial LIGO data from the sixth science run (S6).\footnote{$\phi_0$ is being used here as opposed to $\Phi_{22}^C$ as we are not having to compare to the previous
code.}\footnote{S6 data is publically available at \url{https://losc.ligo.org} \citep{2015JPhCS.610a2021V}.} These are two out of the thirteen total continuous wave
injections and were called ``Pulsar03'' and ``Pulsar05''
respectively. The data have been heterodyned with the known phase evolution of the signals. The extracted posterior distributions can be seen in
Figures~\ref{fig:hwinj03} and \ref{fig:hwinj05} and are in very good agreement with the injection parameters.\footnote{For these hardware injections
exact agreement between the expected injection parameters and those recovered, or between detectors, is not entirely expected. This is due to the
injections needing to be performed using actuation functions (converting the expected \gw strain into a force required to be exerted in the
interferometer test mass) that may not exactly match the actuations functions later required for calibrating the detectors.} For ``Pulsar03''
the signal is recovered with SNRs of 164 and 96 in H1 and L1 respectively, and a coherent SNR of 190, and with
$\log{}_{10}\left(\mathcal{O}_{\text{S}/\text{N}}\right) = 6729$ and $\log{}_{10}\left(\mathcal{O}_{\text{S}/\text{I}}\right) = 9.1$. For ``Pulsar05''
the signal is recovered with SNRs of 9.9 and 6.4 in H1 and L1 respectively, and a coherent SNR of 11.8, and with
$\log{}_{10}\left(\mathcal{O}_{\text{S}/\text{N}}\right) = 22.5$ and $\log{}_{10}\left(\mathcal{O}_{\text{S}/\text{I}}\right) = 4.6$. The odds
would imply a very high significance in distinguishing the coherent signal model from the noise models.

\begin{figure}[!phtb]
\begin{center}
\includegraphics[width=1\columnwidth]{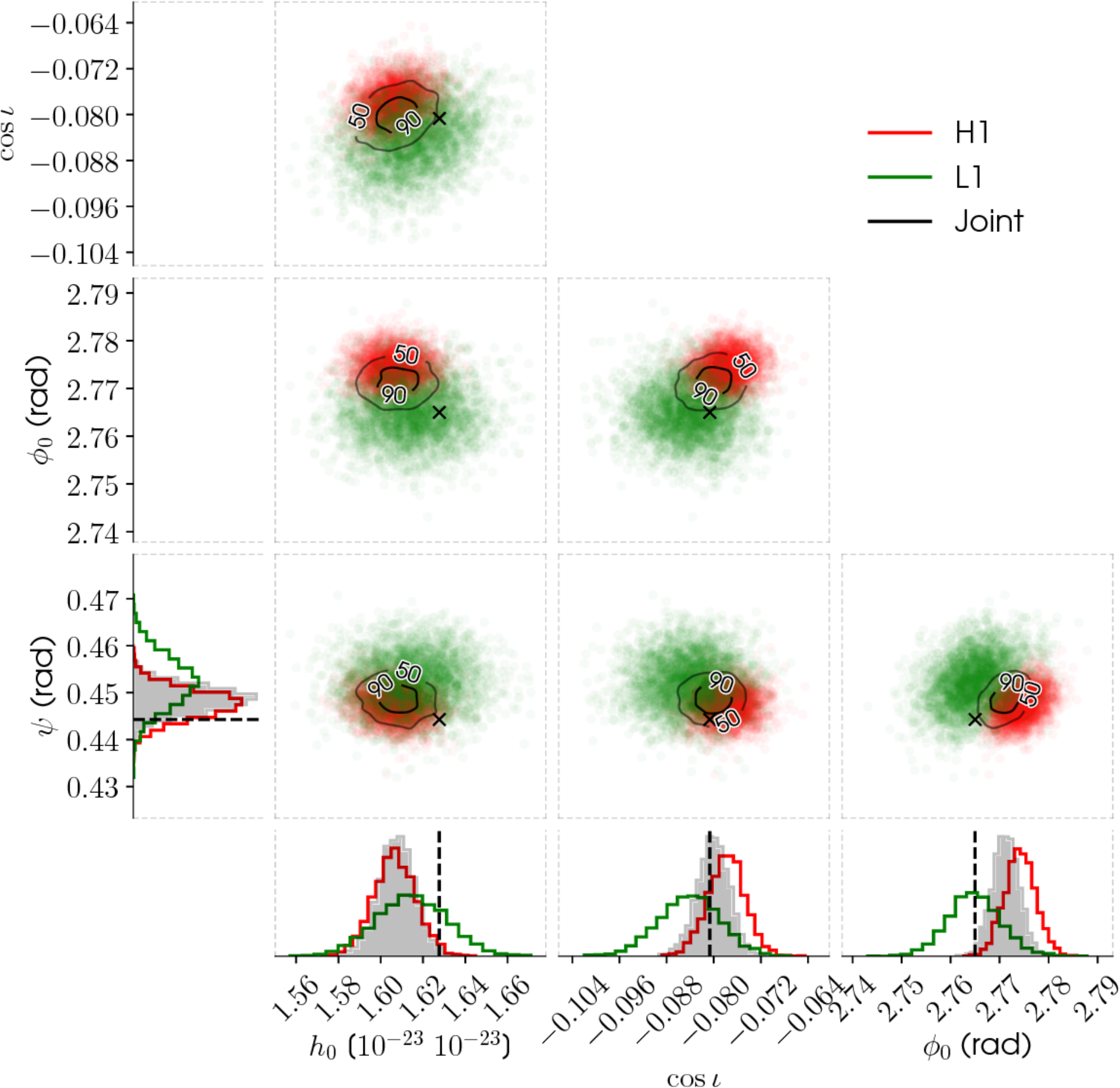}
\caption{ \protect\input{./figures/codeeval/simulations/S6_hwinj/hwinj03/caption}}
\end{center}
\end{figure}

\begin{figure}[!phtb]
\begin{center}
\includegraphics[width=1\columnwidth]{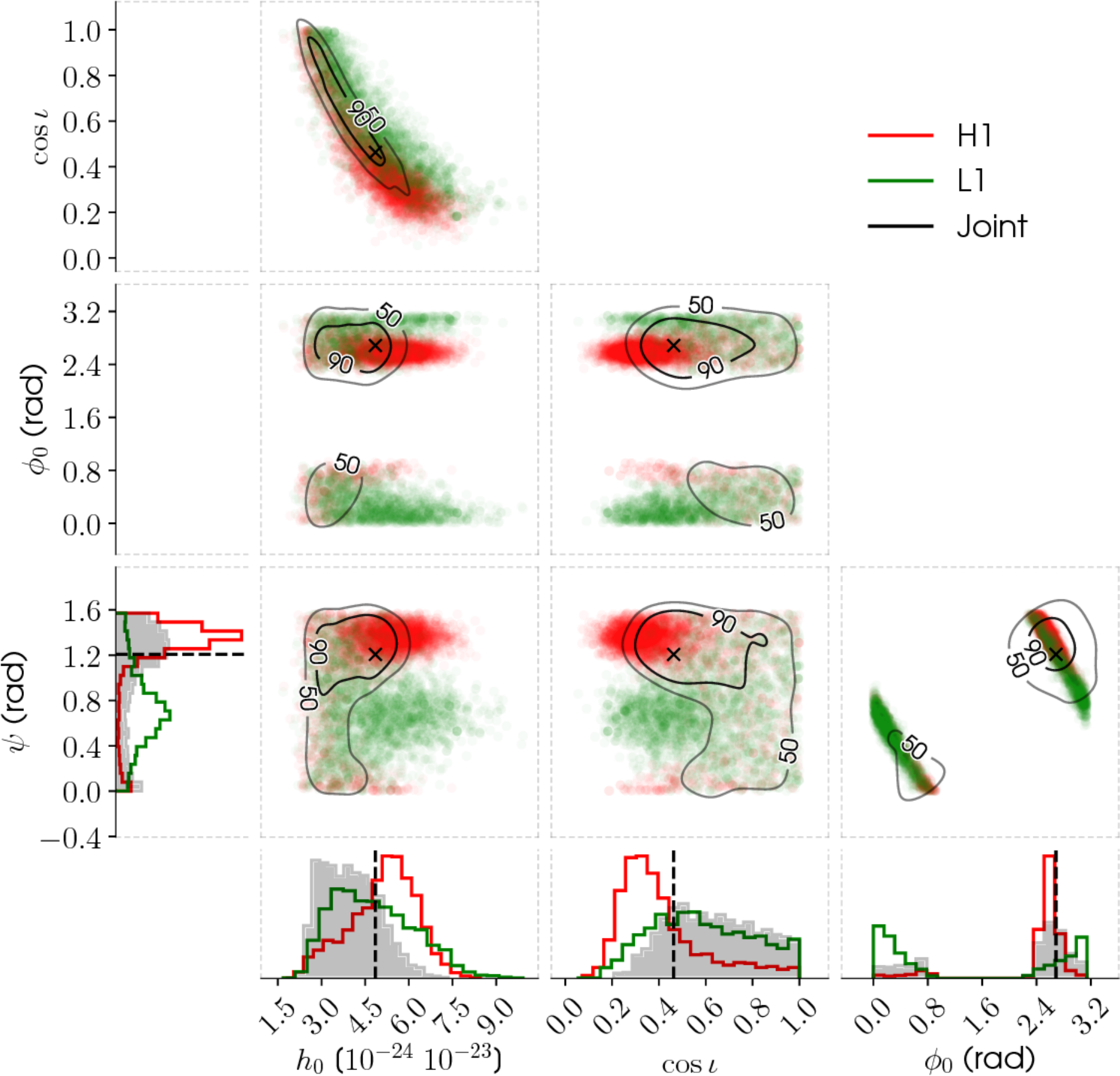}
\caption{ \protect\input{./figures/codeeval/simulations/S6_hwinj/hwinj05/caption}}
\end{center}
\end{figure}

Several thousand simulated signals have been added to the LIGO sixth science run dataset \citep[see Section~IV of][]{2016PhRvD..94l4010W}.
For one particular signal (chosen for its moderate SNR over the run) we have used \lppen to perform parameter estimation on the parameters
$\vec{\theta} = \{h_0, \cos{\iota}, \psi, \phi_{0}, f, \dot{f}, \alpha, \delta\}$. In this case, as for the binary signal discussed in \S\ref{sec:simsignal}, we
have purposely heterodyned the signal using values of $f$ and $\dot{f}$ offset
from their known rotational values by $\Delta f = 5\ee{-8}$\,Hz and $\Delta \dot{f} = -4\ee{-15}$\,Hz\,s$^{-1}$.\footnote{The frequency offset is approximately
two Fourier frequency bins away from the actual signal frequency.} We have used the whole of the S6 dataset,
spanning almost 1.3\,yr, with roughly 50\% duty factors for both H1 and L1. The code was run using the ROQ mode for each individual detector and a joint
detector analysis. Gaussian priors were used on the frequency and sky location parameters. The recovered marginalised posterior distributions can be seen
in Figure~\ref{fig:s6sw_inj}, where all parameters are observed to be correctly recovered at their true values, even $f$ and $\dot{f}$ which we offset (it is
hard to see in the plots, but these values are constrained well within their allowed prior ranges). This signal was recovered with SNR of roughly 29 and 21 in
H1 and L1 individually, and with a coherent SNR of 35. The odds values are $\log{}_{10}\left(\mathcal{O}_{\text{S}/\text{N}}\right) = 251$ and $\log{}_{10}\left(\mathcal{O}_{\text{S}/\text{I}}\right) = 9.5$ showing that the signal is very significantly distinguished from the noise model.

\begin{figure*}[!phtb]
\begin{center}
\includegraphics[width=0.9\textwidth]{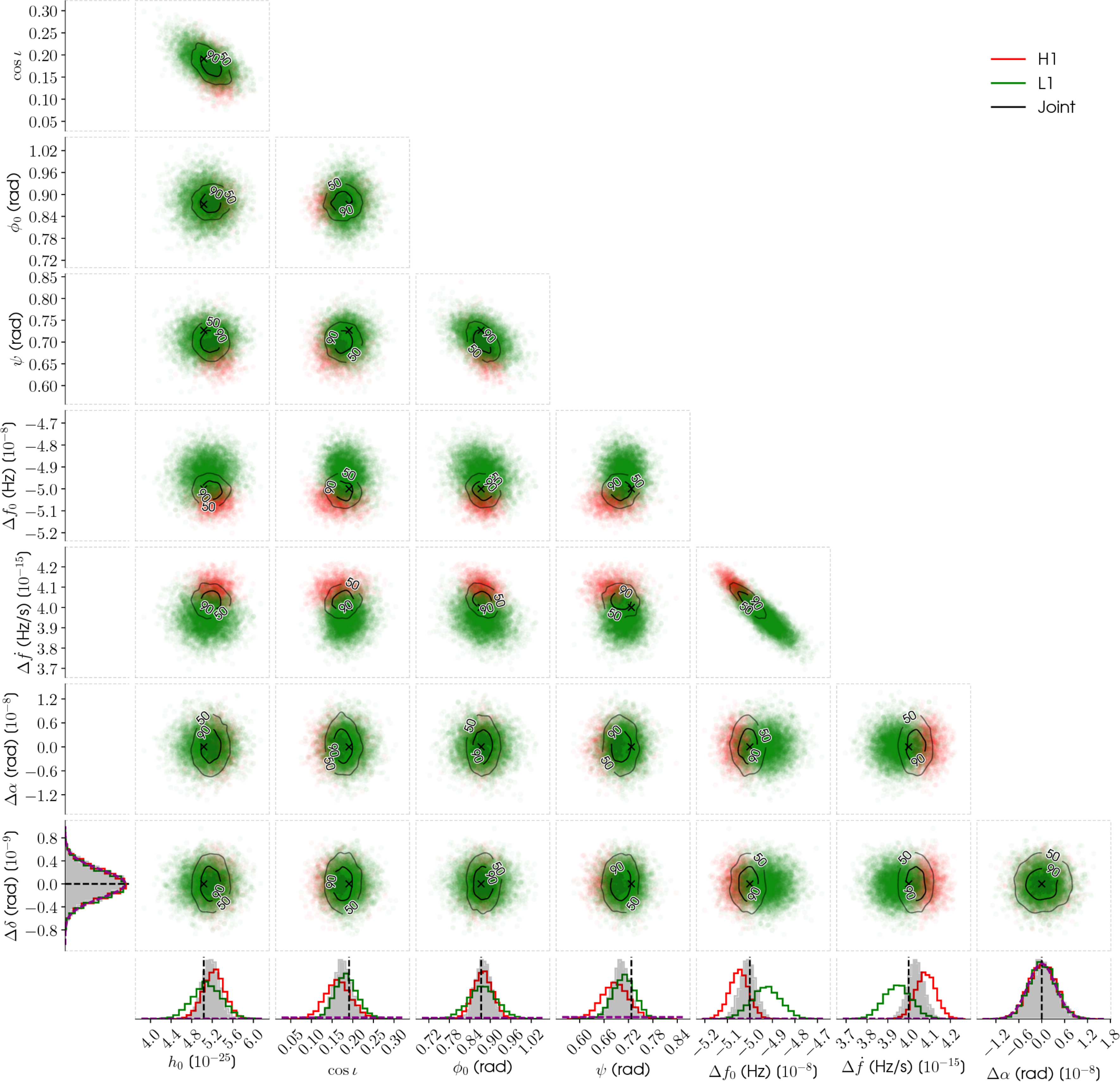}
\caption{ \protect\input{./figures/codeeval/simulations/S6_swinj/caption}}
\end{center}
\end{figure*}

\subsection{Monte-Carlo studies}

In this section we will assess the outputs of \lppen using a range of simulated signals. We have created two sets of 2000 signal parameters with which to
inject simulated signals into Gaussian noise (with the same standard deviation of $1\ee{-22}$) for the H1 and L1 detectors. Both sets have been generated
with $\phi_0$, $\psi$ and $\cos{\iota}$ drawn from uniform distributions over their minimal allowed ranges \citep[Table~1 of][]{2015MNRAS.453.4399P},
but for the first set $h_0$ is calculated such that the coherent SNR (Equation~\ref{eq:snr}) is drawn uniformly between 0 and 20; the second set draws the $h_0$ uniformly between
0 and $3.25\ee{-22}$, such that the maximum coherent SNR (given the known noise level) is $\sim 55$. This latter set is used to evaluate the posterior
distributions as described below in \S\ref{sec:ppplots}, whilst both sets are used to evaluate odds distributions. The sources' sky locations are drawn
uniformly from over the sky-sphere. For these analyses each
simulated data set is one solar day long, consisting of 1440 complex data points (generated as if heterodyned at precisely the known source phase evolution)
sampled once per minute. This data has been used to estimate the four parameters $h_0$, $\cos{\iota}$, $\phi_0$ and $\psi$, with each being given
a flat prior: the angles and cosine of orientation angle over their minimal ranges, and $h_0$ between zero and $1\ee{-20}$ (well above the extent of the
bulk of the likelihood). When running \lppen on these, two parallel runs with 1024 live points where used, along with the default sampler proposals described
in \S\ref{sec:proposals}, and with the samples from both runs being combined to produce posterior samples and evidence estimates.

Another set of 2000 simulations has been created that again create simulated signals and add them to Gaussian noise for the H1 and L1 detectors. In this
case the total SNR shared between the detectors has be drawn from a uniform distribution between 0 and 30, however, the signals have been purposely created
to be incoherent between detectors to simulate instrumental noise features. The parameters $\phi_0$, $\psi$ and $\cos{\iota}$ are drawn from their minimal allowed ranges,
but are different for
the signals input into the two detectors, whilst the sky positions are also independent for both detectors. The signal rotational frequency and frequency
derivative are also offset between detectors, such that the offsets are drawn from Gaussian distributions of ($1\sigma$) width $1/2048$\,Hz and $1\ee{-10}$\,Hz$\,s^{-1}$
respectively. The length of the data is the same as in the previous case, but it is assumed that the data for both detectors was heterodyned using the
known phase evolution for the signal in only the H1 detector. The same prior and run settings as above were used. These simulations have been used for
assessing the odds for a coherent versus incoherent signal (see \S\ref{sec:odds}).

Finally, two sets of simulations of 500 signals have been created to assess the code when estimating additional phase parameters: in this case
the rotational frequency $f_0$, frequency derivative $\dot{f}_0$, and right ascension $\alpha$. Gaussian priors on these three parameters have been
used, with parameters given in Table~\ref{tab:gaussianpriors}. These Gaussian priors are used when generating signals and when estimating them
using \lppen, whilst the simulated data is created such that it was heterodyned using the mean values. As above, one set of these simulations has amplitudes 
calculated such that the coherent SNRs are drawn uniformly between 0 and 20, whilst the other has amplitude drawn uniformly between 0 and $3.25\ee{-22}$.
The latter of these is used to evaluate the posterior distributions, whilst both are used for odds distributions assessment. Again, the angles are drawn
uniformly from their minimal ranges and sky positions are drawn uniformly over the sky-sphere.

\begin{table}[!hptb]
\caption{Gaussian prior parameters for rotational frequency, frequency derivative and right ascension for a set of simulated signals.
\label{tab:gaussianpriors}}
\begin{center}
\begin{tabular}{l | c c}
\hline
Parameter & mean & standard deviation \\                      
\hline
\hline
$f_0$ (Hz) & 100 & $5\ee{-5}$ \\
$\dot{f}_0$ & $-1\ee{-9}$ & $2\ee{-10}$ \\
$\alpha$ (rads) & $\pi$ & $0.0007272$ ($10^{\text{s}}$) \\
\hline
\end{tabular}
\end{center}
\end{table}

\subsubsection{Evaluating the posterior distributions}\label{sec:ppplots}

In \citet{2014PhRvD..89h4060S} a method was developed for evaluating whether sky location credible intervals for compact binary coalescence \gw signals, produced
from posterior distribution output by \lalinf codes \citep{2015PhRvD..91d2003V}, behaved self-consistently. The approach uses injected signals, with parameters
drawn from the prior distributions used when reconstructing them\footnote{In this case the amplitudes of the simulations are not drawn from the full prior range,
but are drawn from a flat distribution within that range, so are valid for this comparison.}, and sees if the credible intervals effectively behave as frequentist
confidence intervals, i.e.\ do 50\% of the known parameter values fall within some pre-specified definition (like the minimal credible region, or credible
region bounded by the lower extent of the prior) of the 50\% credible region. This can test whether the underlying \lalinf parameter sampling is working as
expected, or if there are biases to wider, or narrower, credible regions. The method, and the \href{https://en.wikipedia.org/wiki/P\%E2\%80\%93P\_plot}{``P-P plots''}
it produces through evaluation over a range of credible
intervals, has now become commonly used to assess the \lalinf parameter estimation codes \citep{2015PhRvD..91d2003V} as a way of highlighting any problems.

We have used this method here to evaluate our parameter posteriors produced from the simulations described above for which the amplitudes were drawn from
a uniform distribution. For each simulation we have calculated the minimal credible region from the one-dimensional marginalised posterior samples that the
injected value falls within (using a greedy-binning approach). The cumulative histogram of these values, for each parameter when using the simulations that
just estimate the four \gw parameters, can be seen in Figure~\ref{fig:pp_standard}. Also, shown for each parameter is a Kolmogorov-Smirnov test $p$-value
comparing our observed distribution to a uniform distribution, and a ``cloud'' of cumulative distributions that are random realisations of the expected
cumulative distribution. We see that, overall, our posteriors give credible intervals that behave as expected, without any large deviations.

\begin{figure}[!phtb]
\begin{center}
\includegraphics[width=1\columnwidth]{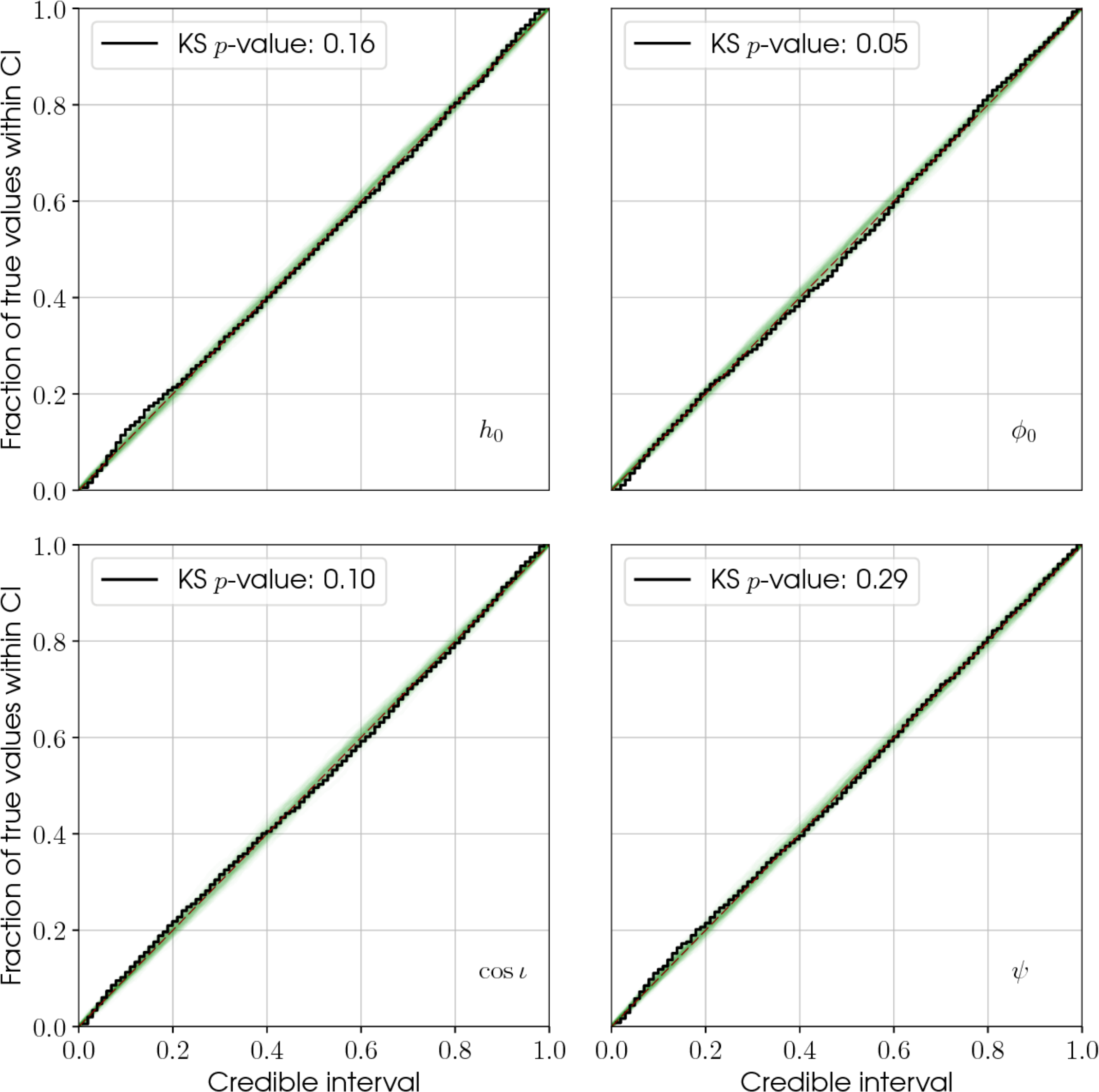}
\caption{ \protect\input{./figures/codeeval/stats/pp_standard/caption}}
\end{center}
\end{figure}

Similarly, we have done the same thing for the posteriors produced from the simulations that included searches over $f_0$, $\dot{f}_0$, and $\alpha$. 
In this case we know that the parameters are $f_0$ and $\phi_0$ are completely correlated (i.e.\ there is a degeneracy), so we have held the value of $\phi_0$
fixed at zero.\footnote{In cases where there is a complete degeneracy between parameters the samplers used in the code can have problems, leading to observable effects in
the ``P-P plots''. In certain cases, where the degeneracies are known, there are ways to deal with this: one parameter of a degenerate pair can be held fixed,
a custom proposal distribution can be hard coded for degenerate parameters that deals with the correlations, or, reparameterisations can be found that do not
have the degeneracies. In the future the latter of these may be implemented in our code to deal with strong degeneracies between the initial phase and derivatives of it
(frequency, and further frequency derivatives). Such a parameterisation could involve using, e.g.\ the start and end phase, rather than $\phi_0$ and $f_0$,
with further derivatives being incorporated through additional phase parameters at fixed times throughout the model. So, as another example, rather than
internally sampling in $\phi_0$, $f_0$, and $\dot{f}_0$, one could use $\phi_{\text{start}}$, $\phi_{\text{mid}}$, and $\phi_{\text{end}}$, and convert between
the two using linear algebra.} These ``P-P plots'' can be seen in Figure~\ref{fig:pp_extra}, where generally we see posterior credible intervals behaving as
expected.

\begin{figure}[!phtb]
\begin{center}
\includegraphics[width=1\columnwidth]{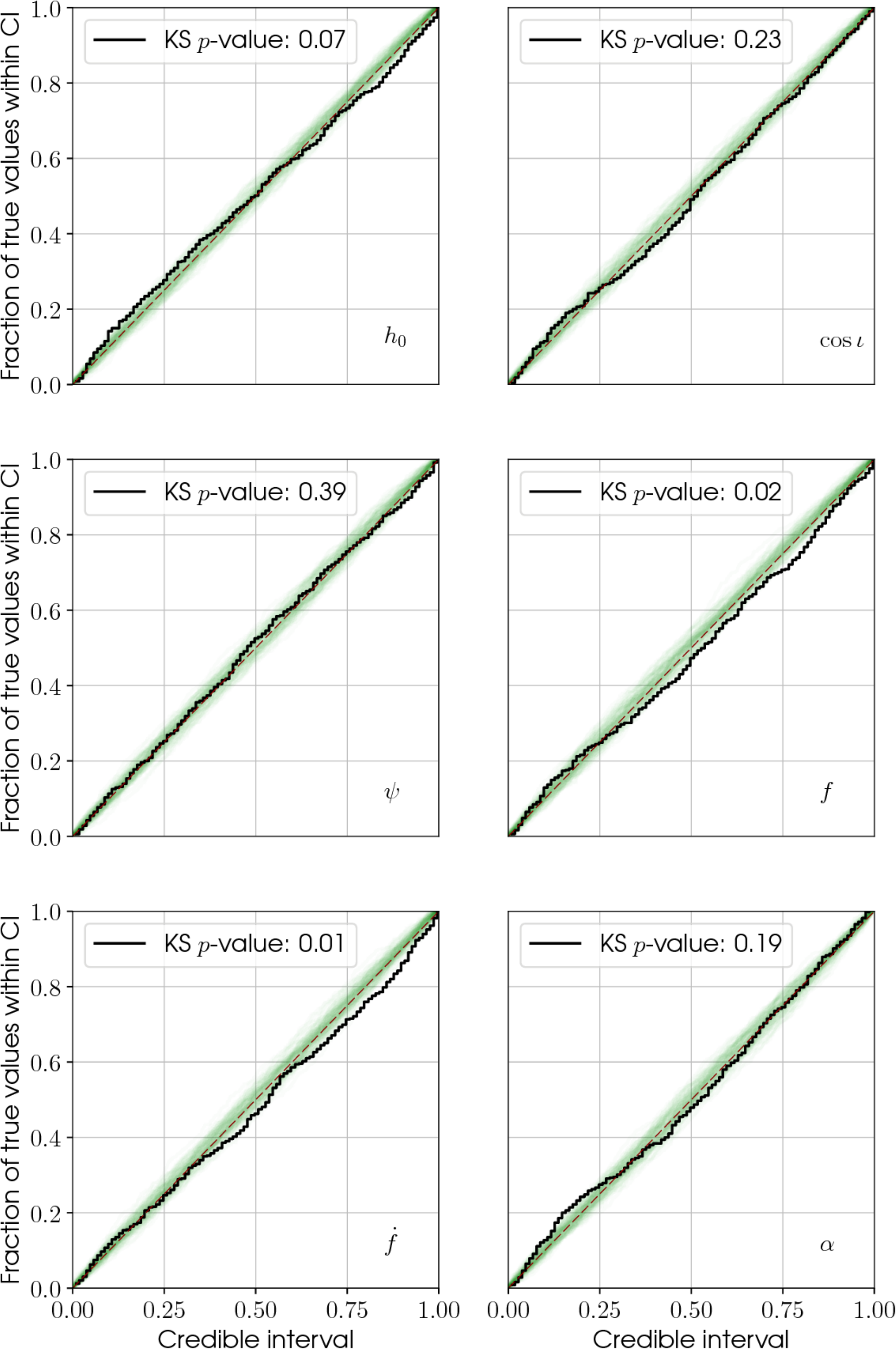}
\caption{ \protect\input{./figures/codeeval/stats/pp_extra/caption}}
\end{center}
\end{figure}

\subsubsection{Evaluating the odds}\label{sec:evalodds}

It is useful to see how the odds values, described in \S\ref{sec:odds}, look when calculated for our simulations. As mentioned above, along with coherent
signals between two detectors, we have purposely created incoherent signals to see how the odds look. Firstly, just considering the coherent simulations,
we can see in Figure~\ref{fig:snrvsodds} how the odds $\mathcal{O}_{\text{S}/\text{N}}$, $\mathcal{O}_{\text{S}/\text{I}}$ and
$\mathcal{O}_{\text{S}/\text{I}_{\text{simple}}}$ (Equations~\ref{eq:sigvsnoise}, \ref{eq:cohvincoh2} and \ref{eq:cohvincoh1}, respectively) vary as a function
of coherent SNR. We see that $\mathcal{O}_{\text{S}/\text{N}}$ rapidly increases, $\mathcal{O}_{\text{S}/\text{I}_{\text{simple}}}$ has a shallow growth,
whilst $\mathcal{O}_{\text{S}/\text{I}}$ transitions between following $\mathcal{O}_{\text{S}/\text{N}}$ at very low SNR to matching 
$\mathcal{O}_{\text{S}/\text{I}_{\text{simple}}}$ at higher SNR. This behaviour is consistent with our expectations described in Appendix~\ref{app:oddratios} with
$\ln{\mathcal{O}_{\text{S}/\text{N}}} \propto \rho_{\text{coh}}^2$ and $\mathcal{O}_{\text{S}/\text{I}_{\text{simple}}} \propto \rho_{\text{coh}}$.
We note that the locations of these distributions on the y-axis can be strongly effected by the prior volume, changes to which will move the distributions
up or down, and can affect where they intersect. However, the general shape of the distributions should be very similar.

\begin{figure}[!phtb]
\begin{center}
\includegraphics[width=1\columnwidth]{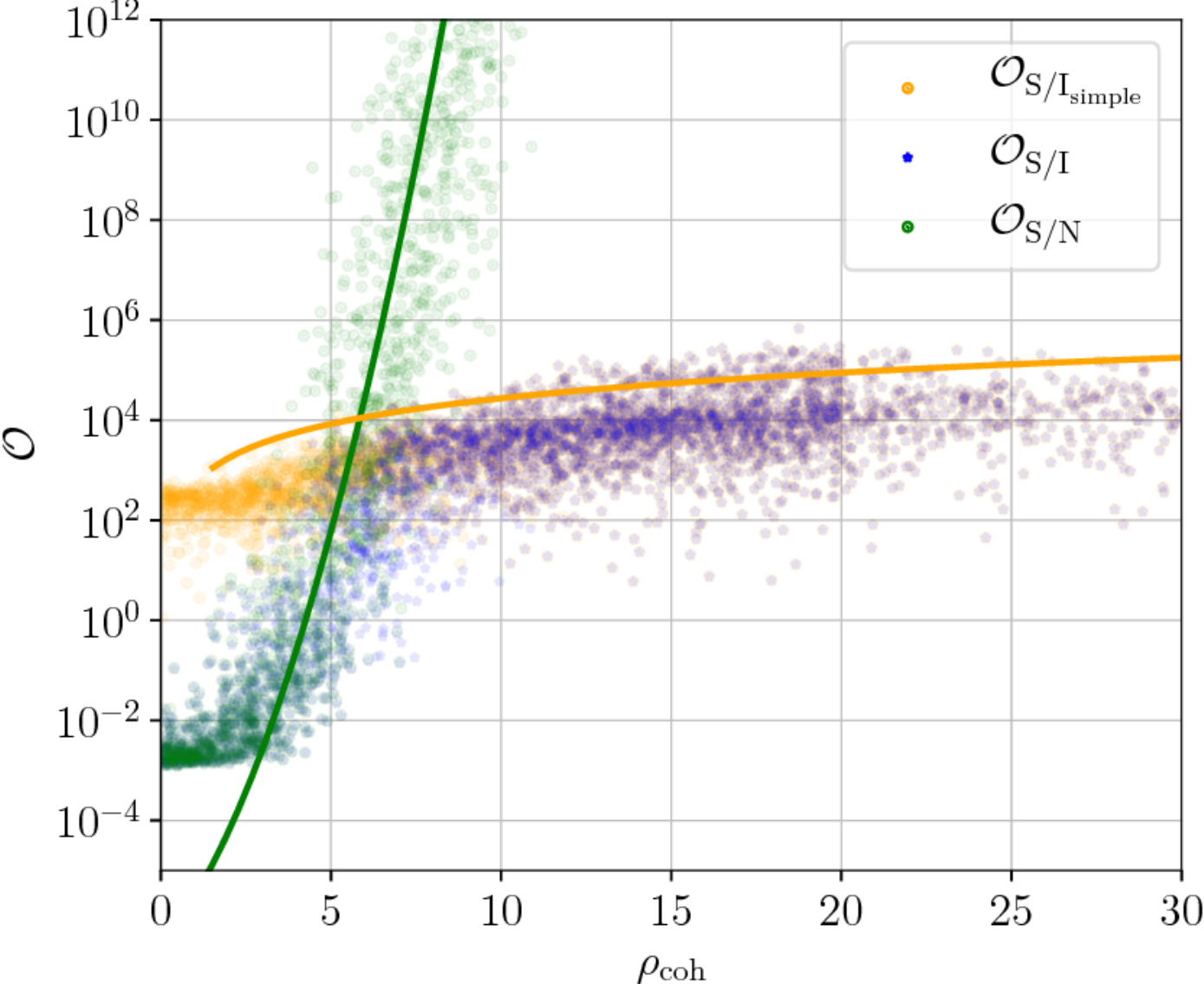}
\caption{ \protect\input{./figures/codeeval/stats/snr_vs_odds/caption}}
\end{center}
\end{figure}

In Figure~\ref{fig:oddsplot} there are two panels showing $\mathcal{O}_{\text{S}/\text{I}}$ as a function of $\mathcal{O}_{\text{S}/\text{N}}$. In the upper
panel (showing the coherent simulations) the colours of points show the signal's coherent SNR. The plateauing of $\mathcal{O}_{\text{S}/\text{I}}$ is just that seen in Figure~\ref{fig:snrvsodds}. The
bottom panel shows a zoomed out version, with the incoherent signals as dark circles (these are also seen in the lower left of the upper panel). Here we see that for incoherent signals
the value of $\mathcal{O}_{\text{S}/\text{I}}$ between the coherent and incoherent signals diverges quite quickly, with $\log{}_{10}{\mathcal{O}_{\text{S}/\text{I}}}$
falling off roughly linearly with $\log{}_{10}{\mathcal{O}_{\text{S}/\text{N}}}$. This shows that, hopefully, for reasonable strength signals the value of
$\mathcal{O}_{\text{S}/\text{I}}$ will be useful for distinguishing a coherent from an incoherent (and therefore not astrophysical) signal.\footnote{The caveats
to this are that the way we created our incoherent signals may not well match true signals caused by instrumental (non-astrophysical) effects in \gw detectors.
However, we hope our simulations are extreme, in a conservative way, in that one detector is actually containing an unadulterated signal, and is therefore
potentially producing a far larger $\mathcal{O}_{\text{S}/\text{N}}$ value than might be expected whereas an instrumental feature would generally not match
our signal template, even in a single detector.}

\begin{figure*}[phtb]
\begin{center}
\includegraphics[width=0.7\textwidth]{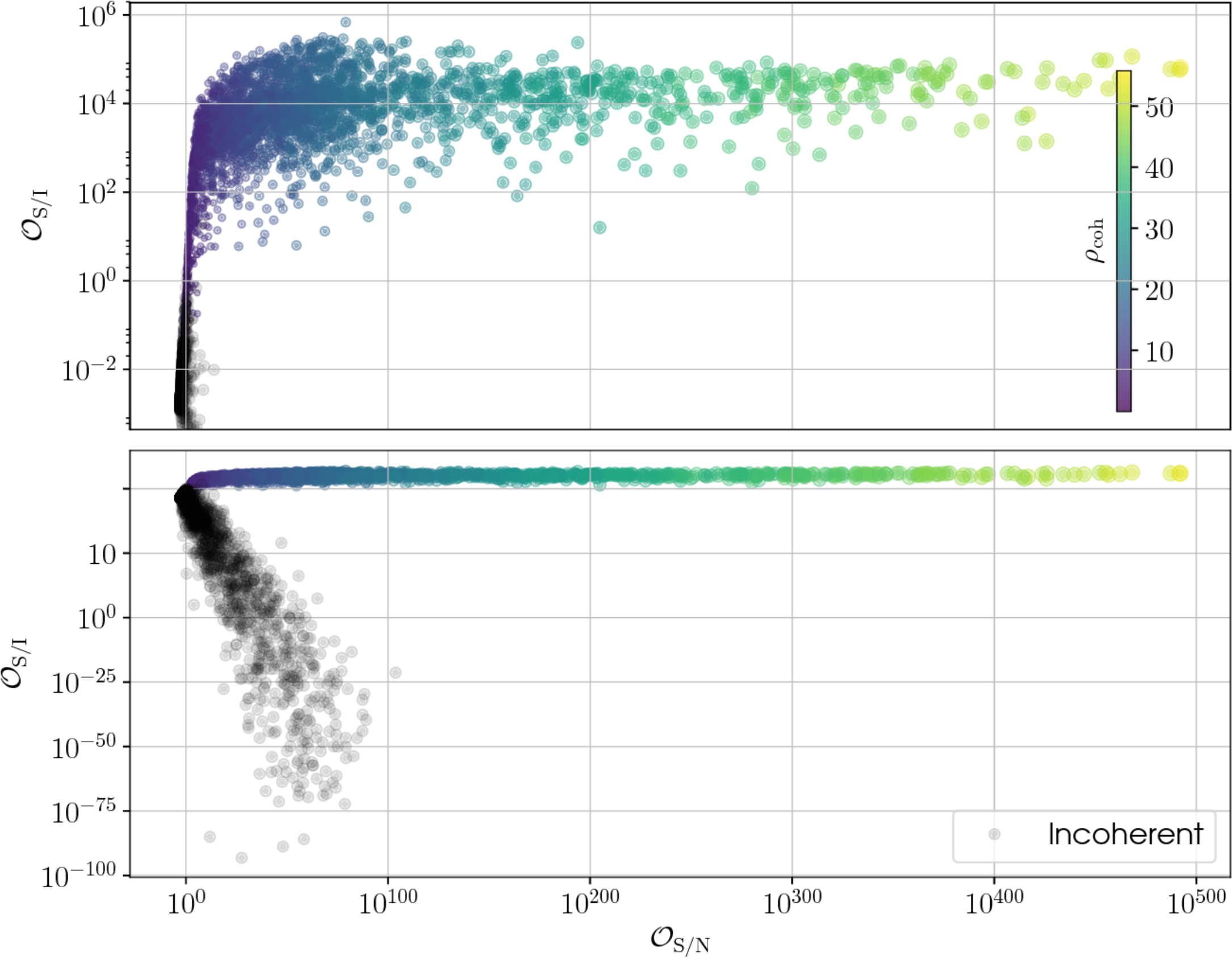}
\caption{ \protect\input{./figures/codeeval/stats/odds/caption}}
\end{center}
\end{figure*}

It is useful to see how these distributions change when the parameter space searched over increases in complexity. So, in Figure~\ref{fig:snrvsodds_larger}
we see similar plots to those above, but for the simulations that included searches over $f_0$, $\dot{f}$, and $\alpha$ (but fixed $\phi_0$). In the left
panel we see $\mathcal{O}_{\text{S}/\text{N}}$ and $\mathcal{O}_{\text{S}/\text{I}}$ as a function of $\rho_{\text{coh}}$ for these simulations, with the
earlier simulations plotted underneath. It is obvious that, for these injections the odds follow the same form as before, however, at low SNR there
is more scatter in $\mathcal{O}_{\text{S}/\text{I}}$ than seen before. This can be interpreted as finding spikes in $f_0\operatorname{-}\dot{f}$-space, that
are not related to the signal, in individual detectors, which are therefore not coherent. For higher SNRs ($\gtrsim 10$) the $\mathcal{O}_{\text{S}/\text{I}}$
value for
this more complex search does tend to give higher values than for the simpler search (shown in Figure~\ref{fig:snrvsodds_larger} as the lighter circles). A hand-wavy explanation for this {\it could be} that a truly coherent signal
found (at large enough SNR) within this larger space is more convincing (more parameters have to match up) than in the simpler case, and this outweighs any
prior Occam factor. The right panel of Figure~\ref{fig:snrvsodds_larger} shows similar information.

\begin{figure*}[phtb]
\begin{center}
\includegraphics[width=0.8\textwidth]{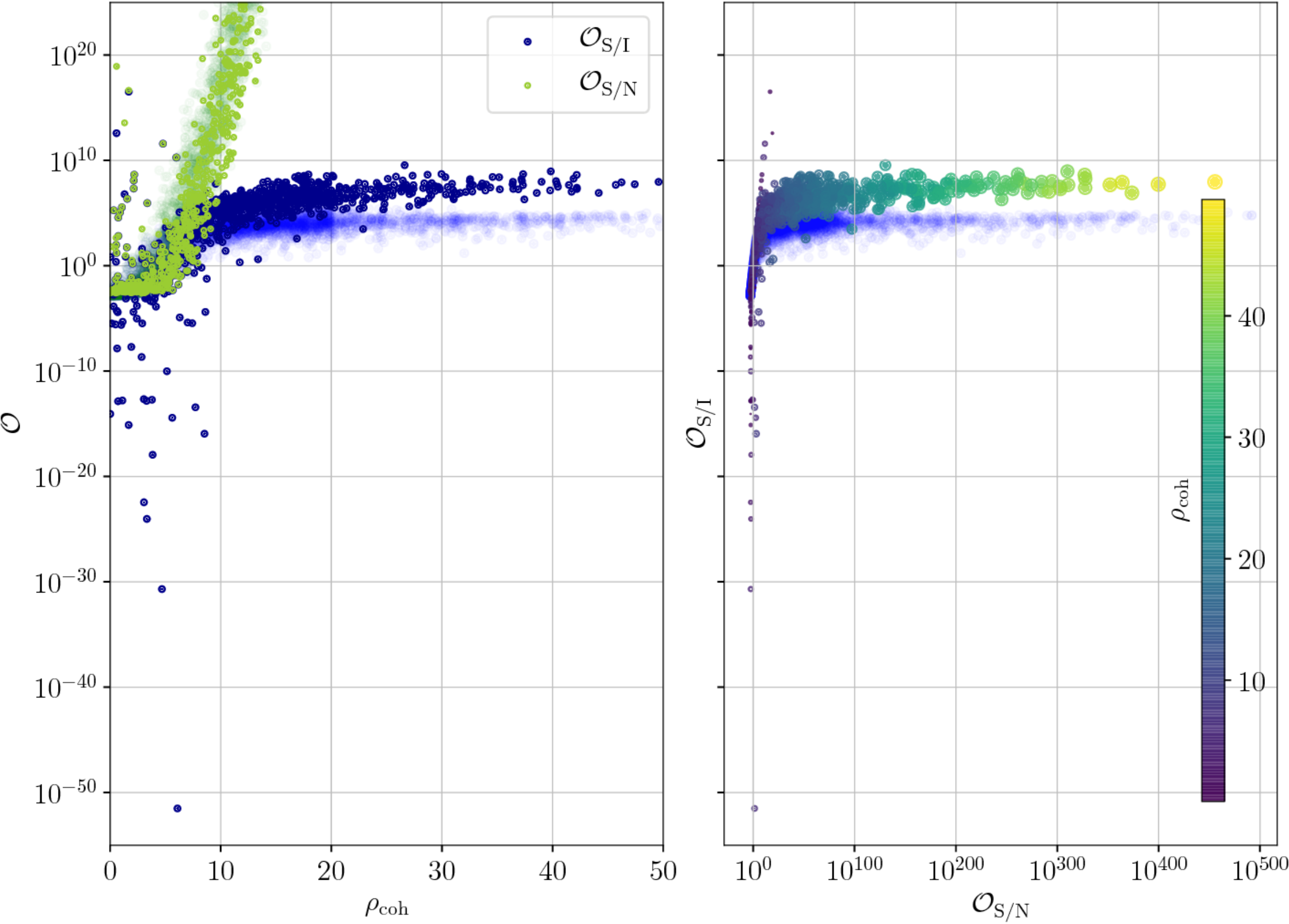}
\caption{ \protect\input{./figures/codeeval/stats/snr_vs_odds_larger/caption}}
\end{center}
\end{figure*}

\subsubsection{Further parameter estimation}

In Figures~\ref{fig:ffdot_inj1} and \ref{fig:ffdot_inj2} we show the posterior distributions for two of the simulations performed for the searches over
$h_0$, $\cos{\iota}$, $\phi_0$, $\psi$, $f_0$, $\dot{f}$, and $\alpha$ above.\footnote{Earlier, during the discussing of the ``P-P plots'' in \S\ref{sec:ppplots},
it was mentioning that $\phi_0$ was held fixed due to the degeneracies between $f_0$ and $\phi_0$. In this case, as we also have $\dot{f}$, the correlations
become more complex, and whilst it would be expected that holding $\phi_0$ fixed would give valid posteriors, it is in such cases that a re-parameterisation
may be more appropriate.} The simulation in Figure~\ref{fig:ffdot_inj1} had injected SNRs of 4.4 and 6.5 for H1
and L1 respectively, with a coherent SNR of 7.9, and the recovered values of $\log{}_{10}\mathcal{O}_{\text{S}/\text{N}}$ and
$\log{}_{10}\mathcal{O}_{\text{S}/\text{I}}$ of 1.55 and 0.95 respectively. It can be seen that the search in H1 alone finds additional structure
in the frequency space, whilst in L1 the signal dominates, and the joint analysis also recovers the signal. The simulation in Figure~\ref{fig:ffdot_inj2}
is a little stronger with injected SNRs of 7.3 and 8.5 for H1 and L1 respectively, and a coherent SNR of 11.2, and recovered values of 
$\log{}_{10}\mathcal{O}_{\text{S}/\text{N}}$ and $\log{}_{10}\mathcal{O}_{\text{S}/\text{I}}$ of 23.8 and 6.3 respectively. The signal is by far the strongest
feature in the data, and no other structure is present in the posteriors. The first signal would probably only rate as marginal if determining whether it was
a true signal or not, whereas the second signal would be a clear detection.

\begin{figure*}[!phtb]
\begin{center}
\includegraphics[width=0.6\textwidth]{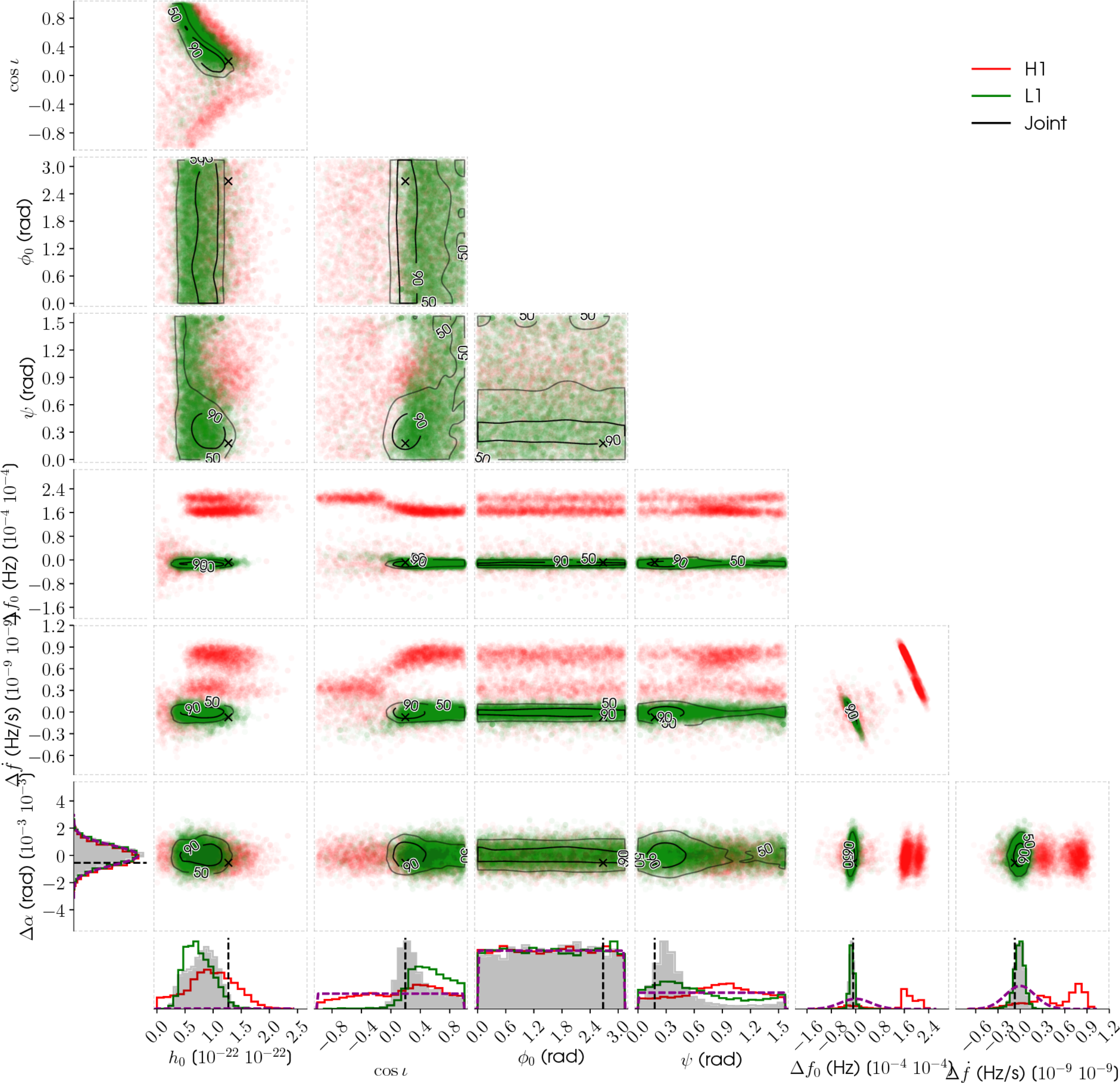}
\caption{ \protect\input{./figures/codeeval/simulations/signal_freq/one/caption}}
\end{center}
\end{figure*}

\begin{figure*}[!phtb]
\begin{center}
\includegraphics[width=0.6\textwidth]{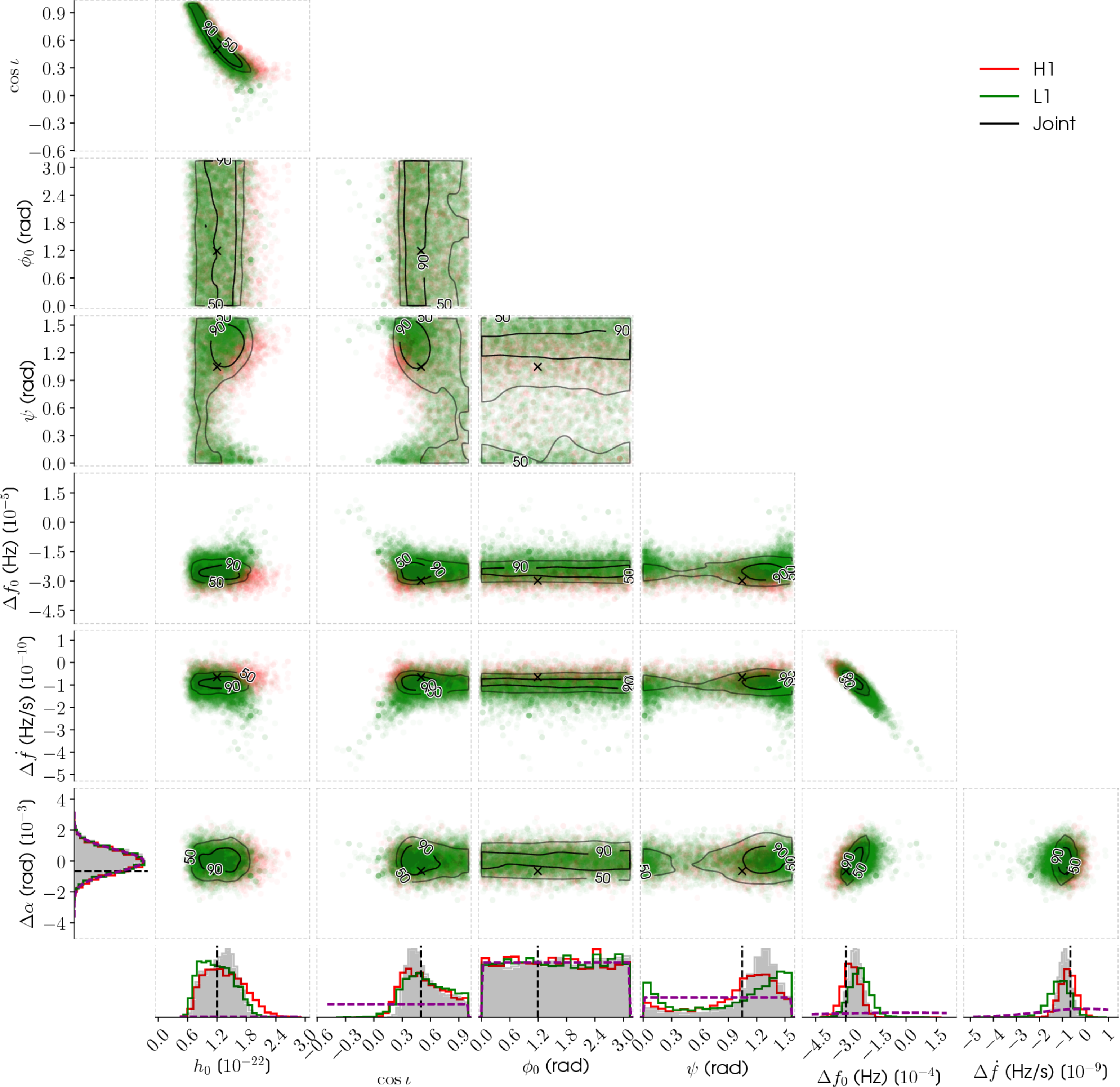}
\caption{ \protect\input{./figures/codeeval/simulations/signal_freq/two/caption}}
\end{center}
\end{figure*}

\section{Conclusions}

In this document we have described some of the workings of a code, \lppenf, that is used for performing parameter estimation and Bayesian model
selection in searches for \gw signals from known pulsars. We have described the available signal models, likelihood functions and prior functions.
We have compared the code to one that had been previously used for the same purpose, \lppef, and found that the results agree very well for a range
of situations. The comparisons have not been completely exhaustive, and there are situations that have not been tested, however, it is worth noting
that \lppe is not necessarily expected to provide complete ground truth. The code has also been validated on additional software and hardware simulated
signals and appears to return posterior distributions that behave as expected, and evidence values that show the expected behaviour.

As a test of the internal nested sampling algorithm, and proposal distributions, used by the code a large number of simulations were performed using
a simple Gaussian likelihood function. These interestingly highlighted that if using, in particular, the ensemble walk proposal distribution the evidence
values output became systematically more biased away from the true value as a function of information gain between the prior and posterior distributions.
The test showed that this bias was alleviated by including a proposal distribution that drew values from the full prior a certain fraction of the time.
The reason for this systematic bias is currently unknown, but in general, given the magnitude of the bias, it should have a relatively minor effect. It was
also shown that with the ensemble walk proposal source amplitude upper limits could be systematically biased by about 2\%, however, again this effect
appears to be alleviated with the flat proposal distribution.

Some example use cases of \lppen are documented in Appendix~\ref{app:usage}.
\appendix

\section{Code usage}\label{app:usage}

Here we document the various command line options for the code and its use in a variety of situations. The majority of the command
line options for \lppen, as shown with the \verb|--help| command, are given (in a slightly edited form) in Table~\ref{tab:commands}.
\begin{footnotesize}
\begin{longtable}{|p{0.15\textwidth}p{0.8\textwidth}|}
\caption{A list of command line arguments for the \lppen code.}\label{tab:commands} \\
\hline
\multicolumn{2}{|l|}{Usage: \lppenf [options]} \\
 {\tt --help}            &  Display these commands. \\
 {\tt --verbose}         &  Display progress information from the code, including details of the nested sampling progress. \\
 {\tt --detectors}       &  Input all interferometers whose data is to be analysed, e.g.\, {\tt H1,L1} (delimited by commas) for the LIGO Hanford and
 LIGO Livingston detectors (see the discussion in \S\ref{sec:example2} for the allowed detector names). If generating fake data, described below, these
                     should not be set, otherwise this is {\bf required}. \\
 {\tt --par-file}        &  The pulsar's parameter ({\tt .par}) file (see \S\ref{sec:parfile}). {\bf Required}. \\
 {\tt --cor-file}        &  The pulsar's parameter correlation matrix file (see \S\ref{sec:priorfile}). {\bf Optional}. \\
 {\tt --harmonics}       &  The signal model frequency harmonics that you want to use (delimited by commas). Currently
                            this can be either a single value (e.g., the {\bf default} is to assume a 'triaxial' star emitting from the $l=m=2$ mode and use a value
                            of 2), or {\tt 1,2} for a model with emission at both the rotation frequency and twice the rotation frequency. \\
 {\tt --input-files}     &  The files containing the pre-processed (heterodyned, or spectrally interpolated) data for each detector and model harmonic specified by
                            {\tt --detectors} and {\tt --harmonics} delimited by commas (these files must be listed in the same order as passed to
                            {\tt --detectors} and {\tt --harmonics}, with the files for different harmonics for individual detectors listed together). If this is not
                            set you can generate fake data (see {\tt --fake-data below}), but it is otherwise this is {\bf required}. \\
 {\tt --outfile}         &  The name for the output data file. This must be a HDF5 file with the extension '.hdf' or '.h5'. {\bf Required}. \\
 {\tt --output-chunks}   &  Output lists of the stationary chunks into which the data has been split (see \S\ref{sec:splitting}). {\bf Optional}. \\
 {\tt --chunk-min}       &  The minimum stationary length of data to be used in the likelihood. This {\bf defaults} to 5, e.g., 5 mins for data
                            sampled at 1/60~Hz. {\bf Optional}. \\
 {\tt --chunk-max}       &  The maximum stationary length of data to be used in the likelihood. This {\bf defaults} to 0, which is the value for no
                            maximum length to be applied, except if the {\tt --oldChunks} flag is set, in which case it defaults to 30. If running with the
                            {\tt --roq} flag it can be worth setting a value here (e.g., 1440) to avoid large amounts of memory being required in training
                            set generation. {\bf Optional}. \\
 {\tt --time-bins}       &  Number of time bins over a sidereal day in the antenna response lookup table. This {\bf defaults} to 2880 bins. {\bf Optional}. \\
 {\tt --prior-file}      &  The file containing the prior ranges for parameters to search over (see \S\ref{sec:priorfile}). {\bf Required}. \\
 {\tt --ephem-earth}     &  The Earth ephemeris file. If not supplied this will attempt to be found in the path. \\
 {\tt --ephem-sun}       &  The Sun ephemeris file. If not supplied this will attempt to be found in the path. \\
 {\tt --ephem-timecorr}  &  The Einstein delay time correction ephemeris file. If not supplied this will attempt to be found in the path. \\
 {\tt --biaxial}         &  Set this if using the waveform model parameters with two harmonics, and specifically for a biaxial star \citep[see, e.g.,][]{2015MNRAS.453.4399P}. \\
 {\tt --gaussian-like}   &  Set this if a Gaussian likelihood is to be used. If the input file(s) contains a column specifying
                            the noise standard deviation of the data then that will be used in the Gaussian likelihood
                            function, otherwise the noise variance will be calculated from the data. \\
 {\tt --nonGR}           &  Set this to allow non-GR polarisation modes and/or a variable speed of gravitational waves. \\
 {\tt --randomise}       &  Set this, with an integer seed, to randomise the data (through permutations of the time stamps) for
                            use in Monte-Carlo studies. Note that this will not work if using the code to create injections. \\
 {\tt --truncate-time}   &  Only analyse data up to the given GPS time. {\bf Optional}. \\
 {\tt --truncate-samples}&  Only analyse data samples up to the number given here. {\bf Optional}. \\
 {\tt --truncate-fraction}& Only analyse the given fraction of data samples (this must be a value between 0 and 1). {\bf Optional}. \\
 ~ & ~ \\
\multicolumn{2}{|l|}{~Nested sampling parameters:} \\
 {\tt --Nlive}           &  Set the integer number of live points for nested sampling. {\bf Required}. \\
 {\tt --Nmcmc}           &  Set the length of the MCMC used to find new live points (if not specified an adaptive number of
                            points is used). \\
 {\tt --Nmcmcinitial}    &  Set the number of MCMC points to use in the initial resampling of the prior. This defaults to 5000, but
                            for our analyses this can be set to zero. \\
 {\tt --tolerance}       &  The tolerance used as the stopping criterion for the nested sampling integrator. This {\bf defaults} to 0.1. \\
 {\tt --randomseed}      &  A seed for the random number generator. By default, if not supplied, this is set by the system clock. \\
 ~ & ~ \\
\multicolumn{2}{|l|}{~MCMC proposal parameters: (see \S\ref{sec:proposals})} \\
 {\tt --diffev}          &  The (integer) relative weight of using differential evolution of the live points as the proposal. The {\bf default} is 0,
                            e.g., 0\%. \\
 {\tt --freqBinJump}     &  The (integer) relative weight of using jumps to adjacent frequency bins as a proposal. The {\bf default} is 0, and this is not
                            required unless searching over frequency. \\
 {\tt --ensembleStretch} &  The (integer) relative weight of the ensemble stretch proposal. The {\bf default} is 0. Note that this
                            proposal greatly increases the parameter autocorrelation lengths, so in general should be avoided. \\
 {\tt --ensembleWalk}    &  The (integer) relative weight of the ensemble walk proposal. The {\bf default} is 3, e.g., 75\%. \\
 {\tt --uniformprop}     &  The (integer) relative weight of uniform proposal. The {\bf default} is 1, e.g., 25\%. \\
 ~ & ~ \\
\multicolumn{2}{|l|}{~Reduced order quadrature (ROQ) parameters:} \\
 {\tt --roq}             &  Set this flag to use reduced order quadrature to compute the likelihood. \\
 {\tt --ntraining}       &  The (integer) number of training models used to generate an orthonormal basis of waveform models. \\
 {\tt --roq-tolerance}   &  The tolerance used during the basis generation. The {\bf default} is {\tt 1e-11}. \\
 {\tt --enrich-max}      &  The (integer) number of times to try and ``enrich'' \citep[see, e.g.,][]{2016PhRvD..94d4031S} the basis set using new training
                            data. The enrichment process stops before this value is reached if three consecutive enrichment steps
                            produce no new bases. The {\bf default} is 100. \\ 
 {\tt --roq-uniform}     &  Set this flag to cause training model parameters, for parameters with Gaussian prior distributions, to be drawn from a
                            uniform distribution spanning $\mu \pm 5 \sigma$. Otherwise, by
                            default, parameters are drawn from their given prior distributions. \\
 {\tt --output-weights}  &  Give an output file name to which to output (a binary version of) the weights will be output. If this is set the
                            programme will exit after outputting the weights. These files could be read in later instead of being regenerated.
                            This allows the potential for the ROQ to be generated on a machine with a large amount of RAM, whilst the full parameter
                            estimation can run on a machine with less RAM. \\
 {\tt --input-weights}   &  A binary file, created using the {\tt --output-weights} command, containing all the weights in a defined format.
                            If this is present then the ROQ will not be recalculated. \\
 ~ & ~ \\
\multicolumn{2}{|l|}{~Signal injection parameters:} \\
 {\tt --inject-file}     &  A pulsar parameter ({\tt .par}) file containing the parameters of a signal to be injected. If this is
                           given a signal will be injected. \\
 {\tt --inject-nonGR}    &  A parameter ({\tt .par}) file containing the values of specific non-GR signal parameters to be injected (this is an alternative to {\tt --inject-file}, so
 both should not be passed at the same time). \\
 {\tt --inject-output}   &  A filename to which the injected signal will be output if specified. \\
 {\tt --inject-only}     &  If this is set, and an {\tt --inject-output} file is set, then do not perform nested sampling on the
                            created injection, but just exit the code after creation of, and writing out of, the injection. \\
 {\tt --fake-data}       &  A list of interferometers for which fake data will be generated, e.g., H1,L1 (delimited by commas). Unless
                            the {\tt --fake-psd} flag is set the power spectral density for the data will be generated from the
                            noise models in described in \S\ref{sec:example2}. The noise will be white across the data bandwidth. \\
 {\tt --fake-psd}        &  If wanting to generate fake data with specific power spectral densities for each detector given
                            by {\tt --fake-data} then they should be specified here delimited by commas (e.g., for {\tt --fake-data H1,L1}
                            then you could use {\tt --fake-psd 1e-48,1.5e-48}) where values are single-sided PSDs in Hz$^{-1}$. \\
 {\tt --fake-starts}     &  The start times (in GPS seconds) of the fake data for each detector separated by commas (e.g.,
                            {\tt 910000000,910021000}). If not specified these will all {\bf default} to 900000000. \\
 {\tt --fake-lengths}    &  The length of each fake data set (in seconds) for each detector separated by commas. If not
                            specified these will all {\bf default} to 86400 (i.e., 1 solar day). \\
 {\tt --fake-dt}         &  The data sample rate (in seconds) for the fake data for each detector. If not specified these will
                            {\bf default} to 60~s. \\
 {\tt --scale-snr}       &  This gives a (multi-detector) SNR value to which you want to scale the injection. By {\bf default} this is 1. \\
 ~ & ~ \\
\multicolumn{2}{|l|}{~Legacy code flags:} \\
 {\tt --oldChunks}       &  Set this if using fixed chunk sizes for dividing the data as in the old code, rather than the
                            calculating chunks using the change point method. \\
 {\tt --source-model}    &  Set this if using both 1 and 2 multiples of the frequency and requiring the use of the original source
                            model parameters from \citet{2015MNRAS.453.4399P}. \\
 ~ & ~ \\
\multicolumn{2}{|l|}{~Benchmarking:} \\
 {\tt --time-it}         &  Set this if wanting to time the various parts of the code. A output file with the ``outfile'' filename
                            appended with ``\_timings'' will contain the timings \\
 {\tt --sampleprior}     &  Set this to output this (integer) number of samples generated from the prior. The nested sampling will not
                            be performed. \\
\hline
\end{longtable}
\end{footnotesize}

\subsection{Examples}\label{app:examples}

Here we show a few real examples of using the code for a variety of situations. In all cases there are a minimum of
three files that must be specified to run the code:
\begin{enumerate}
 \item One, or more, data files containing a complex time series. These will generally be (although do not have to be) the output
 of the heterodyne method described in \citep{2005PhRvD..72j2002D} (and implemented with {\tt lalapps\_heterodyne\_pulsar}), or
 the {\it Spectral Interpolation} method described in \citep{2017CQGra..34a5010D}. Files must be in ascii text format with three, or
 four, whitespace-separated columns. In either case the first three columns specify the data timestamp in GPS seconds, the real part of
 the data, and the imaginary part of the data, e.g., \verb|900000000 1.2564374e-22 -4.5764347e-22|. In the case where a fourth column
 is present this should give the noise standard deviation, and is for use if running with the Gaussian likelihood function. The code will
 automatically determine if the input contains three or four columns. The data does not have to be uniformly sampled and can 
 contain gaps. However, if searching for a signal for which there could still be some frequency modulation it is useful for the data to have a large
 enough bandwidth to contain the modulation. The file can contain comments if the a line starts with a \verb|%| or \verb|#|. Gzipped versions of these
 files can also be input.
 \item A pulsar parameter file. This should be an ascii text {\sc tempo(2)}-style \verb|.par| file containing information on the
 source's parameters, e.g. sky location and frequency, as generated by {\sc tempo}\footnote{http://tempo.sourceforge.net/} or
 {\sc tempo2} \citep{2006MNRAS.369..655H}  from pulsar pulse time-of-arrival observations. An example \verb|.par| file is given below.
 \item A prior distribution file. This should be an ascii text file containing information on the prior distributions
 required for any variable parameters for which the code needs to sample posterior probability distributions. An example
 prior file is given below.
\end{enumerate}

In the examples below we will assume that the LALSuite software package \citep{lalsuite} has
been installed and that the executable binaries for \lppen and {\tt lalapps\_nest2pos} are in your path (i.e.\ the directory path containing those
executables is set in the list of values in your {\tt PATH} environment variable). The code can automatically find solar system ephemeris and timing
files provided they are in a standard location, although these can also be specified explicitly on the command line.

\subsubsection{An example {\tt .par} file}\label{sec:parfile}

The {\tt .par} file input should be the same file that was used when performing the heterodyning (or {\it Spectral Interpolation})
of the data used to produce the input data time series. A {\sc tempo(2)}-style {\tt .par} file could have the following form, e.g.,
for a pulsar in a binary system:
\newsavebox{\Lst}
\begin{lrbox}{\Lst}
\begin{lstlisting}
PSRJ    J0234+5612
RAJ     02:34:12.123485
DECJ    56:12:05.237474
F0      102.764742743786  1  1.2341e-10
F1      -8.854276e-14
PEPOCH  54012.2853
BINARY  BT
ECC     0.023
T0      53424.24435856
OM      12.858534
PB      9.858345
A1      3.474743
\end{lstlisting}
\end{lrbox}
\\[5pt] \indent \fbox{\usebox{\Lst}} \\[5pt]

The majority of the parameters that can be in a {\tt .par} file are listed in Table~A1 of \citet{2006MNRAS.372.1549E}.
Note that our code does not compute effects of dispersion measure (i.e.\ it assumes observations at infinite frequency),
or atmospheric effects, so it is preferable to use a {\tt .par} file for which these parameters have been not
used, or held fixed.

For pulsars in binary systems, the code can currently use the following binary system models \citep[see, e.g.,][for discussion
of some of the models]{1989ApJ...345..434T,2007PhRvD..76d2006P}: {\tt BT}, {\tt BTX},
{\tt BT1P}, {\tt BT2P}, {\tt ELL1}, {\tt DD}, {\tt DDS}, {\tt MSS}, and {\tt T2} (although not all aspects of the
{\tt T2} model are implemented). The code can use the TDB or TCB time systems, with the TCB system being the default if not otherwise
specified (either by supplying the require timing files, or by setting ``{\tt UNITS TDB}'' in the {\tt .par} file).
The code can accept {\tt .par} files calculated using the JPL DE200, DE405, DE414, or DE421 solar system ephemerides, noting
that it only takes into account the motion of the Earth-Moon system and the Sun, but no other solar system bodies.

The {\tt .par} file can also contain a selection of gravitational wave parameters not present in the case of a standard
electromagnetically observed pulsar. A full list of the allowed parameter names for a {\tt .par}, and a prior file described below,
is given in Table~\ref{tab:paramlist}.

\begin{longtable}{p{0.15\textwidth}|p{0.8\textwidth}}
\caption{A list of parameters allowed in a {\tt .par} file or a prior file. The true-type font versions, in uppercase only,
are what is allowed in the files. The specified units are for the {\tt .par} file, whereas the prior file
requires values in SI units.}\label{tab:paramlist} \\
Parameter & Description \\
\hline
\multicolumn{2}{c}{Gravitational wave parameters} \\
\hline
{\tt H0}, $h_0$      & gravitational wave amplitude for the $l=m=2$ mode in the source model \citep[see, e.g.,][]{1998PhRvD..58f3001J} \\
{\tt IOTA}, $\iota$  & pulsar inclination angle to the line-of-sight (rads) ($0^{\circ}$ is gives a source whose rotation axis is pointed directly along the line-of-sight) \\
{\tt COSIOTA}, $\cos{\iota}$ & cosine of $\iota$ \\
{\tt PSI}, $\psi$    & polarisation angle of the gravitational waves (rads) \citep[see, e.g.,][]{1998PhRvD..58f3001J} \\
{\tt PHI0}, $\phi_0$ & rotational phase of the pulsar (rads) \\
{\tt C22}, $C_{22}$  & gravitational wave amplitude for the $l=m=2$ mode in the waveform model \citep[see][]{2015MNRAS.453.4399P} \\
{\tt C21}, $C_{21}$  & gravitational wave amplitude for the $l=2, m=1$ mode in the waveform model \citep[see][]{2015MNRAS.453.4399P} \\
{\tt PHI22}, $\Phi_{22}^C$ & gravitational wave phase for the $l=m=2$ mode in the waveform model (rads) \citep[see][]{2015MNRAS.453.4399P} \\
{\tt PHI21}, $\Phi_{21}^C$ & gravitational wave phase for the $l=2, m=1$ mode in the waveform model (rads) \citep[see][]{2015MNRAS.453.4399P} \\
{\tt Q22}, $Q_{22}$ & $l=m=2$ mass quadrupole moment (kg\,m$^{2}$) \citep[see e.g.][]{2005PhRvL..95u1101O} \\
{\tt I21}, $I_{21}$ & physical gravitational wave amplitude in source model \citep[see][]{2015MNRAS.453.4399P} \\
{\tt I31}, $I_{31}$ & physical gravitational wave amplitude in source model \citep[see][]{2015MNRAS.453.4399P} \\
{\tt LAMBDA}, $\lambda$ & Euler angle in source model \citep[see][]{2015MNRAS.453.4399P} \\
{\tt THETA}, $\theta$ & Euler angle in source model \citep[see][]{2015MNRAS.453.4399P} \\
{\tt COSTHETA}, $\cos{\theta}$ & cosine of $\theta$ \\
\hline
\multicolumn{2}{c}{Signal phase parameters} \\
\hline
{\tt F}$X$, $f_X$ & rotational frequency, and frequency derivatives (starting at 0), (Hz$\,s^{-X}$) \\
{\tt PEPOCH} & the epoch of the frequency (MJD, defined in the inertial frame of the pulsar, with often this being the solar system barycenter) \\
\hline
\multicolumn{2}{c}{Signal positional parameters} \\
\hline
{\tt RA}/{\tt RAJ}, $\alpha$ & pulsar right ascension ({\tt hh:mm:ss.s}) \\
{\tt DEC}/{\tt DECJ}, $\delta$ & pulsar declination ({\tt dd:mm:ss.s}) \\
{\tt PMRA} & proper motion in right ascension (mas\,yr$^{-1}$) \\
{\tt PMDEC} & proper motion in declination (mas\,yr$^{-1}$) \\
{\tt POSEPOCH} & the epoch of the sky position (MJD) \\
{\tt EPHEM} & the JPL solar system ephemeris \\
{\tt PX} & pulsar parallax (mas) \\
{\tt DIST} & pulsar distance (kpc) \\
\hline
\multicolumn{2}{c}{Binary system oribital parameters} \\
\hline
{\tt BINARY} & binary system model \citep[see e.g.][and references therein]{1989ApJ...345..434T} \\
{\tt A1}, $a\sin{i}$ & projected semi-major axis (lt-s) \\
{\tt PB}, $P_{\text{b}}$ & binary period (days) \\
{\tt OM}, $\omega_0$ & longitude of periastron (degs) \\
{\tt T0} & time of periastron (MJD) \\
{\tt ECC}, $e$ & orbital eccentricity \\
{\tt EPS1}, $\varepsilon_1$/$\eta$ & $e\sin{\omega_0}$ for {\tt ELL1} model \citep[see Appendix in][]{2001MNRAS.326..274L} \\
{\tt EPS2}, $\varepsilon_2$/$\kappa$ & $e\cos{\omega_0}$ for {\tt ELL1} model \citep[see Appendix in][]{2001MNRAS.326..274L} \\
{\tt TASC}, $T_{\text{asc}}$ & time of the ascending node for {\tt ELL1} model \citep[see Appendix in][]{2001MNRAS.326..274L} \\
{\tt GAMMA}, $\gamma$ & relativistic gravitational redshift and time dilation term in {\tt BT} model (s) \\
{\tt OMDOT}, $\dot{\omega}_0$ & time derivative of $\omega_0$ (degs\,yr$^{-1}$) \\
{\tt XDOT} & time derivative of $a\sin{i}$ (lt\,s\,s$^{-1}$ or $10^{-12}$lt-s\,s$^{-1}$ if $|\dot{a\sin{i}}| > 10^{-7}$)  \\
{\tt PBDOT}, $\dot{P}_{\text{b}}$ & time derivative of period ($10^{-12}$ if $|\dot{P_{\text{b}}}| > 10^{-7}$) \\
{\tt EDOT}, $\dot{e}$ & time derivative of eccentricity (s$^{-1}$ or $10^{-12}$\,s$^{-1}$ if $|\dot{e}| > 10^{-7}$) \\
{\tt EPS1DOT}, $\dot{\varepsilon}_1$ & time derivative of $\varepsilon_1$ (s$^{-1}$ or $10^{-12}$\,s$^{-1}$ if $|\dot{\varepsilon_1}| > 10^{-7}$) \\
{\tt EPS2DOT}, $\dot{\varepsilon}_2$ & time derivative of $\varepsilon_2$ (s$^{-1}$ or $10^{-12}$\,s$^{-1}$ if $|\dot{\varepsilon_2}| > 10^{-7}$) \\
{\tt XPBDOT} & time derivative of period minus GR prediction ($10^{-12}$ if $|\dot{P_{\text{b}}}| > 10^{-7}$) \\
{\tt SINI}, $\sin{i}$ & sine of orbital inclination angle \\
{\tt MTOT}, $M$ & total binary system mass (M$_{\odot}$) \\
{\tt M2}, $m_2$ & binary companion mass (M$_{\odot}$) \\
{\tt DR}, $d_r$ & relativistic deformation of the orbit \\
{\tt DTHETA}, $d_{\theta}$ & relativistic deformation of the orbit \\
{\tt SHAPMAX}, $s_x$ & defined as $-\ln{(1-\sin{i})}$ \\
% {\tt KIN}, $i$ & inclination angle (Kopeikin term) (degs) \\
{\tt KOM}, $\Omega$ & longitude of ascending node (Kopeikin term) (degs) \\
{\tt D\_AOP} & Kopeikin term (arcsec$^{-1}$) \citep[see, e.g., Section~2.7.1 of][]{2006MNRAS.372.1549E} \\
{\tt A1\_}$X$ & projected semi-major axes for multiple orbital components ($X=2,3$ for {\tt BT1P} and {\tt BT2P} models) (lt-s) \\
{\tt PB\_}$X$ & periods for multiple orbital components ($X=2,3$ for {\tt BT1P} and {\tt BT2P} models) (days) \\
{\tt OM\_}$X$ & longitudes of periastron for multiple orbital components ($X=2,3$ for {\tt BT1P} and {\tt BT2P} models) (degs) \\
{\tt T0\_}$X$ & times of periastron for multiple orbital components ($X=2,3$ for {\tt BT1P} and {\tt BT2P} models) (MJD) \\
{\tt ECC\_}$X$ & eccentricities for multiple orbital components ($X=2,3$ for {\tt BT1P} and {\tt BT2P} models) \\
{\tt FB}$X$ & orbital frequencies for multiple components (in {\tt BTX} model) (Hz) \\
{\tt A0}, $A$ & first aberration parameter \citep[see, e.g., Section~2.7.3 of][]{2006MNRAS.372.1549E} \\
{\tt B0}, $B$ & second aberration parameter \citep[se, e.g., Section~2.7.3 of][]{2006MNRAS.372.1549E} \\
\hline
\multicolumn{2}{c}{Glitch parameters \citep[see][]{2006MNRAS.369..655H,2013MNRAS.429..688Y}} \\
\hline
{\tt GLEP\_}$X$ & glitch epochs (MJD) \\
{\tt GLPH\_}$X$ & glitch rotational phase offsets (rads) \\
{\tt GLF0\_}$X$ & glitch frequency offsets (Hz) \\
{\tt GLF1\_}$X$ & glitch first frequency derivative offsets (Hz\,s$^{-1}$) \\
{\tt GLF2\_}$X$ & glitch second  frequency derivative offsets (Hz\,s$^{-2}$) \\
{\tt GLF0D\_}$X$ & glitch decaying frequency components (Hz) \\
{\tt GLTD\_}$X$ & glitch time constant for decaying components (s) \\
\hline
\multicolumn{2}{c}{Non-GR parameters} \\
\hline
{\tt CGW}, $c_{\text{gw}}$ & speed of gravitational waves as a fraction of the speed of light \\
{\tt HPLUS}, $H_+$ &  gravitational wave amplitude for the tensor $+$-polarisation \citep[see, e.g.,][for definitions of this and the subsequent
parameters]{MaxCWpolariations} \\
{\tt HCROSS}, $H_{\times}$ &  gravitational wave amplitude for the tensor $\times$-polarisation \\
{\tt PHI0TENSOR}, $\Phi_{\text{t}}$ & tensor mode phase parameter \\
{\tt PSITENSOR}, $\psi_{\text{t}}$ & tensor mode phase parameter \\
{\tt HVECTORX}, $H_{\text{x}}$ &  gravitational wave amplitude for the vector x-polarisation \\
{\tt HVECTORY}, $H_{\text{y}}$ &  gravitational wave amplitude for the vector y-polarisation \\
{\tt PHI0VECTOR}, $\Phi_{\text{v}}$ & vector mode phase parameter \\
{\tt PSIVECTOR}, $\psi_{\text{v}}$ & vector mode phase parameter \\
{\tt HSCALARB}, $H_{\text{x}}$ &  gravitational wave amplitude for the scalar $b$-polarisation \\
{\tt HSCALARL}, $H_{\text{y}}$ &  gravitational wave amplitude for the scalar $l$-polarisation \\
{\tt PHI0SCALAR}, $\Phi_{\text{s}}$ & scalar mode phase parameter \\
{\tt PSISCALAR}, $\psi_{\text{s}}$ & scalar mode phase parameter \\
\hline
\hline
\end{longtable}

\subsubsection{Example prior file}\label{sec:priorfile}

The prior functions used by the code are described in \S\ref{sec:priorfuncs}. To implement them in the code requires a prior
file specifying the prior function require for each parameter that will be varied. A an example prior file containing all the
allowed prior types (for descriptive purposed only) is:
\begin{lstlisting}[frame=single]
H0      fermidirac 4.316e-24 9.1625
PHI0    uniform    0.0       3.141593
PSI     gaussian   0.6764    0.16532
A1      loguniform 1e-3      1e6
F0:F1   gmm        2  [[123.4,-1e-9],[123.4,-1.5e-9]] [[[1e-16,1e-18],[1e-18,1e-10]],[[1e-16,1e-18],[1e-18,1e-10]]] [0.5,1]
COSIOTA gmm        2  [[1.5896]]  [[[0.0219]]]  [1.0]  [[1.5519]]  [[[0.0219]]]  [1.0] [-1,1]
\end{lstlisting}
The units required for parameters in a prior file are all SI units, whereas for the {\tt .par} file they often follow a different
convention (see Table~\ref{tab:paramlist}).

Going through each line we have:
\begin{description}
 \item[{\tt{H0}}] this parameter has been assigned a Fermi-Dirac prior (\S\ref{sec:fdprior}) using the {\tt fermidirac} tag, which requires two values: $\sigma$ and $r$.
 \item[{\tt{PHI0}}] this parameter has been assigned a uniform prior (\S\ref{sec:uniformprior}) using the {\tt uniform} tag, which requires two values: a lower limit and an upper limit.
 \item[{\tt{PSI}}] this parameter has been assigned a Gaussian prior (\S\ref{sec:gaussianprior}) using the {\tt gaussian} tag, which requires two values: a mean and a standard deviation.
 \item[{\tt{A1}}] this parameter has been assigned a prior uniform in the logarithm of the parameter (\S\ref{sec:loguniform}) using the {\tt loguniform} tag, which
 required two values: a lower limit and an upper limit on the parameter ({\it not} the logarithm of the parameter).
 \item[{\tt{F0:F1}}] this pair of parameters has been assigned a Gaussian mixture model prior (\S\ref{sec:gmmprior}). This prior can be used for any number of
 parameters if given as a colon separated list in the prior file. Following the {\tt gmm} assignment, a number gives the number of modes for the model. This is then
 followed by a square-bracketed list of lists, with each mode giving a comma separated main list entry containing a sub-list of comma separated means for each parameter 
 (no additional whitespace is allowed in the lists). The list of means is then followed by a similar list of parameter covariance matrices for each mode (again no
 additional whitespace is allowed). This is then followed by a final bracketed list giving the relative weights for each mode (the weights do not have to be
 normalised to unity, as the code will automatically normalise them). If required this can then be followed by pairs of bracketed, comma separated values, for each
 parameter giving their minimum and maximum allowed ranges. If these are given then they have to be given for all parameters, not just some subset. These are not
 shown in this example.
 \item[{\tt{COSIOTA}}] this single parameter has been assigned a Gaussian mixture model prior (\S\ref{sec:gmmprior}). This shows how to input a Gaussian mixture
 mode for a single parameter, and also shows at the end minimum and maximum ranges for the parameter have also been input.
\end{description}

If parameters in the prior file are specified with the {\tt gaussian} tag then a multivariate Gaussian prior may also be used for them if a correlation matrix
is passed to \lppen (using the {\tt --cor-file} flag) containing correlation coefficients for the required parameters. Only the lower diagonal of the
correlation matrix is required, e.g.\ if we wanted a multivariate prior on $f_0$ and $\dot{f}$ we could have a correlation coefficient file containing:
\begin{lrbox}{\Lst}
\begin{lstlisting}
      F0   F1
F0    1
F1    0.5  1
\end{lstlisting}
\end{lrbox}
\\[5pt] \indent \fbox{\usebox{\Lst}} \\[5pt]
where the whitespace can be tabs or spaces, whilst the prior file contains the means and standard deviations of the parameters as noted above. It is worth noting
that, for fully (anti)correlated parameters correlation coefficients of one can lead to numerical issues during matrix inversion. So, in such cases the
(anti)correlation could be reduced to, say, 99.99\%. However, better solutions to such a degeneracy would be to fix one of the correlated values (i.e.\ do not include
it in the prior file) and only searching over the other, or finding a non-degenerate re-parameterisation.

\subsubsection{Example 1: single detector, single harmonic, input data}\label{sec:example1}

One of the simplest cases for running the code is on input data from a single detector and at a single frequency harmonic (in this
case at twice the rotation frequency). Let us assume a search that is just interested in a signal model described by the four parameters
$h_0$, $\phi_0$, $\cos{\iota}$ and $\psi$, and set a prior file, called {\tt prior.txt}, containing
\begin{lrbox}{\Lst}
\begin{lstlisting}
H0      uniform 0 1e-20
PHI0    uniform 0 3.14159265359
PSI     uniform 0 1.57079632679
COSIOTA uniform -1 1
\end{lstlisting}
\end{lrbox}
\\[5pt] \indent \fbox{\usebox{\Lst}} \\[5pt]
Given a pulsar {\tt .par} file, called, for example, {\tt pulsar.par} containing\footnote{For a search that does not include any phase evolution parameters
the only required values in the {\tt .par} file are the right ascension and declination, but in general the file passed to the code should be the one used
to perform the heterodyning that produced the input data file.}
\begin{lrbox}{\Lst}
\begin{lstlisting}
PSRJ   J1234+5601
RAJ    12:24:00.00
DECJ   56:01:00.00
F0     100.01
F1     -1.1e-10
PEPOCH 56789
EPHEM  DE405
\end{lstlisting}
\end{lrbox}
\\[5pt] \indent \fbox{\usebox{\Lst}} \\[5pt]
and assuming the input data (from the LIGO Hanford detector, H1, in this case) is in a file called {\tt data\_H1.txt}, then one would run:
\begin{lstlisting}[frame=single]
$ lalapps_pulsar_parameter_estimation_nested --detectors H1 --input-files data_H1.txt --par-file pulsar.par --prior-file prior.txt --outfile nest_H1.hdf --Nmcmcinitial 0 --Nlive 1024 --tolerance 0.1
\end{lstlisting}
where we have set an output HDF5-format file called {\tt nest\_H1.hdf} in which the nested samples will be output. This setup has used 1024 nested sampling
live points and a tolerance of 0.1 for stopping the code. The {\tt --Nmcmcinitial 0} flag tells the code that the initial live points drawn do not have to
be resampled.\footnote{This option is present for cases when the initial live points may not be easily sampled from their prior distributions, and therefore
an MCMC procedure is required to try and evolve them to be drawn from the actual prior. However, for all our priors there are simple ways to draw the initial
live points from the prior, so this resampling is not required, and this input argument can be set to zero.} In this case the solar system ephemeris files
will be found automatically. If the {\tt EPHEM} or {\tt UNITS} parameters are not explicitly set in the {\tt .par} file then these will default to the
DE405 ephemeris and the TCB time system. The progress of the above code can be monitored by including the {\tt --verbose} flag, and watching how the
reported {\tt dZ} value approaches the required tolerance.

Along with the file of nested samples two additional files called {\tt nest\_H1\_SNR} and {\tt nest\_H1\_Znoise} will be output. The former contains the
recovered optimal SNR for the maximum likelihood sample. The later contains the null likelihood, or noise, evidence for the data, described in
\S\ref{sec:nulllike}. The noise evidence is also contained in the {\tt nest\_H1.hdf} file, but this additional file is useful in cases when multiple
harmonics are used as it will contain noise evidences for individual harmonic datastreams, as well as that for the multiple combined streams.

The nested sample file can be converted to a file containing posterior samples, called, say, {\tt post\_H1.hdf}, using:
\begin{lrbox}{\Lst}
\begin{lstlisting}
$ lalapps_nest2pos -p post_H1.hdf nest_H1.hdf
\end{lstlisting}
\end{lrbox}
\\[5pt] \indent \fbox{\usebox{\Lst}} \\[5pt]
where the {\tt -p} flag give the name of the posterior output file. If the above use of \lppen had been run more than once on the same data and prior file,
with output names given by e.g.\ {\tt nest\_H1\_01.hdf}...{\tt nest\_H1\_$N$.hdf}, then all these could be listed as input to {\tt lalapps\_nest2pos} to
combined all the samples and evidence estimates. As well as the posterior samples, this file also contains (as does the {\tt nest\_H1.hdf}),
the signal model evidence, the noise model evidence, and the information gain (amongst other things).

The posterior samples, and noise and signal model (natural logarithm) evidences, can be extracted using a python function within {\tt lalapps}, via
\begin{lrbox}{\Lst}
\begin{lstlisting}
$ python
>>> from lalapps import pulsarpputils as pppu
>>> postsamples, sigevidence, noiseevidence = pppu.pulsar_nest_to_posterior('post_H1.hdf')
\end{lstlisting}
\end{lrbox}
\\[5pt] \indent \fbox{\usebox{\Lst}} \\[5pt]
The {\tt postsamples} variable is an object from which samples from different parameters can be extracted, and plotted as a marginalised posterior
distribution, e.g.,
\begin{lrbox}{\Lst}
\begin{lstlisting}
>>> from matplotlib import pyplot as pl
>>> # plot the posterior samples for the H0 parameter
>>> pl.hist(postsamples['H0'].samples, bins=20, normed=True, histtype='stepfilled')
>>> pl.xlabel('$h_0$')
>>> pl.ylabel('Posterior Probability Density')
>>> pl.show()
\end{lstlisting}
\end{lrbox}
\\[5pt] \indent \fbox{\usebox{\Lst}} \\[5pt]
where here, for the $h_0$ parameter it is contained in {\tt postsamples} as the fully uppercase {\tt 'H0'} value. 

Alternatives to extract the (natural logarithm) evidence values are using the {\tt h5py} python module, or HDF5 command line utilities. For the former one could
use, e.g.,
\begin{lrbox}{\Lst}
\begin{lstlisting}
>>> import h5py
>>> hdf = h5py.File('post_H1.hdf', 'r')
>>> a = hdf['lalinference']['lalinference_nest']
>>> sigevidence = a.attrs['log_evidence']
>>> noiseevidence = a.attrs['log_noise_evidence']
>>> information = a.attrs['information_nats']
\end{lstlisting}
\end{lrbox}
\\[5pt] \indent \fbox{\usebox{\Lst}} \\[5pt]
or for the latter one could use, e.g.\
\begin{lrbox}{\Lst}
\begin{lstlisting}
$ h5ls -v post_H1.hdf/lalinference/ | grep -A2 evidence
    Attribute: log_evidence scalar
        Type:      native double
        Data:  141816
--
    Attribute: log_noise_evidence scalar
        Type:      native double
        Data:  141822
\end{lstlisting}
\end{lrbox}
\\[5pt] \indent \fbox{\usebox{\Lst}} \\[5pt]
However, note that in this latter example the evidence values are truncated to integer values.

\subsubsection{Example 2: single detector, single harmonic, simulated data}\label{sec:example2}

As well as taking real data as an input the code can simulate Gaussian noise, and/or create and add a simulated signal to the data. To create
a signal to inject into the data requires a {\tt .par} file for the injected signal parameters. This can be the same, or different, to the file input
using the {\tt --par-file} flag, but to produce a non-zero signal it must contain at least a \gw amplitude parameter (any parameters not
included will be assumed to be zero). Any differences in the phase evolution parameters between the two files will be incorporated into the
signal via Equation~\ref{eq:deltaphi}. Given an injection {\tt .par} file, called, say, {\tt injection.par}, containing
\begin{lrbox}{\Lst}
\begin{lstlisting}
PSRJ    J1234+5601
RAJ     12:24:00.00
DECJ    56:01:00.00
F0      100.01
F1      -1.1e-10
PEPOCH  56789
EPHEM   DE405
H0      1e-25
COSIOTA -0.43
PHI0    0.6
PSI     1.2
\end{lstlisting}
\end{lrbox}
\\[5pt] \indent \fbox{\usebox{\Lst}} \\[5pt]
(which we see is the same as in \S\ref{sec:example1}, but with \gw amplitude parameters), one would run the following (assuming various other
names are the same as in \S\ref{sec:example1})
\begin{lstlisting}[frame=single]
$ lalapps_pulsar_parameter_estimation_nested --fake-data H1 --inject-file injection.par --scale-snr 10 --par-file pulsar.par --prior-file prior.txt --outfile nest_H1.hdf --Nmcmcinitial 0 --Nlive 1024 --tolerance 0.1
\end{lstlisting}
Here, rather than inputting a real data file, the {\tt --fake-data} flag has been used to specify fake data from the LIGO Hanford detector, where the
design sensitivity \citep{SRD}\footnote{\url{https://labcit.ligo.caltech.edu/~jzweizig/distribution/LSC_Data/}} power spectral density noise curve
would be used to set the Gaussian noise level for an initial LIGO detector at the appropriate \gw frequency (twice the value of {\tt F0} given in the {\tt .par}
file for the standard search).\footnote{Other
named detectors that can be provided are L1 for the LIGO Livingston detector, G1 for the GEO600 detector \citep[from Table~IV of][]{2001PhRvD..63d4023D},
V1 for the initial Virgo detector, and T1 for the TAMA300 detector \citep[from Table~IV of][]{2001PhRvD..63d4023D}. For the two LIGO detectors
\citep{aLIGOSRD} and Virgo \citep[see Equation~6 of][]{2012arXiv1202.4031M}, design curve noise levels for their Advanced configurations can be obtained
by prefixing the names with an `A'.} Setting the noise level from the design sensitivity for a given detector can be overridden by also using the
{\tt --fake-psd} flag to input the required single sided power spectral density value. The simulated data will by default start at a GPS time of 900000000
(13 Jul 2088, 15:59:46 UTC), be sampled at a rate of once per minute, and last for one solar day. However, each of these can be set with the appropriate flags:
{\tt --fake-starts}, {\tt --fake-lengths}, and {\tt --fake-dt}, respectively. In this example a value is also set for {\tt --scale-snr}, which means that after the
creating of a signal with parameters as given in the {\tt injection.par} file, it will be re-scaled so that it has the given SNR - in this case a value of 10. This
enables signals to be injected with a known SNR without needing to know the noise level a priori. If the {\tt --scale-snr} flag is not set, or has a value of zero,
then no re-scaling takes place. In this case the output {\tt nest\_H1\_SNR} file would also contain the signal's injected SNR and the recovered SNR, e.g.,
\begin{lrbox}{\Lst}
\begin{lstlisting}
$ cat nest_H1_SNR
# Injected SNR
H1	2.000	2.642255e+01	1.000000e+01
# Recovered SNR
H1	2.000	1.092211e+01
\end{lstlisting}
\end{lrbox}
\\[5pt] \indent \fbox{\usebox{\Lst}} \\[5pt]
where, after the detector name the harmonic frequency scaling factor is given, and for the injected signal the two SNR values are the pre-scaled value
and the actually injected value after rescaling via {\tt --scale-snr} value.

A simulated signal can also be injected into real data in the same way.

\subsubsection{Example 3: multiple detectors, single harmonic}

Running with multiple detectors requires a simple change to the example given in \S\ref{sec:example1}. Given two input files, e.g.\ {\tt data\_H1.txt}
and {\tt data\_L1.txt}, for the two LIGO detectors H1 and L1, one would simply use, e.g.\
\begin{lstlisting}[frame=single]
$ lalapps_pulsar_parameter_estimation_nested --detectors H1,L1 --input-files data_H1.txt,data_L1.txt --par-file pulsar.par --prior-file prior.txt --outfile nest_H1L1.hdf --Nmcmcinitial 0 --Nlive 1024 --tolerance 0.1
\end{lstlisting}
where the order of listing the detectors with the {\tt --detectors} flag should match the order of listing the data files with the {\tt --input-files} flag.
No whitespace is allowed between detector values or input file values, and they must be separated by a comma. In this case the output {\tt nest\_H1L1\_SNR} file
will contain the individual detector recovered SNRs and the coherent SNR, whilst the {\tt nest\_H1L1\_Znoise} file will contain both the individual detector
noise model evidences and the combined noise model evidence.

\subsubsection{Example 4: multiple detectors, two harmonics}

It is simple to extend the analysis to include the two harmonics from the $l=m=2$ mode (with a \gw frequency at twice the rotation rate) and the $l=2$, $m=1$
harmonic (with a \gw frequency at the rotation rate). To search at these two harmonics requires that the parameters searched over are either the signal or
waveform parameters given in \citet{2015MNRAS.453.4399P}. Using the (simpler and non-degenerate) waveform parameters $C_{21}$, $C_{22}$, $\Phi_{21}^C$ and
$\Phi_{22}^C$, from Equations~\ref{eq:hf} and \ref{eq:h2f}, could lead to the following ({\tt prior.txt}) prior file
\begin{lrbox}{\Lst}
\begin{lstlisting}
C22     uniform 0 1e-20
C21     uniform 0 1e-20
PHI22   uniform 0 6.28318530718
PHI21   uniform 0 6.28318530718
PSI     uniform 0 1.57079632679
COSIOTA uniform -1 1
\end{lstlisting}
\end{lrbox}
\\[5pt] \indent \fbox{\usebox{\Lst}} \\[5pt]
Assuming that for each detector (H1 and L1) the data has been heterodyned at both harmonics, leading to files {\tt data\_H1\_1f.txt}, {\tt data\_H1\_2f.txt},
{\tt data\_L1\_1f.txt}, {\tt data\_L1\_2f.txt}, where {\tt 1f} and {\tt 2f} note the harmonic rotation frequency scaling, then one would use
\begin{lstlisting}[frame=single]
$ lalapps_pulsar_parameter_estimation_nested --detectors H1,L1 --harmonics 1,2 --input-files data_H1_1f.txt,data_H1_2f,data_L1_1f.txt,data_L1_2f.txt --par-file pulsar.par --prior-file prior.txt --outfile nest_H1L1.hdf --Nmcmcinitial 0 --Nlive 1024 --tolerance 0.1
\end{lstlisting}
Here the ordering of the input files is important, with the files for each harmonic for a given detector listed together, and in the same order as given by the
{\tt --harmonics} flag. The noise evidence and SNRs for each harmonic and each detector will be output to the files {\tt nest\_H1L1\_Znoise} and {\tt nest\_H1L1\_SNR}.

\subsubsection{Example 5: single detectors, additional parameters}

Finally, we show an example when searching over additional phase evolution parameters. We show two options for this: the first is using the standard
model and likelihood method, and the second performs the same search, but sped-up by using the ROQ likelihood. If one were searching over frequency and
frequency derivative, $f_0$ and $\dot{f}$, then an example {\tt .par} file ({\tt pulsar.par}) could be
\begin{lrbox}{\Lst}
\begin{lstlisting}
PSRJ   J1234+5601
RAJ    12:24:00.00
DECJ   56:01:00.00
F0     100.01
F1     -1.1e-10
PEPOCH 54660
EPHEM  DE405
\end{lstlisting}
\end{lrbox}
\\[5pt] \indent \fbox{\usebox{\Lst}} \\[5pt]
with a prior file ({\tt prior.txt}) containing
\begin{lrbox}{\Lst}
\begin{lstlisting}
H0      uniform 0 1e-20
PHI0    uniform 0 3.14159265359
PSI     uniform 0 1.57079632679
COSIOTA uniform -1 1
F0      gaussian 100.01   1e-5
F1      gaussian -1.1e-10 2e-11
\end{lstlisting}
\end{lrbox}
\\[5pt] \indent \fbox{\usebox{\Lst}} \\[5pt]
Note that the frequency and frequency derivative values in the prior file are for the rotation frequency, so if searching for the standard $l=m=2$
harmonic then the actual frequencies and derivatives (and in this case their standard deviation) covered will be twice these values. This would also
be the case for any other form of the prior. As, in this example, the {\tt F0} and {\tt F1} values have be assigned Gaussian priors, if they have a
known correlation then a correlation coefficient file ({\tt pulsar.cor}) could also be set, e.g.\
\begin{lrbox}{\Lst}
\begin{lstlisting}
    F0  F1
F0  1.0
F1  0.7 1.0
\end{lstlisting}
\end{lrbox}
\\[5pt] \indent \fbox{\usebox{\Lst}} \\[5pt]
However, if this is not provided, then the parameters will be assumed to a priori be uncorrelated.

If running without ROQ, for a data from a single detector, H1 ({\tt data\_H1.txt}), the following commands could be used
\begin{lstlisting}[frame=single]
$ lalapps_pulsar_parameter_estimation_nested --detectors H1 --input-files data_H1.txt --par-file pulsar.par --prior-file prior.txt --cor-file pulsar.cor --outfile nest_H1.hdf --Nmcmcinitial 0 --Nlive 1024 --tolerance 0.1
\end{lstlisting}

To run with ROQ enabled, in the simplest case, requires the user to set a number of training waveforms and tolerance to use the produce a reduced order
model.\footnote{For now we do not attempt to explain these parameters, which we are leaving to a future publication, other than to say that for broader
parameter space searches (or searches over frequency derivatives, for which the epoch is further from the data epoch) more training waveforms are required
for a given tolerance. For a fixed tolerance, if the parameter space is too broad more templates may be required than are computational feasible (due to, e.g.,
memory constraints) to use.} In this example we will use 2500 training waveforms, and a tolerance of $5\ee{-12}$, and use data that starts at the same
epoch as given in the {\tt .par} file. The following commands could be used
\begin{lstlisting}[frame=single]
$ lalapps_pulsar_parameter_estimation_nested --detectors H1 --input-files data_H1.txt --par-file pulsar.par --prior-file prior.txt --cor-file pulsar.cor --outfile nest_H1.hdf --Nmcmcinitial 0 --Nlive 1024 --tolerance 0.1 --roq --roq-tolerance 5e-12 --ntraining 2500
\end{lstlisting}

In this example, for a single run on a random realisation of Gaussian noise lasting one day and sampled at a rate of one per minute, we find that the ROQ
version runs about 1.8 times faster. Within the given tolerance it reduces the prior parameter space to 32 orthogonal waveform templates, thus reducing the
the model and likelihood evaluations from using 1440 to 32 points. Due to various overheads the speed-up we see between the ROQ and regular run is only a factor
of $\sim 1.8$, rather than the $1440/32 = 45$ factor that might be hoped for. However, more impressive speed-up can be observed for longer data sets. It should be
noted that this observed speed-up is not a general rule, but is specific to the data and set up we used, but similar speed-ups for similar set ups would not be
unexpected. We also find that the likelihoods produced by the ROQ method agree very well with that produced using the full likelihood evaluation, with the
maximum-likelihood-template likelihoods agreeing to within $6\ee{-10}\%$.

\subsubsection{Example 6: single detector, non-GR model}

The standard analysis considers two hypotheses: GR signal and Gaussian noise. On top of these, we can also include non-GR signal hypotheses consisting of all
possible combinations of tensor, vector and scalar modes: GR+s, GR+v, GR+sv, s, v, sv \citep[see][for definitions of these]{MaxCWpolariations}.\footnote{If the
orientation of the source is known, we can also make a distinction between GR and a ``free-tensor'' hypothesis, corresponding to a signal model with $+$ and $\times$
polarisations with unrestricted amplitudes and phases.} For any set of data, we can compute Bayes factors for each of these hypotheses using \lppen and then combine
them to produce odds for any-signal vs.\ noise and non-GR vs.\ GR.

The \lppen code accepts the {\tt --nonGR} and {\tt --inject-nonGR} arguments, which can be used to search for and inject non-GR signals, respectively. Both options may
be used as simple flags or with a specific argument. In the former case, the code will use a generic signal model that includes all polarisations mentioned in the prior
file. For instance, for a single detector (H1 in this case) the most generic search (one including all 5 non-degenerate polarisations freely) would be carried out by calling (without injections):
\begin{lstlisting}[frame=single]
lalapps_pulsar_parameter_estimation_nested --par-file pulsar.par --input-files data.txt --outfile output.hdf --prior-file prior.txt --Nlive 1000 --detectors H1 --nonGR
\end{lstlisting}
with a prior file like:
\begin{lrbox}{\Lst}
\begin{lstlisting}
HPLUS       loguniform  1e-28  1e-21
HCROSS      loguniform  1e-28  1e-21
PHI0TENSOR  uniform     0      6.283185307179586
PSITENSOR   uniform     0      6.283185307179586
HVECTORX    loguniform  1e-28  1e-21
HVECTORY    loguniform  1e-28  1e-21
PHI0VECTOR  uniform     0      6.283185307179586
PSIVECTOR   uniform     0      6.283185307179586
HSCALARB    loguniform  1e-28  1e-21
PHI0SCALAR  uniform     0      6.283185307179586
\end{lstlisting}
\end{lrbox}
\\[5pt] \indent \fbox{\usebox{\Lst}} \\[5pt]
where {\tt HPLUS} and {\tt HCROSS} are the tensor amplitudes, {\tt HVECTORX} and {\tt HVECTORY} are the vector amplitudes and {\tt HSCALARB} is the scalar amplitude
(we could equivalently use {\tt HSCALARL}, since the breathing and longitudinal modes are degenerate to quadrupolar antennas like LIGO or Virgo); the {\tt PHI0TENSOR},
{\tt PHI0VECTOR} and {\tt PHI0SCALAR} parameters determine the overall complex phase offset between different ranks, while {\tt PSITENSOR} and {\tt PSIVECTOR} are 
phase offsets between modes of the same rank. Relevant parameters that do not explicitly show up in the prior file are set to zero; as a result, the same command above
can be used to search only over, say, vector modes by changing the prior file to read:
\begin{lrbox}{\Lst}
\begin{lstlisting}
HVECTORX loguniform 1e-28 1e-21
HVECTORY loguniform 1e-28 1e-21
PHI0VECTOR uniform 0 6.283185307179586
PSIVECTOR uniform 0 6.283185307179586
\end{lstlisting}
\end{lrbox}
\\[5pt] \indent \fbox{\usebox{\Lst}} \\[5pt]
This can be used to produce any of the ``free'' non-GR searches (s, v, t, sv, st, vt, svt). Note that, if any of the prior files are used without the {\tt --nonGR}
flag, the code will fail, as it will expect GR-specific parameters (like {\tt H0} and {\tt COSIOTA}). A generic non-GR injection may be produced by using the
{\tt --inject-nonGR} flag with an injection file that gives the values of non-GR parameters desired, e.g.,
\begin{lrbox}{\Lst}
\begin{lstlisting}
HPLUS 3e-24
HVECTORX  4e-24
PHI0VECTOR 3
\end{lstlisting}
\end{lrbox}
\\[5pt] \indent \fbox{\usebox{\Lst}} \\[5pt]
would produce an injection with `+' and vector-x components of the given strain amplitudes, with no phase offset for plus ({\tt PHI0TENSOR} is 0) and 3 radians for
vector-x.

The non-GR options also accept an argument that specifies a particular non-GR model; currently accepted options are {\tt G4V} and {\tt EGR}. The former corresponds
to a vector-only model \citep[proposed in][]{2015arXiv150304866M} that we usually use for testing; this model includes vector x and y modes with a particular
weighting given by the pulsars inclination as in equations~7 and 8 of \citet{2015PhRvD..91h2002I}. An example of a prior file for a {\tt --nonGR G4V} analysis
would be:
\begin{lrbox}{\Lst}
\begin{lstlisting}
PSI         uniform    -0.785398163397448  0.785398163397448
IOTA        uniform     0                  6.283185307179586
H0          loguniform  1e-28              1e-21
PHI0VECTOR  uniform     0                  6.283185307179586
\end{lstlisting}
\end{lrbox}
\\[5pt] \indent \fbox{\usebox{\Lst}} \\[5pt]
The second option accepted by {\tt --nonGR} and {\tt --inject-nonGR} is {\tt EGR}, which stands for ``enhanced GR''. This corresponds to a model made up of a GR signal
(parametersied in the usual way by $h_0$, $\cos{\iota}$, $\phi_0$ and $\psi$) plus contributions from any non-tensorial modes specified in the prior file. For instance, passing {\tt --nonGR EGR} with a prior like:
\begin{lrbox}{\Lst}
\begin{lstlisting}
PSI         uniform    -0.785398163397448 0.785398163397448
COSIOTA     uniform    -1                 1
H0          loguniform  1e-28             1e-21
PHI0TENSOR  uniform     0                 6.283185307179586
HVECTORX    loguniform  1e-28             1e-21
HSCALARB    loguniform  1e-28             1e-21
PHI0SCALAR  uniform     0                 6.283185307179586
\end{lstlisting}
\end{lrbox}
\\[5pt] \indent \fbox{\usebox{\Lst}} \\[5pt]
will produce a GR+s search; thus, by appropriately modifying the prior file, the {\tt --nonGR EGR} option can be used to also search for GR+v and GR+sv signals.
Non-GR injection files are constructed similarly. 

\subsection{The pipeline}

The process of performing searches for known pulsars does not just use \lppen. It actually requires a whole pipeline of code to: a) gather the \gw detector
data and operational segments (``science mode''); b) pre-processing the raw data via heterodyning (or spectral interpolation) to give the complex down-sampled
time series; c) set up the required prior distribution files; d) pass the data and priors to \lppen; e) convert the output nested samples into posterior samples;
and, finally, f) display those results (including odds values) in a useful fashion, e.g.\ on a webpage. This process can be performed manually in a step by step
manner, but there also exists a pipeline script called {\tt lalapps\_knope} that gathers together all these operations in a way that can be run on a computer
cluster running the HTCondor\footnote{\url{https://research.cs.wisc.edu/htcondor/}} job management system.

To run this pipeline requires just a single configuration file to be written in the {\tt INI}\footnote{\url{https://en.wikipedia.org/wiki/INI_file}} format. An
example configuration file is given below containing comments on the necessary input fields:
\begin{lstlisting}[frame=single,basicstyle=\tiny\ttfamily]
# Example configuration file for the full known pulsar search pipeline

; general inputs for the whole analysis
[analysis]
# a list of the inteferometers to analyse (REQUIRED)
ifos = ['H1', 'L1']

# a GPS start time (if a single value is set then this is used for all detectors), or a dictionary of GPS start times for the analysis; one for each detector (e.g. {'H1': 1129136415, 'L1': 1129137415})
starttime = 1129136415

# a GPS end time (if a single value is set then this is used for all detectors), or a dictionary of GPS end times for the analysis; one for each detector (e.g. {'H1': 1129136415, 'L1': 1129137415})
endtime = 1129284015

# choose whether to use lalapps_heterodyne_pulsar (heterodyne) or lalapps_SplInter (splinter) (value is case insensitive)
preprocessing_engine = heterodyne

# a flag to set if just wanting to do the data processing (e.g. heterodyne or splinter) and not parameter estimation
preprocessing_only = False

# a flag to set if just wanting to do the postprocessing (parameter estimation and webpage page creation)
postprocessing_only = False

# if just doing the postprocessing (i.e. 'postprocessing_only = True') a pickled version of the class used for the
# preprocessing (or entire previous run is required) to obtain information on the pre-processed data
preprocessed_pickle_object =

# flag to set whether to run the analysis for individual detectors only (default is to run on individual detectors AND coherently with all detectors)
incoherent_only = False

# flag to set whether to only run the coherent multi-detector analysis
coherent_only = False

# set the number of background odds ratio studies when doing parameter estimation
num_background = 0

# a list of multiplicative factors of the pulsar's rotation frequency to analyse (e.g. [1., 2.] for analysing both the 1f and 2f modes)
# This can be a list of either 1 or 2 values, but if it is a list of two values they can currently only be [1., 2.] (or [2., 1.]),
# whereas if there is only one value it can take any positive number.
freq_factors = [2.0]

# the path for timing and solar system ephemeris file
ephem_path = /usr/share/lalpulsar

# the directory in which the Condor DAGs are created and run from
run_dir = /home/user/analysis/dags

# the name of the DAG file to create (if not set this will be generated automatically)
dag_name =

# set this flag to automatically submit the Condor DAG created by the script
submit_dag = True

# this flag sets whether running in autonomous mode
# If running in autonomous mode then information on the current state of the run for each pulsar must be kept
# (e.g. a json file must be created for each pulsar containing the end time that has currently been run until,
# it will also contain the previously used segment list and cache file). If no file is found to exist for a
# particular pulsar, e.g. that pulsar has newly been added to the pulsar directory, then it will start from
# scratch for that pulsar (based on the start time in this file). The script (run e.g. with cron once per week)
# that uses this file for the automation should update the endtime on each run.
# A script will be required to produce the the cache files for each pulsar.
autonomous = False

# the initial start time of any autonomous run (as starttime will be updated each time the code is run)
autonomous_initial_start =

# a base directory (for each detector) for preprocessing outputs (structure for the parameter estimation is still required).
# Within each the following structure structure is assumed:
# -> base_dir/PSRname
#           |       '-> /segments.txt (science segments file for that pulsar)
#           |       '-> /data (directory for processed data files)
#           |               '-> /coarse (directory for coarse heterodyned data)
#           |               |         '-> /freqfactor (directory for heterodyne at freq factor times rotation frequency)
#           |               '-> /fine (directory for fine heterodyned data)
#           |               |       '-> /freqfactor (directory for heterodyne at rotation frequency)
#           |               '-> /splinter (directory for spectrally interpolated data)
#           |                           '-> /freqfactor (directory for heterodyne at rotation frequency)
#           '-> /splinter (a directory to output files from splinter before moving them into the above structure)
#           '-> /cache.lcf (a general frame/SFT cache file used by all pulsar if their directory does not containing their own cache)
preprocessing_base_dir = {'H1': '/home/username/analysis/H1', 'L1': '/home/username/analysis/L1'}

# path to directory containing pulsar parameter (.par) files, or an individual parameter file
# (once the analysis script has been run at least once each .par file will have an associated file (.mod_parfilename) with
# it modification time - if the file is updated, and therefore has a different modification time, then
# the full analysis will be re-run for that pulsar)
pulsar_param_dir = /home/username/analysis/pulsar

# path to Condor log files
log_dir = /home/username/analysis/log

# set to true if running on software/hardware injections
injections = False

# file to output a pickled version of the KnownPulsarPipelineDAG class
pickle_file = /home/username/analysis/run.p

# email address for job completion notication (if no email is given the no notications will be sent)
email =

; Condor information
[condor]
# Condor accounting group
accounting_group =

# Condor accounting group user
accounting_group_user = user.name

# the data find executable (e.g. /usr/bin/gw_data_find), or a dictionary of pre-found cache files (one entry for each detector) if not wanting to use gw_data_find e.g. {'H1': '/home/username/analysis/H1cache.txt'}
datafind = /usr/bin/gw_data_find

; inputs for running a data find job (NOTE: this can be used to search for frame if using the heterodyne, or SFTs if using spectral interpolation)
[datafind]

# a dictionary of frame types to be returned; one for each detector
type = {'H1': 'H1_HOFT_C00', 'L1': 'L1_HOFT_C00'}

# a string to match in the URL paths returned
match = localhost

; inputs for running a science segment finding job
[segmentfind]
# path to segment database query script, or a dictionary of pre-found segment files (one entry for each detector) if not wanting to use segment finding script e.g. {'H1': '/home/username/analysis/H1segments.txt'}
segfind = /usr/bin/ligolw_segment_query_dqsegdb

# path to ligolw_print
ligolw_print = /usr/bin/ligolw_print

# URL of segment database server
server = https://segments.ligo.org

# a dictionary of the required segment types
segmenttype = {'H1': 'H1:DMT-ANALYSIS_READY:1', 'L1': 'L1:DMT-ANALYSIS_READY:1'}

; inputs for running the lalapps_heterodyne_pulsar code
[heterodyne]
# condor universe
universe = vanilla

# path to lalapps_heterodyne_pulsar
heterodyne_exec = /usr/bin/lalapps_heterodyne_pulsar

# path to directory containing pulsar parameter file (.par) updates, or an individual parameter file
# pulsar_update_dir = /home/username/analysis/pulsar_update # DONT USE FINE HETERODYNE UPDATES - JUST REDO WHOLE HETERODYNE SO THAT THERE AREN'T MULTIPLE PAR FILES/HETERODYNE FILES

# low-pass filter (9th order Butterworth) knee frequency (Hz)
filter_knee = 0.25

# the frame data sample rate (for the coarse heterodyne) (Hz)
coarse_sample_rate = 16384

# the re-sampling rate for the coarse heterodyne (Hz)
coarse_resample_rate = 1

# a dictionary of frame channel names; one for each detector
channels = {'H1': 'H1:GDS-CALIB_STRAIN', 'L1': 'L1:GDS-CALIB_STRAIN'}

# the fine heterodyne re-sampling rate (the sample rate is taken from coarse_resample_rate) (Hz)
fine_resample_rate = 1/60

# the standard deviation threshold for removing outliers
stddev_thresh = 3.5

# set to output the coarse heterodyne data in binary files (if true then a binary input will be assumed for the fine heterodyne)
binary_output = True

# gzip the coarse output files rather than outputting as binary
gzip_coarse_output = False

# gzip the fine output files
gzip_fine_output = True

; inputs for running the lalapps_SplInter code
[splinter]
# condor universe
universe = vanilla

# path to SplInter executable
splinter_exec = /usr/bin/lalapps_SplInter

# list with the start and end frequency ranges for the SFTs
freq_range = [30., 2000.]

# the standard deviation threshold for removing outliers
stddev_thresh = 3.5

# the bandwidth (Hz) around the signal frequency to use in interpolation
bandwidth = 0.3 # this is the default value from the code

# minimum length (seconds) of science segments to use
min_seg_length = 1800 # this is the default value (half an hour) from the code

; inputs for running the parameter estimation code lalapps_pulsar_parameter_estimation_nested
[pe]
# condor universe
universe = vanilla

# path to the parameter estimation executable
pe_exec = /usr/bin/lalapps_pulsar_parameter_estimation_nested

# the base output directory for the nested samples (directories for each detector/joint analysis will be created)
pe_output_dir = /home/username/analyses/nested_samples

# a pre-made prior file for the analysis (if this is not set then the prior file will be generated for each source based on the par file and/or other information) - this has priority over any other options in this configuration file
premade_prior_file =

# a set of prior options in dictionary form e.g.
# prior_options = {'H0': {'priortype': 'uniform', 'ranges': [0., 1e-22]}, 'PHI0': {'priortype': 'uniform', 'ranges': [0., 3.14]}}
prior_options =

# if 'derive_amplitude_prior' is true (and the 'premade_prior_file' is not set) then amplitude priors will be derived
# from the heterodyned/spectrally interpolated data based on the either previous upper limits or those
# estimated from ASDs of previous runs.
# The 'amplitude_prior_type' can also be set to either 'fermidirac' (default) or 'uniform'. All other parameters
# will have their ranges taken from 'predefined_prior_options'.
derive_amplitude_prior = True

amplitude_prior_scale = 5

amplitude_prior_type = 'fermidirac'

amplitude_prior_model_type = 'waveform'

# a JSON file with pulsar name keys associated with previous posterior samples files for use as priors in current analysis
previous_posteriors_file = path_to_file_of_previous_posterior_files

# a JSON file with amplitude upper limits from previous runs
amplitude_prior_file = path_to_file_of_previous_upper_limits

# a file, or dictionary or files, containing paths to amplitude spectral density files
amplitude_prior_asds =

# a value, or dictionary of values, containing the observation times (in days) for use with the above ASD files
amplitude_prior_obstimes =

# go through the pulsar parameter file and use the errors to set Gaussian priors in the prior file (also create a correlation coefficient file for these, setting everything to be uncorrelated except the extremely highly correlated binary parameters)
use_parameter_errors = False

# the number of parallel lalapps_pulsar_parameter_estimation_nested runs for a given pulsar/detector combination
n_runs = 5

# the number of live points for each run
n_live = 2048

# the number of MCMC samples for each nested sample update (if not set this will be automatically calculated)
n_mcmc =

# the number of MCMC samples for initial shuffling of the prior points (shouldn't be needed for pulsar code)
n_mcmc_initial =

# the tolerance (stopping criterion) for the runs
tolerance = 0.1

# a random seed for the RNG (if not set then this will default to it's standard method)
random_seed =

# flag to set whether running with non-GR parameterisation
non_gr = False

# flag to say whether to use the 'waveform' or 'source' parameterisation (defaults to 'waveform')
model_type = 'waveform'

# flag to set whether using a Gaussian likelihood, or the default Student's t-likelihood (if using Splinter for the pre-processing engine then a Gaussian likelihood will automatically be used, and override anything set here)
gaussian_like = False

# flag to set whether the model under consideration is a biaxial model
biaxial = False

# path to lalapps_nest2pos for nested sample -> posterior conversion
n2p_exec = /usr/bin/lalapps_nest2pos

# the base output directory for posteriors
n2p_output_dir = /home/username/analyses/posterior_samples

## information for background runs
# the number of live points for background analyses
n_live_background = 1024

# the number of of parallel runs for each
n_runs_background = 2

# the base output directory for the background nested samples (directories for each detector/joint analysis will be created)
pe_output_dir_background = /home/username/analyses/background/nested_samples

# the base output directory for the background posteriors
n2p_output_dir_background = /home/username/analyses/background/posterior_samples

# flag to set whether to clean (i.e. remove) all the nested sample files (keeping the posteriors)
clean_nest_samples = False

# set to true if wanting the output to use the l=m=2 gravitational wave phase as the initial phase, rather than the default rotational phase
use_gw_phase = False

# flags for use of Reduced Order Quadrature (ROQ)
# Any generated ROQ interpolant files will be placed in structure of the pe_output_dir
use_roq = False

# set the number of training sets for using the in reduced basis and interpolant generation
roq_ntraining = 2500

# set the maximum data chunk length for when using ROQ
roq_chunkmax = 1440

# set the tolerance for producing the ROQ bases
roq_tolerance = 5e-12

# set if wanting the bases produced over a uniform parameter space even for parameters with Gaussian priors
roq_uniform = False

; inputs for creating results pages
[results_page]
# condor universe
universe = local

# results page (Condor) log directory
log_dir = /usr1/username/logs

# results page creation executable
results_exec = /usr/bin/lalapps_knope_result_page.py

# results collation executable
collate_exec = /usr/bin/lalapps_knope_collate_results.py

# the output base web directory for the results
web_dir = /home/username/public_html/results

# the equivalent output base URL for the above path
base_url = https://myurl/~username/results

# the upper limit credible interval to use (default to 95%)
upper_limit = 95

# value on which to sort the results table
sort_value = name

# direction on which to sort the results table
sort_direction = ascending

# list upper limits to show in the results table
results = ['h0ul', 'ell', 'sdrat', 'q22', 'bsn']

# list of source values to output
parameters = ['f0rot', 'f1rot', 'ra', 'dec', 'dist', 'sdlim']

# set whether to show posterior plots for all parameters
show_all_posteriors = False

# set whether to subtract the true/heterodyned value from any phase parameters in a search for plotting
subtract_truths = False

# set whether to show the priors on the 1D posterior plots
show_priors = True

# set whether to copy par file, prior files, heterodyne files, and posterior files into results page directory
copy_all_files = True
\end{lstlisting}

Unless specified as otherwise within a {\tt .par} file the pipeline assumes the use of the TCB timing convention (as is the default for TEMPO2), and the DE405
solar system ephemeris.

\section{Derivation of the Student's {\it t}-likelihood function}\label{app:likelihood}

Here we derive, in detail, the form of the Student's {\it t}-likelihood given in Section~\ref{sec:likelihood}.
This derivation can, in part, be found in \citet{Dupuisthesis} and \citet{2005PhRvD..72j2002D}, but we correct a
slight error from those references.

If a stretch of data of length $m$ is assumed to consist of a signal
defined by a set of parameters $\vec{\theta}$ and Gaussian noise with zero mean and standard deviation
$\sigma$, then the standard deviation can be included as an unknown parameter in the likelihood function and
marginalised over. The likelihood can therefore be given by
\begin{equation}\label{eq:likesigma}
p(\mathbf{B}|\vec{\theta},I) = \int_0^{\infty} p(\mathbf{B},\sigma|\vec{\theta},I) \text{d}\sigma =
\int_0^{\infty} p(\mathbf{B}|\vec{\theta},\sigma,I)p(\sigma|I) \text{d}\sigma,
\end{equation}
where $p(\mathbf{B}|\vec{\theta},\sigma,I)$ is the likelihood function for the data given the unknown signal
parameters and the noise standard deviation, and $p(\sigma|I)$ is the prior on $\sigma$.

Given that the noise is assumed Gaussian the likelihood within the integral is given by
\begin{equation}\label{eq:likesigma2}
p(\mathbf{B}|\vec{\theta},\sigma,I) = \left(\frac{1}{2\pi\sigma^2}\right)^{m}
\exp{\left( -\frac{\sum_{k=1}^m|B_k-y(\vec{\theta})_k|^2}{2\sigma^2} \right)}.
\end{equation}
for our complex data $B$ and model $y$, where for a complex number $x$ we have used $|x|^2 =
\Re{(x)}^2+\Im{(x)}^2$, and the noise in the real and imaginary parts is assumed to have the same
distribution.

We chose a scale invariant prior on $\sigma$ of $p(\sigma|I) = 1/\sigma$, for which the marginalisation of
Eqn.~\ref{eq:likesigma} can be performed analytically as we will show.
Substituting this prior and Eqn.~\ref{eq:likesigma2} into Eqn.~\ref{eq:likesigma} we can begin the
integration with the substitution
\begin{equation}\label{eq:u}
u = \sqrt{\frac{\sum_{k=1}^m|B_k-y(\vec{\theta})_k|^2}{2\sigma^2}},
\end{equation}
giving
\begin{equation}\label{eq:du}
\left|\frac{\text{d}u}{\text{d}\sigma}\right| = \sqrt{\frac{\sum_{k=1}^m|B_k-y(\vec{\theta})_k|^2}{2}}\sigma^{-2}.
\end{equation}
Here we note that Eqn.~\ref{eq:du} is different from the equivalent Eqn.~2.28 in \citet{Dupuisthesis} by a
factor of two. Rearranging Eqns.~\ref{eq:u} and~\ref{eq:du}, and for simplicity using the substitution $X =
\sum_{k=1}^m |B_k-y(\vec{\theta})_k|^2$, gives
\begin{equation}
\sigma = \left(\frac{X}{2u^2}\right)^{1/2},
\end{equation}
and
\begin{align}
\text{d}\sigma = & \sigma^2 \left(\frac{2}{X}\right)^{1/2} \text{d}u
\nonumber \\
 = & \frac{X}{2u^2} \left(\frac{2}{X}\right)^{1/2} \text{d}u \nonumber \\
 = & 2^{-1/2} X^{1/2} u^{-2} \text{d}u.
\end{align}
Putting everything into Eqn.~\ref{eq:likesigma} gives
%\begin{widetext}
\begin{align}\label{eq:integral1}
p(\mathbf{B}|\vec{\theta})= & (2\pi)^{-m} \int_0^{\infty} 2^{(m+\frac{1}{2})}
X^{-(m+\frac{1}{2})} u^{2m+1} e^{-u^2} 2^{-1/2} X^{1/2} u^{-2} \text{d}u, \nonumber \\
 = & \pi^{-m} X^{-m} \int_0^{\infty} u^{2m-1} e^{-u^2} \text{d}u.
\end{align}
%\end{widetext}
The integral in Eqn.~\ref{eq:integral1} can then be performed by making the substitutions $x = u^2$ and $\text{d}x
= 2u\,\text{d}u$, giving
\begin{equation}
p(\mathbf{B}|\vec{\theta}) = \pi^{-m} X^{-m} \int_0^{\infty} \frac{1}{2} x^{m-1} e^{-x}
\text{d}x.
\end{equation}
This integral is the definition of the Gamma function
\begin{equation}
 \int_0^{\infty} x^{m-1} e^{-x} \text{d}x = \Gamma(m).
\end{equation}
For positive integers, Stirling's formula shows that the Gamma function is given by $\Gamma(n+1) = n!$, so our
likelihood becomes
\begin{equation}\label{eq:complex}
p(\mathbf{B}|\vec{\theta}) = \frac{(m-1)!}{2\pi^m} \left(\sum_{k=1}^m
|B_k-y(\vec{\theta})_k|^2\right)^{-m}.
\end{equation}
Note that, again, this differs from Eqn.~2.31 in \citet{Dupuisthesis} by some constant factors.

For our analysis the data is split into chunks for each of which the assumption of noise stationarity is
thought to be valid, but between chunks is thought to be invalid (see discussion in \S\ref{sec:splitting}). The joint likelihood of all data can be
obtained from the product of the likelihood for each chunk given by
\begin{equation}\label{eq:prod}
p(\mathbf{B}|\vec{\theta},I) = \prod_{j=1}^M \frac{(m_j-1)!}{2\pi^{m_j}}
\left(\sum_{k=k_0}^{k_0+(m_j-1)} |B_k-y(\vec{\theta})_k|^2\right)^{-m_j},
\end{equation}
where $M$ is the total number of independent data chunks with lengths $m_j$ and $k_0 = 1+\sum_{i=1}^{j}
m_{i-1}$ (with $m_0 = 0$) is the index of the first data point in each chunk.

\section{Fast likelihood evaluation}\label{app:fle}

In \S\ref{sec:fastlike} we stated that the likelihood can be evaluated quickly in cases where the $\Delta\phi_{\mathcal{K}}$ values in Equations~\ref{eq:hf}, \ref{eq:h2f}
and \ref{eq:hnongr} are zero, i.e., no phase evolution parameters are required. This can be done by pre-summing over data, $B$, and combinations of the time-varying
antenna response function, after which only coefficients of these pre-summed terms need to be calculated. Here we will explicitly write out the derivation
of these coefficients for the case of a signal in GR, for which the antenna pattern functions for the `$+$' and `$\times$' components are given by Equation~\ref{eq:antenna}.

Assuming a signal in GR, we start with a model of the form
\begin{align}\label{eq:appmodel}
h(t) =& C_+ F_+(\psi,t)e^{i\Phi_{lm}} + iC_{\times}F_{\times}(\psi,t)e^{i\Phi_{lm}}, \nonumber \\
=& C_+ F_+(\psi,t)\cos{\Phi_{lm}} - C_{\times}F_{\times}(\psi,t)\sin{\Phi_{lm}} + i\left(C_+ F_+(\psi,t)\sin{\Phi_{lm}}+ C_{\times}F_{\times}(\psi,t)\cos{\Phi_{lm}} \right), \nonumber \\
=& C_{R,+}F_+(\psi,t) + C_{R,\times}F_{\times}(\psi,t) + i\left(C_{I,+}F_+(\psi,t) + C_{I,\times}F_{\times}(\psi,t) \right),
\end{align}
where, setting $\zeta = 90^{\circ}$ in Equation~\ref{eq:antenna}, the antenna pattern functions are
\begin{align}\label{eq:antennanew}
F_+(\psi,t) &=a(t)\cos{2\psi} + b(t)\sin{2\psi}, \nonumber \\
F_{\times}(\psi,t) &= b(t)\cos{2\psi} - a(t)\sin{2\psi},
\end{align}
and we have set the real and imaginary model amplitude coefficients for the two polarisations to be $C_{R,+} =
C_+\cos{\Phi_{lm}}$, $C_{R,\times} = -C_{\times}\sin{\Phi_{lm}}$, $C_{I,+} =
C_+\sin{\Phi_{lm}}$, $C_{I,\times} = C_{\times}\cos{\Phi_{lm}}$.\footnote{For our GR signal model emitting at twice
the rotation frequency, given by Equation~\ref{eq:h2f}, we would have $C_+ = -\frac{C_{22}}{2}\left(1+\cos{}^2{\iota}\right)$ and $C_{\times} = C_{22}\cos{\iota}$.}
If we just take the likelihood (either the Student's $t$-likelihood of Equation~\ref{eq:stlikelihood}, or the
Gaussian likelihood of Equation~\ref{eq:gausslikelihood})\footnote{If using the Gaussian likelihood there is a slight difference to what is 
described here to take account of the data noise standard deviation, $\sigma$, 
such that we require the substitutions $B_i \rightarrow B_i/\sigma_i$, $a(t_i) \rightarrow a(t_i)/\sigma_i$ and $b(t_i) \rightarrow b(t_i)/\sigma_i$.} for a single detector,
data stream, and data chunk (see \S\ref{sec:likelihood}) then we see that it contains the summation over time samples
\begin{equation}
L = \sum_{i=1}^N \left|B(t_i) - h(t_i)\right|^2 = \sum_{i=1}^N \left[ \left(\Re{[B(t_i)]} - \Re{[h(t_i)]}\right)^2 + \left(\Im{[B(t_i)]} - \Im{[h(t_i)]}\right)^2 \right].
\end{equation}
We can expand this out to give
\begin{equation}\label{eq:likepart}
L = \sum B_R^2 + \sum B_I^2 + \sum h_R^2 + \sum h_I^2 - 2\sum\left(B_R h_R + B_I h_I \right),
\end{equation}
where for convenience we have removed the summation subscripts and explicit time dependence, and used $R$ and $I$ subscripts to represent the real and imaginary components respectively. We are now interested in the summations that contain parts of the model function $h$. If we take the summation over the square of the real part of the model we get
\begin{align}\label{eq:exlikepart}
\sum h_R^2 =& \sum \left(C_{R,+} F_+ + C_{R,\times} F_{\times} \right)^2, \nonumber \\
 = & \sum \left(C_{R,+}^2 F_+^2 + C_{R,\times}^2 F_{\times}^2 + 2C_{R,+}C_{R,\times}\right), \nonumber \\
 = & C_{R,+}^2 \left(a^2 \cos{}^2{2\psi} + b^2 \sin{}^2{2\psi} + 2ab\sin{2\psi}\cos{2\psi}\right) + \nonumber \\
 & C_{R,\times}^2 \left(b^2 \cos{}^2{2\psi} + a^2\sin{}^2{2\psi} - 2ab\sin{2\psi}\cos{2\psi} \right) + \nonumber \\
 & 2C_{R,+}C_{R,\times}\left( (b^2 - a^2)\sin{2\psi}\cos{2\psi} + ab(\cos{}^2{2\psi} - \sin{}^2{2\psi}) \right),
\end{align}
where the summation is now implicit in the $a$ and $b$ terms, i.e., $a^2 = \sum a(t)^2$, $b^2 = \sum b(t)^2$, and $ab = \sum a(t)b(t)$, which are the only time varying components
and can be pre-computed. If we now say
\begin{equation}
\sum h_R^2 = K_{a^2}^Ra^2 + K_{b^2}^Rb^2 + K_{2ab}^R2ab,
\end{equation}
where the $K^R$s are the coefficients of $a^2$, $b^2$ and $2ab$, we find them to be
\begin{align}
K_{a^2}^R & = C_{R,+}^2 \cos{}^2{2\psi} + C_{R,\times}^2\sin{}^2{2\psi} - 2C_{R,+}C_{R,\times}\sin{2\psi}\cos{2\psi}, \nonumber \\
K_{b^2}^R & = C_{R,+}^2 \sin{}^2{2\psi} + C_{R,\times}^2\cos{}^2{2\psi} + 2C_{R,+}C_{R,\times}\sin{2\psi}\cos{2\psi}, \nonumber \\
K_{2ab}^R & = C_{R,+}C_{R,\times}\left(\cos{}^2{2\psi} - \sin{}^2{2\psi} \right) + \sin{2\psi}\cos{2\psi}\left(C_{R,+}^2 - C_{R,\times}^2\right).
\end{align}
Factorising these out we find that
\begin{align}\label{eq:realcoeffs}
K_{a}^R &= C_{R,+}\cos{2\psi} - C_{R,\times}\sin{2\psi}, \nonumber \\
K_{b}^R &= C_{R,+}\sin{2\psi} + C_{R,\times}\cos{2\psi},
\end{align}
where $K_{a^2}^R = \left(K_{a}^R\right)^2$, $K_{b^2}^R = \left(K_{b}^R\right)^2$, and $K_{2ab}^R = K_{a}^RK_{b}^R$.
It is easy to see from above that things will be identical for the imaginary components and the equations hold just
by swapping $R$ for $I$.
It is worth noting that the coefficients given in Equation~\ref{eq:realcoeffs} take the opposite form (covariant vs.\ contravariant rotation) than
the antenna pattern functions given in Equation~\ref{eq:antennanew}, and this difference is present in the model
in the \lppen code, but should not be unexpected.

We can also see that the coefficients given in Equation~\ref{eq:realcoeffs} are those required in the components of
Equation~\ref{eq:exlikepart} that sum over the product of the data and model. For the real part we have
\begin{align}
\sum B_Rh_R &= \sum B_R\left(C_{R,+}F_+ + C_{R,\times}F_{\times} \right), \nonumber \\
 &=\sum B_R\left( C_{R,+}\left(a\cos{2\psi}+b\sin{2\psi}\right) + C_{R,\times}\left(b\cos{2\psi} - a\sin{2\psi} \right) \right), \nonumber \\
 &= d^a_R\left(C_{R,+}\cos{2\psi} - C_{R,\times}\sin{2\psi}\right) + d^b_R\left(C_{R,+}\sin{2\psi} + C_{R,\times}\cos{2\psi} \right),
\end{align}
where we have set $d^a_R = \sum B_R a$ and $d^b_R = \sum B_R b$. We can therefore see that the coefficients
of $d^a_R$ and $d^b_R$ are indeed $K_a^R$ and $K_b^R$ respectively, i.e.,
\begin{equation}
\sum d_Rh_R = d^a_RK_a^R + d^b_RK_b^R.
\end{equation}
Again, this holds for the imaginary parts by just swapping $R$ for $I$.

The same process can be applied for likelihood evaluations for the non-GR polarisation modes. For the `x' and `y'
modes the antenna patterns can be defined as \citep[Equations~32 and 33 of][]{2015PhRvD..91h2002I}
\begin{align}
F_{\text{x}}(\psi,t) &= A_{\text{x}}(t)\cos{\psi} + A_{\text{y}}\sin{\psi}, \nonumber \\
F_{\text{y}}(\psi,t) &= A_{\text{y}}(t)\cos{\psi} - A_{\text{x}}\sin{\psi}.
\end{align}
In this case we find, in an identical fashion to above, that the required coefficients of $A_{\text{x}}$ and $A_{\text{y}}$ are
\begin{align}\label{eq:realcoeffsxy}
K_{A_{\text{x}}}^{R/I} &= C_{R/I,\text{x}}\cos{\psi} - C_{R/I,\text{y}}\sin{\psi}, \nonumber \\
K_{A_{\text{y}}}^{R/I} &= C_{R/I,\text{y}}\cos{\psi} + C_{R/I,\text{x}}\sin{\psi},
\end{align}
where $C_{R/I,\text{x}/\text{y}}$ are the real/imaginary signal amplitude components for the two modes.
Finally, for the breathing and longitudinal modes, where the antenna patterns are defined as \citep[Equations~34 and 35 of][]{2015PhRvD..91h2002I}
\begin{align}
F_{b}(t) &= A_{b}(t), \nonumber \\
F_{l}(t) &= A_{l}(t),
\end{align}
and there is no $\psi$ dependence, the coefficients of $A_{b}$ and $A_l$ are just the real/imaginary signal amplitude coefficients $C_{R/I,b/l}$ for the modes.
\section{Relation between odds and SNRs}\label{app:oddratios}

It is interesting to look at how we might expect odds values (ratios of evidences) to vary with signal-to-noise ratio (SNR) \citep[also
see similar discussions in][]{MaxCWpolariations}. We will take the simple case of two Gaussian
datasets both of the same length ($N = 100$) and standard deviation, $\sigma$. Both datasets contains a mean offset of $\mu_h = 5$, and we are concerned with a
search over the mean parameter. We want to calculate two things: the odds for a mean offset versus the data being Gaussian noise with a known mean of zero; and,
the odds for both datasets containing the same mean offset (a coherent signal) versus them containing two independent mean offsets (an incoherent signal). We can
try and work out what to expect using approximations, and then see if that's actually the case when calculating the values analytically.

To set up the situation, we have the two datasets $d_1$ and $d_2$, which have likelihoods given by
\begin{equation}
 p(d_j|\mu,M,I) = \left(\frac{1}{2\pi\sigma^2}\right)^{(N/2)} \exp{\left(-\frac{\sum_i^N (d_{j,i}-\mu_i)^2}{2\sigma^2}\right)},
\end{equation}
and priors on $\mu$ given by $p(\mu|M,I)$, which we will take as constant (i.e.\ a flat prior). We will also use the fact that $d_{j,i} = \mu_t + n_{j,i}$
where $n_{j,i}$ is a value drawn from a Gaussian distribution of zero mean and standard deviation $\sigma$, and $\mu_t$ is the true offset.

To re-cap from earlier in this paper, the evidence is given by
\begin{equation}
\mathcal{Z}_M = \int_{\mu_{\text{min}}}^{\mu_{\text{max}}} p(d|\mu, M, I) p(\mu|M,I) {\text{d}\mu}
\end{equation}
for hypothesis $M$.

\subsection{Coherent signal vs.\ noise odds}

For large SNRs the relation between signal ($S$) versus noise ($N$) evidence odds for Gaussian noise is well known, but we'll write it out explicitly here.
Formally, for our two data steams, we have
\begin{equation}
 \mathcal{O}_{\text{S}/\text{N}} = \frac{\mathcal{Z}_{S}}{\mathcal{Z}_{N}} = (2\pi\sigma^2)^{-N} \frac{p(\mu)}{\mathcal{Z}_{N}} \int_{\mu_{\text{min}}}^{\mu_{\text{max}}} p(d_1|\mu,I) p(d_2|\mu,I) {\text{d}\mu}.
\end{equation}
For high SNR signals the we find that
\begin{equation}
\int_{\mu_{\text{min}}}^{\mu_{\text{max}}} p(d|\mu,M,I) \text{d}\mu \approx p(d|\mu=\mu_{t/\text{ML}},M,I) \left(\frac{2\pi\sigma^2}{N} \right)^{1/2},
\end{equation}
where $p(d|\mu=\mu_{t/\text{ML}},M,I)$ is the likelihood at the true/maximum likelihood value of $\mu$. This value, for the joint datasets, can be seen to be
\begin{equation}
p(\{d\}|\mu=\mu_{t/\text{ML}},M,I) = \left(\frac{1}{2\pi\sigma^2}\right)^{N} \exp{\left(-\frac{1}{2\sigma^2}\sum_i^N \left[(d_{1,i} - \mu_t)^2 + (d_{2,i} - \mu_t)^2 \right]\right)},
\end{equation}
which in turn (with $d_{j,i} = \mu_t + n_{j,i}$) becomes
\begin{align}
p(\{d\}|\mu=\mu_{t/\text{ML}}, M, I) &= \left(\frac{1}{2\pi\sigma^2}\right)^{N} \exp{\left(-\frac{1}{2\sigma^2}\sum_i^N \left[ n_{1,i}^2 + n_{2,i}^2 \right]\right)} \nonumber \\
&\approx \left(\frac{1}{2\pi\sigma^2}\right)^{N} \exp{(-N)},
\end{align}
where we have used $\sum_i^N n_i^2/\sigma^2 \approx N \pm \sqrt{N}$. Similarly, we have the noise evidence
\begin{align}
\mathcal{Z}_N &= \left(\frac{1}{2\pi\sigma^2}\right)^{N} \exp{\left(-\frac{1}{2\sigma^2}\sum_i^N \left[2\mu_t^2 + (n_{1,i}^2 + n_{2,i}^2) - 2\mu_t(n_{1,i} + n_{2,i}) \right] \right)} \nonumber \\
&\approx 
\left(\frac{1}{2\pi\sigma^2}\right)^{N}\exp{\left(-N - \sum_i^N \frac{\mu_t^2}{\sigma^2} \right)},
\end{align}
where we use the approximation that $\sum_i^N \mu_t n_{j,i} \approx 0$. So, the odds becomes
\begin{equation}
\mathcal{O}_{\text{S}/\text{N}} \approx \left(\frac{2\pi\sigma^2}{N} \right)^{1/2}p(\mu|M,I)\exp{\left(\sum_i^N \frac{\mu_t^2}{\sigma^2}\right)},
\end{equation}
where we can also say that the single dataset SNR is defined as $\rho^2 = \sum_i^N \mu_t^2/\sigma^2 = N\left(\mu_t/\sigma\right)^2$, and in this case
$\rho_{\text{coh}} = \sqrt{2}\rho$, and we can see that if we take the natural logarithm of this that
\begin{equation}
\ln{\mathcal{O}_{\text{S}/\text{N}}} = \ln{\left(p(\mu|M,I)\right)} + \frac{1}{2}\ln{\left(\frac{2\pi\sigma^2}{N}\right)} + \frac{1}{2}\rho_{\text{coh}}^2.
\end{equation}
We see here that the log odds is proportional to $\rho^2$, with a minor influence from the $\ln{\left(2\pi\sigma^2/N\right)}$ term which is roughly equivalent
to $\ln{\left(2\pi\rho^{-2}\right)}$. Empirically (and probably with some mathematical justification that can be derived) this approximation seems to give the
median odds value on random realisations of noise, even down to low SNR (see Figure~\ref{fig:approx_odds}).

\subsection{Coherent signal vs.\ incoherent signal}

We can define a coherent signal versus incoherent signal odds ratio as
\begin{equation}
\mathcal{O}_{\text{S}/\text{I}_{\text{simple}}} = \frac{\mathcal{Z}_{\text{coh}}}{\mathcal{Z}_{\text{incoh}}} = \frac{\mathcal{Z}_{\{d\}}}{\mathcal{Z}_{d_1}\mathcal{Z}_{d_2}},
\end{equation}
where $\mathcal{Z}_{\{d\}}$, $\mathcal{Z}_{d_1}$, and $\mathcal{Z}_{d_2}$, are the evidences for a signal in the full dataset, and two individual
datasets (this is equivalent to Equation~\ref{eq:cohvincoh1}). Using the same approximations as above, we find that
\begin{equation}
\mathcal{O}_{\text{S}/\text{I}_{\text{simple}}} \approx \frac{1}{2\sqrt{\pi}}\frac{1}{p(\mu|M,I)}\frac{\sqrt{N}}{\sigma}.
\end{equation}
In this we see that there is the Occam factor of $1/p(\mu|M,I)$ accounting for the fact that the incoherent model has an extra parameter. We also see that this is
proportional to $\sqrt{N}/\sigma$ which is roughly equivalent to $\rho$.

So we see that for the coherent signal versus noise odds roughly scales with $e^{\rho^2}$, whilst the coherent signal versus incoherent signal odds roughly scales with
$\rho$, i.e.\ its scaling is a lot shallower. Empirically (and probably with some mathematical justification that can be derived) this approximation seems to give the
maximum odds value on random realisations of noise down to low SNR (see Figure~\ref{fig:approx_odds}).

In Figure~\ref{fig:approx_odds} we see how these two approximations compare to analytical calculations of the odds values as the SNR changes. For a range of SNRs
we created multiple realisations of two dataset of Gaussian random noise (with SNR altered by altering the standard deviation of the noise) with a given
mean offset. In each case we have analytically calculated the evidences for a coherent $\mu$ for both datasets, and independent $\mu\text{s}$ in both datasets.
These are then compared to the approximations above. In this case we have purposely chosen a prior on $\mu$ such that the distributions of odds values for
$\mathcal{O}_{\text{S}/\text{I}_{\text{simple}}}$ and $\mathcal{O}_{\text{S}/\text{N}}$ cross, which is similar to the case we observed in our actual analysis shown in \S\ref{sec:simsignal}.

\begin{figure}[!phtb]
\begin{center}
\includegraphics[width=0.5\textwidth]{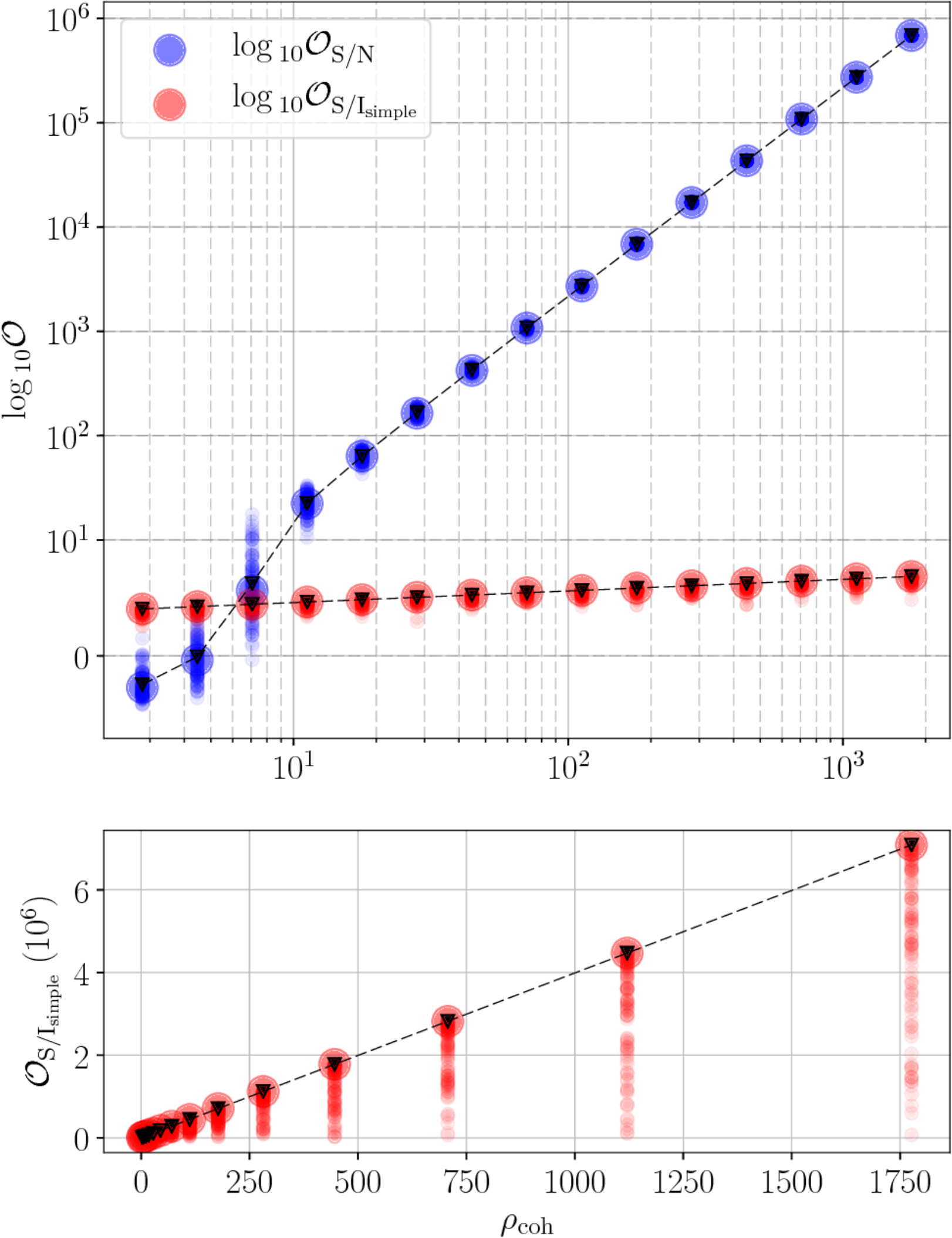}
\caption{ \protect\input{./figures/appendix4/caption}}
\end{center}
\end{figure}

\acknowledgements

We are grateful for computational resources provided by Cardiff University, and funded by an STFC grant (ST/I006285/1) supporting UK Involvement in
the Operation of Advanced LIGO. MP is funded by the STFC under grant number ST/N005422/1. JV is funded by the STFC under grant number ST/K005014/1. The versions of \lppenf and \lppef used in the tests for this document
can be found in the {\tt lalapps\_knope\_O2} branch of the \href{https://github.com/lscsoft/lalsuite/tree/lalapps_knope_O2}{LALSuite git repository} with 
the {\tt git} hash \href{https://github.com/lscsoft/lalsuite/tree/fe9b7d4779cd7fbfbb57af946766601dfd270b23}{{\tt fe9b7d4}}. The various
corner plots, showing posterior parameter distributions, in this document have been produced using \url{https://github.com/mattpitkin/scotchcorner}, and all
plots have used Matplotlib \citet{Hunter:2007,michael_droettboom_2017_248351}.
This is expected
to be a living document that will be updated with any changes to the code. The document has LIGO DCC number \href{https://dcc.ligo.org/P1700086}{LIGO-P1700086}.

%\bibliography{./bibliography/biblio}{}
\bibliography{"main.tex"}{}

\end{document}